\documentclass[aip,jcp,amsmath,reprint,superscriptaddress,a4paper,floatfix,longbibliography]{revtex4-1}
\usepackage[english]{babel}
\usepackage{graphicx}
\usepackage{amssymb}
\usepackage{amsmath}
\usepackage{amsfonts}
\usepackage{color}
\usepackage[T1]{fontenc} 
\usepackage[hidelinks]{hyperref}
\usepackage[update,prepend]{epstopdf}
\usepackage{placeins}
\usepackage{float}
\usepackage{multirow}
\usepackage{import}

\newcommand{\vv}[1]{\mathbf{#1}}
\usepackage[normalem]{ulem}

\newcommand{\la}{\left\langle}
\newcommand{\ra}{\right\rangle}

\newcommand{\disav}[1]{\left\langle #1 \right\rangle_{\text{dis}}}
\newcommand{\ensav}[1]{\left\langle #1 \right\rangle}

\newcommand{\comment}[1]{}

\newcommand{\cob}{\color{black}}

\begin{document}

\author{Yannick Witzky}
\affiliation{Institute of Physics, Johannes Gutenberg University Mainz, Staudingerweg 7-9, 55128 Mainz, Germany}

\author{Friederike Schmid}
\email{friederike.schmid@uni-mainz.de}
\affiliation{Institute of Physics, Johannes Gutenberg University Mainz, Staudingerweg 7-9, 55128 Mainz, Germany}

\author{Arash Nikoubashman}
\email{anikouba@ipfdd.de}
\affiliation{Leibniz-Institut f{\"u}r Polymerforschung Dresden e.V., Hohe Stra{\ss}e 6, 01069 Dresden, Germany}
\affiliation{Institut f{\"u}r Theoretische Physik, Technische Universit{\"a}t Dresden, 01069 Dresden, Germany}
\affiliation{Cluster of Excellence Physics of Life, Technische Universit{\"a}t Dresden, 01062 Dresden, Germany}

\title{From Heteropolymer Stiffness Distributions to Effective Homopolymers. Part 1: Theoretical Modeling and Computational Verification}

\begin{abstract}  
Synthetic copolymers and biopolymers, such as polypeptides and double-stranded DNA, often exhibit strong variations in bending stiffness along their contour, which can significantly impact conformational behavior at larger scales. To investigate these effects, we employ a discretized heterogeneous worm-like chain model, where the local persistence lengths are drawn from a Gaussian distribution. We develop a theoretical model that maps such heterogeneous chains to homogeneous chains with a single effective persistence length. For uncorrelated disorder, our model predicts that this effective stiffness is systematically smaller than the arithmetic mean of the local persistence lengths, indicating that flexible segments have a bigger influence on the overall chain stiffness than rigid segments. We validate our model predictions using off-lattice Monte Carlo simulations, considering both ideal and self-avoiding chains in good solvent, and find excellent agreement in the regime, where the persistence lengths are on the order of a few bond lengths, consistent with typical values observed in polypeptides.
\end{abstract}

\maketitle

\section{Introduction}
Heterogeneous polymers, composed of chemically or structurally distinct subunits, are ubiquitous in both technology and nature. For example, acrylonitrile butadiene styrene (ABS) is a common thermoplastic polymer that combines the chemical resistance of acrylonitrile, the toughness of polybutadiene, and the rigidity of polystyrene.\cite{abs} In biological systems, (disordered) proteins, composed of various amino acids, are a prime example of heterogeneous polymers, where the specific sequence can lead to distinct secondary and tertiary structures (or the lack of them). The effects of chemical interaction heterogeneity on polymer conformations have been studied extensively through experiments and simulations.\cite{lau:mm:1989, khokhlov:prl:1999, ahmad:poly:2011, lin:bpj:2017, perry:ml:2020, statt:jcp:2020, rekhi:jpcb:2023} Likewise, the role of spatial variations in chain stiffness has attracted significant attentions, particularly in the context of biopolymers.\cite{miller:jmb:1967, olson:csb:2000} However, a key question that remains open is whether such heteropolymers with spatially varying bending stiffness can be described as (semiflexible) homopolymers with a single effective persistence length. Although this simplification is clearly inadequate for heteropolymers with very blocky stiffness distribution, e.g., diblock copolymers with one stiff block attached to a flexible block,\cite{david:mm:1995, olsen:mm:2005, olsen:mm:2006, wang:sm:2011} it seems reasonable for more disordered chains.

In this work, we will focus on cases, where the local stiffness (i.e., parameters of the bending potential) exhibits little or no correlation along the chain contour. A biologically relevant example of such heterogeneous polymers is double-stranded DNA, where the local bending stiffness depends on the base-pair sequence.\cite{zhurkin:pnas:1991, bolshoy:pnas:1991, schellman:bpc:1995, olson:csb:2000, beveridge:bpj:2004, lavery:nar:2010} Several theoretical models have been developed in polymer physics to describe chain conformations, and we begin with a concise review of previous efforts, followed by a detailed discussion of our theoretical approach in Sec.~\ref{sec:theory}. In Sec.~\ref{sec:results}, we compare the our theory with simulation data to assess the validity and limitations of our model.

\begin{figure}[htbp]
    \begin{center}
    \includegraphics[width=6cm]{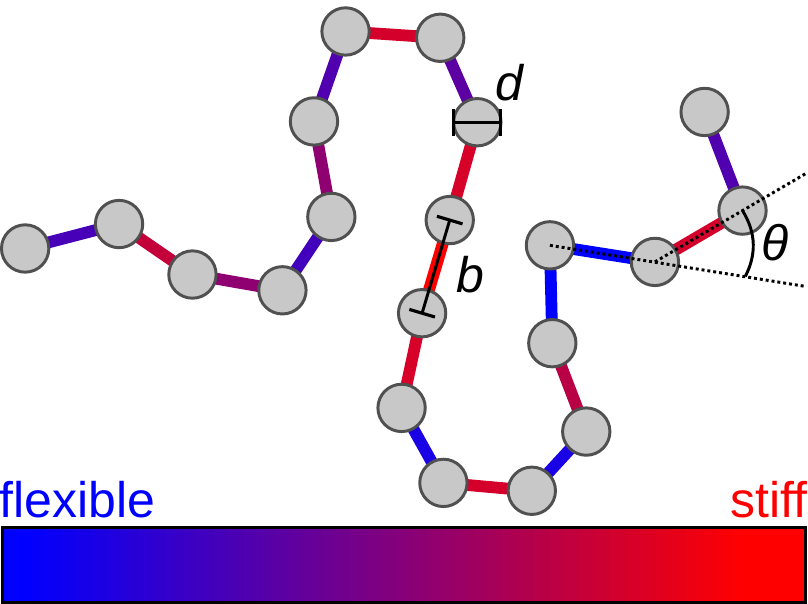}
    \end{center}
    \caption{Schematic representation of a bead-spring polymer model, consisting of $N=20$ spherical monomers of diameter $d$, connected by harmonic springs of rest length $b$. The angle $\theta$ indicates the bending angle between three consecutive beads along the chain, while the segment color indicates the local bending stiffness.}
    \label{fig:schematic} 
\end{figure}

Linear polymers can be modeled as a string of $N$ monomers connected by bonds of length $b$ (see Fig.~\ref{fig:schematic}). We define the (local) persistence length $\ell_\text{p}$ of the polymer through the decay of the orientational correlation\cite{rubinstein:book:2003}
\begin{equation}
    \ensav{\cos(\theta(s))} = \exp(-sb/\ell_\text{p}) ,
    \label{eq:l_p_def}
\end{equation}
where $\theta(s)$ is the angle between unit bond vectors that are $s$ steps apart along the chain contour. For $s=1$, Eq.~\eqref{eq:l_p_def} can be rewritten as
\begin{equation}
    \ell_{\text{p},i} = -\frac{b}{\log\left[\ensav{\cos(\theta_i)}\right]} ,
\end{equation}
with $\theta_i$ being the angle between two subsequent unit bond vectors $\vv{u}_i$ and $\vv{u}_{i+1}$, connecting monomers $i$, $i+1$, and $i+2$ of a chain ($\theta_i = 0$ when the three monomers lie on a line). If $\ell_{\text{p},i}$ is much smaller than the polymer's contour length $L = (N-1)b$, so that the overall chain conformation is still a random walk, then the bending stiffness can also be characterized through the Kuhn length $b_\text{K}$.

The freely rotating chain is the simplest polymer model with non-zero persistence length. For a homopolymer with fixed bending angle $\theta$, the Kuhn length is given by
\begin{equation}
b_\text{K}^\text{hom} = b\frac{1+\cos(\theta)}{1-\cos(\theta)} .
\end{equation}

Flory {\it et al.} determined the Kuhn length for alternating copolymers,\cite{flory:jacs:1952} $b_\text{K}^\text{alt}$, with fixed bond length $b$ and two different bending angles $\theta_1$ and $\theta_2$
\begin{equation}
    b_\text{K}^\text{alt}  = \left[ \frac{1}{2} \left(\frac{1}{b_\text{K}^{(1)}} + \frac{1}{b_\text{K}^{(2)}}\right)\right]^{-1}
    \!\!\!\!\!\!\!
    = b \: \frac{\left[1+\cos(\theta_1)\right]\left[1+\cos(\theta_2)\right]}{1-\cos(\theta_1)\cos(\theta_2)}  .
    \label{eq:bKalt}
\end{equation}
Jung and Depner extended this analysis to block copolymers with blocks of equal length, and to random copolymers, where each bending angle is either $\theta_1$ or $\theta_2$ with equal probability, finding\cite{jung:mc:1991}
\begin{equation}
    b_\text{K}^\text{block}
    =\frac{1}{2} \left(b_\text{K}^{(1)} + b_\text{K}^{(2)}\right)
     = b \:\frac{1-\cos(\theta_1)\cos(\theta_2)}{[1-\cos(\theta_1)][1-\cos(\theta_2)]}
    \label{eq:bBlock}
\end{equation}
and
\begin{equation}
    b_\text{K}^\text{ran} = b\frac{1+\frac{1}{2}\left[\cos(\theta_1)+\cos(\theta_2)\right]}{1-\frac{1}{2}\left[\cos(\theta_1)+\cos(\theta_2)\right]} .
    \label{eq:bKran}
\end{equation}
Thus, the Kuhn length of a random copolymer is identical to that of a freely rotating homopolymer with an effective bending angle defined by $\cos(\theta_\text{eff}) = \frac{1}{2}[\cos(\theta_1) + \cos(\theta_2)]$. By comparing Eqs.~\eqref{eq:bKalt}-\eqref{eq:bKran}, we find $b_\text{K}^\text{block} > b_\text{K}^\text{ran} > b_\text{K}^\text{alt}$ for $\theta_1 \neq \theta_2 < \pi/2$, with the discrepancy between the three lengths increasing as the difference in $\theta_1$ and $\theta_2$ increases.

Long macromolecules with a broad distribution of bending angles can also be described using the worm-like chain (WLC) model, where the polymer conformation is characterized by the tangent vector $\vv{u}(s)$, with $0 \leq s \leq L$ being the position along the contour. Jung and Depner introduced heterogeneity in bending stiffness into the WLC model by assuming a random distribution of persistence lengths,\cite{jung:mc:1991} $\ell_\text{p}(s)^{-1} = \ell_{\text{p},0}^{-1}(1+v(s))$, where $\ell_{\text{p},0}$ is the mean persistence length and $v(s)$ is a Gaussian white noise process with zero mean and variance $\Delta$. Assuming this form of spatially uncorrelated noise, we can calculate the mean-square end-to-end distance of the heteropolymer as
\begin{equation}
    \ensav{R_\text{e}^2} = 2L\ell_\text{p,eff}\left(1-\frac{\ell_\text{p,eff}}{L}\left[1-\exp\left(-\frac{L}{\ell_\text{p,eff}}\right)\right]\right) ,
    \label{eq:Re2WLC}
\end{equation}
which is identical to the standard WLC result, but with effective persistence length
\begin{equation}
    \ell_\text{p,eff} = \left(\frac{1}{\ell_\text{p,0}} - \frac{\Delta}{2\ell_\text{p,0}^2}\right)^{-1} \geq \ell_\text{p,0} .
\end{equation}
Note that Eq.~\eqref{eq:Re2WLC} differs slightly from Eq. (34) in Ref.~\citenum{jung:mc:1991}, as we did not use a first-order power series approximation in $\Delta$. Nevertheless, both expressions exhibit the same qualitative trend, where bending stiffness heterogeneity \textit{increases} the overall chain stiffness. 

Bensimon {\it et al.} also studied the effect of bending heterogenity in the WLC model, but they used a different approach, where they introduced a preferred random orientation between successive segments $i$ and $i+1$ along the chain.\cite{bensimon:epl:1998} Their analysis also showed that the overall elastic response of such a disordered chain resembles that of a homogeneous WLC, but with an effective elastic constant that \textit{decreases} as the degree of disorder increases. This theoretical prediction aligns with experimental and simulation studies of double-stranded DNA, which observed a (modest) reduction in the overall persistence length with increasing local variations in bending stiffness.\cite{olson:csb:2000}

\section{Theory}
\label{sec:theory}
We consider ideal (i.e., non-interacting) chains of length $N$ with fixed bond length $b$ and the angular potential
\begin{equation}
    \beta U_\text{bend}(\kappa_i) = \kappa_i (1- \vv{u}_i \cdot \vv{u}_{i+1}) = \kappa_i\left[1-\cos(\theta_i)\right] ,
    \label{eq:Ubend}
\end{equation}
where $\beta = 1/(k_\text{B}T)$ and $\kappa_i$ controls the bending stiffness between the unit bond vectors $\vv{u}_i$ and $\vv{u}_{i+1}$ of the polymer. $k_\text{B}$ is Boltzmann's constant, and $T$ is the temperature. In our model, the bending potential parameters $\kappa_i$ are Gaussian distributed
\begin{equation}
    P(\kappa) = \frac{1}{\sqrt{2 \pi} \sigma_\kappa} \exp \left[ - \frac{(\kappa - \kappa_0)^2}{2 \sigma_\kappa^2} \right] ,
    \label{eq:Pkappa}
\end{equation}
with mean $\kappa_0$ and variance $\sigma_\kappa^2$. As discussed above, we assume the different angle potential parameters to be uncorrelated, that is, $\ensav{\kappa_i \kappa_j} - \kappa_0^2 = \delta_{ij} \sigma_\kappa^2$. 

The partition function of an ideal chain with a fixed set of $\kappa_i$ is given by
\begin{equation}
    \label{eq:Z}
    \mathcal{Z} = \int \exp\left[-\sum_i^{N-1} \beta U_\text{bend}(\kappa_i)\right] \text{d}\mathcal{R},
\end{equation}
where $\int\text{d}\mathcal{R} = \int\text{d}^2u_1 \dots \text{d}^2u_{N-1}$ denotes the integration over all degrees of freedom, since the bond lengths are fixed. Further, the disorder average of an observable $O$ is then generally defined as
\begin{eqnarray}
     &&\disav{O} = \int  \text{d} \kappa_1 \dots \text{d} \kappa_{N-1}
     P(\kappa_1) \dots P(\kappa_{N-1}) 
     \nonumber \\ &&
     \times \int O \:
       \frac{\exp\left[- \sum_i^{N-1} \beta U_\text{bend}(\kappa_i)\right] }{\mathcal{Z}}\text{d}\mathcal{R} .
     \label{eq:dis_avg}
\end{eqnarray}
Leveraging that the bending potentials along the chain are independent of each other, Eqs.~\eqref{eq:Z} and \eqref{eq:dis_avg} can be factorized, yielding, e.g.
\begin{eqnarray}
    \nonumber
    \mathcal{Z} &=& \prod_i^{N-1} \int \exp\left[-\kappa_i(1-\cos(\theta))\right]\sin(\theta)\text{d}\theta\\
    &=& \prod_i^{N-1} \frac{1-\exp(-2\kappa_i)}{\kappa_i} .
\end{eqnarray}
The Helmholtz free energy of the system, $F \equiv - k_\text{B}T\log(\mathcal{Z})$, then becomes
\begin{equation}
    F = -k_\text{B}T \sum_i^{N-1} \log\left[\frac{1-\exp(-2\kappa_i)}{\kappa_i}\right] .
\end{equation}
As the system is made of independent subsystems, we can consider the Helmholtz free energy of a single bond angle, $f$, for simplifying the calculation
\begin{equation}
    f(\kappa) = -k_\text{B}T \log\left[\frac{1-\exp(-2\kappa)}{\kappa}\right]
    \label{eq:f}
\end{equation}
with disorder average
\begin{equation}
    \disav{f(\kappa)} =  \int P(\kappa) f(\kappa) \text{d}\kappa 
    \label{eq:disav_f}
\end{equation}
after inserting Eq.~\eqref{eq:f} into Eq.~\eqref{eq:dis_avg}. By expanding $f(\kappa)$ around $\kappa = \kappa_0 + \xi_i$ up to second order and invoking the Gaussian integral identities
\begin{equation}
\int_{-\infty}^\infty P(\kappa) (\kappa-\kappa_0)^{n} \text{d}\kappa = 
\begin{cases}
0 & \text{, $n$ is odd}\\
1 & \text{, $n=0$}\\
\sigma_\kappa^2 & \text{, $n=2$}
\end{cases}
\end{equation}
Eq.~\eqref{eq:disav_f} can be solved
\begin{equation}
    \disav{f(\kappa)} \approx f(\kappa_0) + \frac{\sigma_\kappa^2}{2}f''(\kappa_0) ,
    \label{eq:disav_f_taylor}
\end{equation}
with derivatives $f'= \text{d}f/\text{d}\kappa$ and $f''= \text{d}^2f/\text{d}\kappa^2$. Now, we construct a (semiflexible) homopolymer with effective bending stiffness parameter $\kappa_\text{eff}$ that has the same free energy, i.e., $\disav{f(\kappa)} = f(\kappa_\text{eff})$. Using $f(\kappa_\text{eff}) \approx f(\kappa_0) + (\kappa_\text{eff}-\kappa_0)f'(\kappa_0)$ with Eq.~\eqref{eq:disav_f_taylor} results in
\begin{equation}
    \kappa_\text{eff} = \kappa_0 - \frac{\sigma_\kappa^2}{2}h(\kappa_0)
    \label{eq:kappa_eff}
\end{equation}
with
\begin{equation}
    \label{eq:h}
    h(\kappa) = -\frac{f''(\kappa)}{f'(\kappa)} = \frac{1 - \kappa^2/\sinh^2(\kappa)}{\kappa + \kappa^2[1 - \coth(\kappa)]} .
\end{equation}

The function $h(\kappa)$ is strictly positive for $\kappa > 0$, reaching a maximum value of $h(\kappa) \approx 0.4$ at $\kappa \approx 1.18$. Its asymptotic behavior is given by
\begin{equation}
    h(\kappa) \approx
    \begin{cases}
        \frac{1}{3} + \frac{\kappa}{9} & (\kappa \ll 1) \\
        \frac{1}{\kappa}  & (\kappa \gg 1)
    \end{cases} .
 \end{equation}
Thus, the disorder-averaged effective stiffness parameter $\kappa_\text{eff}$ of a heteropolymer is always smaller than the mean of the underlying distribution $P(\kappa)$. Note that Eq.~\eqref{eq:kappa_eff} can yield negative values of $\kappa_\text{eff}$ for $\kappa_0 < \sigma_\kappa^2 h(\kappa_0)/2$, i.e., broad $P(\kappa)$ centered around small $\kappa_0$. {\cob This inequality delineates the regime where our calculation breaks down due to limitations of the second-order expansion of $f(\kappa)$; a fully self-consistent solution to $f(\kappa_\text{eff}) = \int P(\kappa)f(\kappa)\text{d}\kappa$, with $P(\kappa)$ restricted to positive values of $\kappa$, will always result in positive $\kappa_\text{eff}$, but generally requires numerical evaluation of the integral.}

We can rephrase these results in terms of heterogeneously distributed persistence lengths $\ell_{\text{p},i} \equiv \ell_\text{p}(\kappa_i)$ with
\begin{equation}
\label{eq:lp_kappa}
    \ell_\text{p}(\kappa) = - \frac{b}{\ln\left[\coth(\kappa) - 1/\kappa\right]},
\end{equation}
(see SI) which has a distribution $P(\ell_\text{p})$ with average $\disav{\ell_\text{p}}$ and variance $\sigma_\text{p}^2$. After some further calculations (see SI), we finally obtain an expression for the effective persistence length $\ell_\text{p,eff}$ as a function of  $\disav{\ell_\text{p}}$ and $\sigma_\kappa^2$ :
\begin{equation}
 \ell_\text{p,eff} = \disav{\ell_\text{p}}- \frac{\sigma_\kappa^2}{2} \big(
\ell_\text{p}'(\kappa_0) \: h(\kappa_\textrm{0}) + \ell_\text{p}''(\kappa_0) \big) ,\label{eq:lpeffpre_kappa} 
\end{equation}
with derivatives $\ell_\text{p}' = \text{d}\ell_\text{p}/\text{d}\kappa$ and $\ell_\text{p}'' = \text{d}^2\ell_\text{p}/\text{d}\kappa^2$.
Here, $\kappa_0$ depends on $\disav{\ell_\text{p}}$ and $\sigma_\text{p}$ via the implicit equation
\begin{equation}
    \disav{\ell_\text{p}} = \ell_\text{p}(\kappa_0) + \frac{1}{2} \frac{\ell_\text{p}''(\kappa_0)}{\ell_\text{p}'(\kappa_0)^2} \: \sigma_\text{p}^2.
    \label{eq:l_p0_implicit}
\end{equation}

Note that the expression for $\ell_\text{p,eff}$ in Eq.~\eqref{eq:lpeffpre_kappa} depends on the variance $\sigma_\kappa^2$ of the angle potential parameter distribution $P(\kappa)$, which might not always be directly accessible. In such cases, it is convenient to replace $\sigma_\kappa^2$ with the corresponding variance of the persistence length, $\sigma_\text{p}^2$. By using the relation $\sigma_\text{p}^2 = \ensav{\ell_\text{p}^2} - \ensav{\ell_\text{p}}^2 \approx \ell_\text{p}'(\kappa_0)^2 \sigma_\kappa^2$, which has been evaluated up to the order of $\sigma_\kappa^2$ (see SI), we can rewrite Eq.~\eqref{eq:lpeffpre_kappa} as
\begin{eqnarray}
    \ell_\text{p,eff} & = & \disav{\ell_\text{p}} - \frac{\sigma_\text{p}^2}{2} \;
   \left(\frac{h(\kappa_0)}{\ell_\text{p}'(\kappa_0)} + \frac{\ell_\text{p}''(\kappa_0)}{\ell_\text{p}'(\kappa_0)^2}
   \right) \label{eq:lpeffpre} \\
   & = & \ell_\text{p}(\kappa_0)- \frac{\sigma_\text{p}^2}{2} \: \frac{h(\kappa_0)}{\ell_\text{p}'(\kappa_0)}.
\end{eqnarray}

In the limit of large  persistence lengths, $\disav{\ell_{\text{p}}} > 4\,b$, which corresponds to $\kappa_0 > 4$, Eq.~\eqref{eq:lpeffpre} simplifies to
\begin{equation}
    \ell_\text{p,eff} \approx \disav{\ell_\text{p}} - \frac{\sigma_\text{p}^2}{2 \la \ell_\text{p} \ra + b}.
    \label{eq:lpeff}
\end{equation}

The above considerations assume chains to be ideal. Alternatively, we can address the problem more generally by mapping the ensemble of disorder averaged chains onto an ensemble of reference homopolymers using the so-called replica trick.\cite{chen_2000_localization, Denesyuk_2000_adsorption, Polotsky_2004_polymer} This approach is also valid for self-avoiding (disordered and reference) chains and yields the same equations. Details can be found in SI.

\begin{figure}[htb]
    \centering
    \includegraphics[width=7cm]{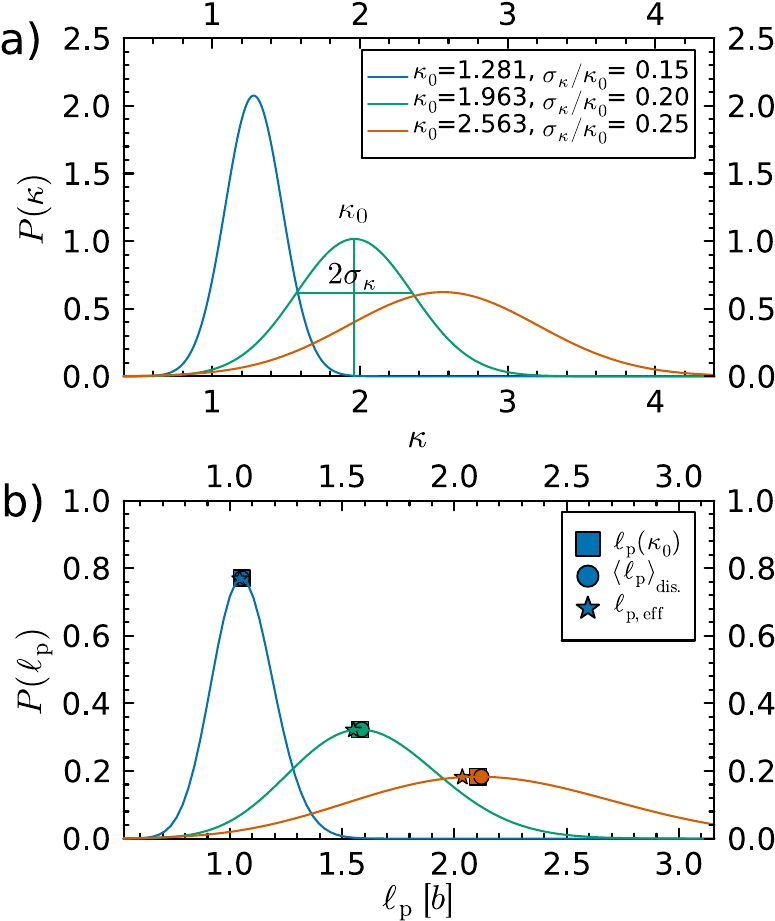}
    \caption{(a) Probability distributions of the bending potential parameter, $P(\kappa)$, for selected mean $\kappa_0$ and standard deviations $\sigma_\kappa$. (b) Corresponding distributions of the persistence length, $P(\ell_\text{p})$.}
     \label{fig:Pdistrib}
\end{figure}

Figure~\ref{fig:Pdistrib} shows $P(\kappa)$ and the resulting $P(\ell_\text{p})$ distributions for three different combinations of $\kappa_0$ and $\sigma_\kappa$. Here, we have also marked the locations of the persistence length taken at the center of $P(\kappa)$, $\ell_\text{p}(\kappa_0)$, the disorder-average persistence length $\disav{\ell_\text{p}}$, and the effective persistence length $\ell_\text{p,eff}$. For narrow distributions, these three values coincide, as expected from Eqs.~\eqref{eq:lpeffpre_kappa} and \eqref{eq:lpeff}. With increasing $\sigma_\kappa$, however, $\ell_\text{p,eff}$ becomes distinctly smaller than $\disav{\ell_\text{p}}$, while $\ell_\text{p}(\kappa_0)$ lies between these two values. 

\begin{figure}[htb]
    \centering
    \includegraphics[width=8cm]{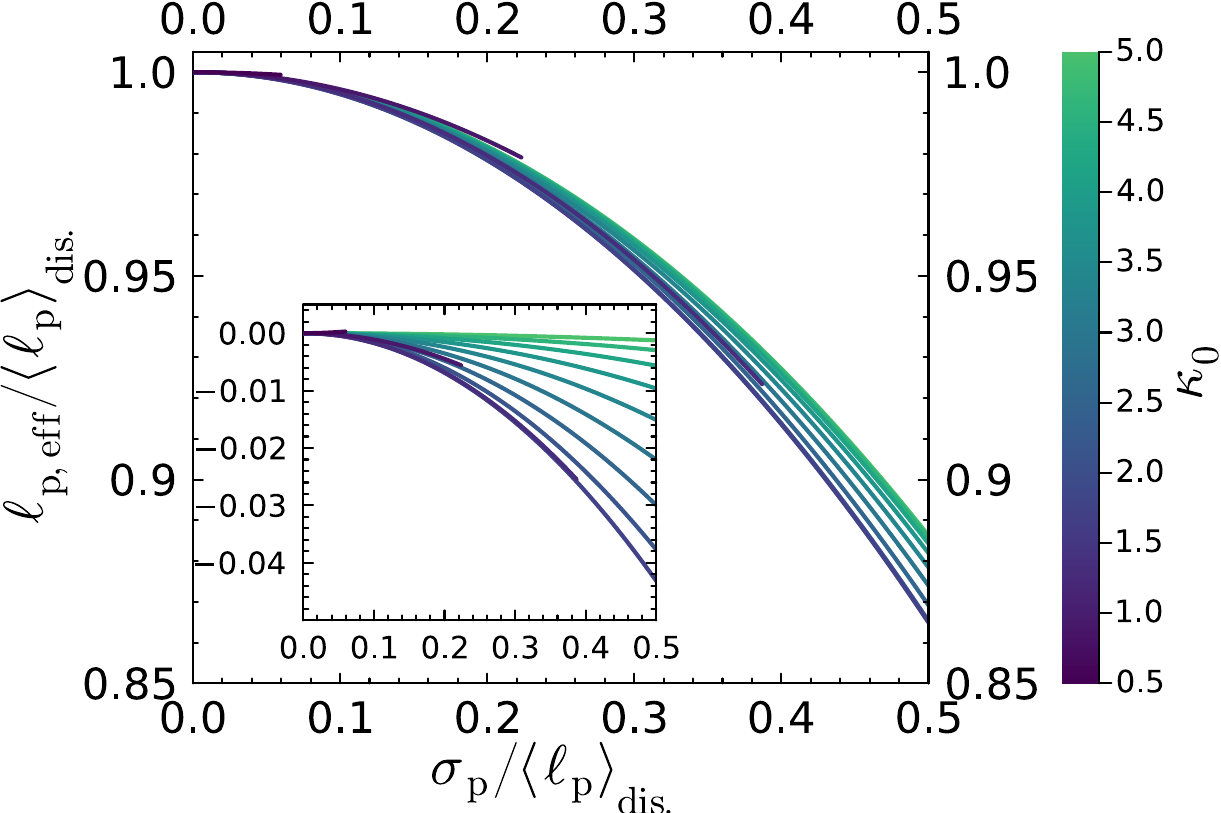}
    \caption{Predicted effective persistence length $\ell_\textrm{p,eff}$ according to Eq.~\eqref{eq:lpeffpre} normalized by the average persistence length $\disav{\ell_\textrm{p}}$  versus reduced standard deviation $\sigma_\text{p}/\disav{ \ell_\text{p}}$ for varying means $\kappa_0$ (color coding). The inset shows the difference between predictions of $\ell_\text{p,eff}$ from the full theory, Eq.~\eqref{eq:lpeffpre}, and from the linearized theory, Eq.~\eqref{eq:lpeff}, normalized by $\disav{\ell_\textrm{p}}$.}
     \label{fig:lp_eff}
\end{figure}

Figure~\ref{fig:lp_eff} shows the ratio between the effective persistence length $\ell_\text{p,eff}$ and the (disorder) average persistence length $\disav{\ell_\text{p}}$ for different means $\kappa_0$ of the $P(\kappa)$ distribution. Except for very flexible chains [i.e., for all $\kappa_0 >0.343$ according to Eq.~\eqref{eq:lpeffpre_kappa}], the effective persistence length $\ell_\text{p,eff}$ is smaller than the simple disorder averaged value $\disav{\ell_\text{p}}$. Intuitively, this result aligns with the notion that flexible bonds have a stronger influence on the overall persistence length of chains than stiff bonds. For example, in a chain with alternating stiff ($\ell_{\text{p}, i} = \infty$) and flexible bonds ($\ell_{\text{p}, i} \approx b$), one expects the effective persistence length to be closer to $\ell_{\text{p,eff}} \approx 2\,b$ than to $\ell_{\text{p,eff}} \approx \infty$. Furthermore, the theory predicts that the leading correction term is proportional to the variance $\sigma_\text{p}^2$ of the persistence length distribution $P(\ell_\text{p})$. 

Using the reference homopolymer approach (see SI), we can also study the effect of correlations between bending parameters $\kappa_i$ or persistence lengths $\ell_{\text{p},i}$ along the chain. As shown in the SI, these correlations have the effect of renormalizing the function $h(\kappa)$. In the presence of correlations, Eq.~\eqref{eq:h} must be replaced by
\begin{equation}
    \label{eq:h_corr}
    h(\kappa) = \frac{1 - \kappa^2/\sinh^2(\kappa)}{\kappa + \kappa^2[1 - \coth(\kappa)]} - c_\text{eff} \left(1+\frac{1}{\kappa} - \coth(\kappa)\right),
\end{equation}
where the correlation factor $c_\text{eff}$ is defined as
\begin{eqnarray}
\nonumber
c_\text{eff} & = &\frac{1}{N} \sum_{ij} \frac{\disav{\kappa_i \kappa_j} - \kappa_0^2}{\sigma_\kappa^2} - 1
\\
&\approx& \frac{1}{N} \sum_{ij} \frac{\disav{\ell_{\text{p},i} \ell_{\text{p},j}} - \disav{\ell_\text{p}}^2}{\sigma_\text{p}^2} - 1.
\end{eqnarray}
For uncorrelated stiffness parameters, $c_\text{eff}=0$. A value $c_\text{eff}>0$ indicates blockiness, and a value $c_\text{eff}<0$ a preference for alternating stiffness. Looking at the limiting behaviors of the modified function $h(\kappa)$, 
\begin{equation}
 h(\kappa) \approx \left\{
 \begin{array}{ll}
 \frac{1}{3} - c_\text{eff} + \kappa
    (\frac{1}{9}+\frac{1}{3} c_\text{eff}) & (\kappa  \ll 1) \\
 \frac{1}{\kappa} (1- c_\text{eff}) & (\kappa \gg 1) \: ,
 \end{array}
 \right.
 \end{equation}
one can see that $h(\kappa)$ decreases with increasing $c_\text{eff}$ and  becomes negative for large positive $c_\text{eff}$. For blocky heteropolymers, the effective persistence length [Eq.~\eqref{eq:lpeffpre}] therefore increases and may even exceed the mean persistence length. By contrast, a correlation bias towards alternating sequences amplifies the trend that heterogeneity effectively softens a chain. We note that these results are in qualitative agreement with the results of Jung and Depner for heterogeneous freely rotating chains with two rotation angles\cite{jung:mc:1991} discussed earlier.

In the following, we will focus on heteropolymers with uncorrelated bending parameters.

\section{Simulation model and analysis procedure}
\label{sec:model}
To evaluate the accuracy of our theory, we performed Monte Carlo (MC) simulations using a bead-spring model. We simulated heteropolymers with a locally varying stiffness distribution $P(\kappa)$ [Eq.~\eqref{eq:Pkappa}] as well as homopolymers with effective persistence length $\ell_{\text{p,eff}}$ determined via Eq.~\eqref{eq:lpeffpre}. Our analytical approach assumes that the polymers are ideal, which is a reasonable approximation when the (average/effective) persistence length is comparable to or exceeds the polymer's contour length.\cite{nikoubashman:jcp:2016} To evaluate the effect of this approximation, we conducted simulations with and without excluded volume interactions. 

All chains consist of $N=100$ monomers with a fixed bond length of $b$, while bending rigidity is imparted using the potential $U_\text{bend}$ [Eq.~\eqref{eq:Ubend}]. Excluded volume interactions between monomers are included through the purely repulsive Weeks-Chandler-Andersen (WCA) potential\cite{weeks:jcp:1971}
\begin{equation}
    U_\text{WCA}(r) = \begin{cases}
    4\varepsilon \left[\left(\frac{d}{r}\right)^{12}-\left(\frac{d}{r}\right)^6\right] + \varepsilon,& r \le 2^{1/6} d \\
    0,& {\rm otherwise}
\end{cases},
\label{eq:Uwca}
\end{equation}
with interaction strength $\varepsilon=k_\text{B}T$ and bead diameter $d = b$. For homopolymers and heteropolymers, all monomer masses are taken to be equal.

Chain configurations were generated using Rosenbluth sampling.\cite{frenkelSmit, rosenbluthMonteCarloCalculation1955} Here, we drew the angle $\theta$ between subsequent bonds randomly according to their corresponding Boltzmann weight [{\it cf.} Eq.~\eqref{eq:Ubend}]. Torsion angles were drawn randomly from a uniform distribution in the interval $[0, 2\pi]$. Following this procedure, we created 32 trial positions for each bead, computed the energy due to excluded volume interactions between non-bonded beads, and chose a trial position accordingly.

We characterize the shape of the simulated polymers by calculating their gyration tensor
\begin{equation}
   \mathbf{G} = \frac{1}{N}\sum_{i=1}^{N} (\mathbf{r}_i-\mathbf{r}_\text{cm})(\mathbf{r}_i-\mathbf{r}_\text{cm})^\text{T},
   \label{eq:Rg2}
\end{equation}
where the sum runs over all $N$ monomers. The eigenvalues $\lambda_1 \leq \lambda_2 \leq \lambda_3$ of the gyration tensor $\mathbf{G}$ characterize the quadratic extension of the chain along its three principal axes, with $R_\text{g} = \left(\lambda_1 + \lambda_2 + \lambda_3\right)^{1/2}$. Note that in what follows, we report mean values of $R_\text{g}$ and $R_\text{ee}$, following the convention used in Ref.~\citenum{Calvados2}, rather than the root-mean-square values $\ensav{O^2}^{1/2}$. Further, disorder averages $\disav{O}$ of an observable $O$ incorporate both an ensemble average over many configurations with the same specific $\kappa_i$ sequence along the heteropolymer and an additional average over a set of $\kappa_i$ sequences sampled from a given probability distribution $P(\kappa)$.

In addition, we can use the eigenvalues of $\mathbf{G}$ to describe the symmetry properties of the polymers:
\begin{align}
    \text{AS} &= \la \left(\lambda_3-\frac{\lambda_1+\lambda_2}{2}\right) / (\lambda_1+\lambda_2+\lambda_3)\ra \label{eq:AS}\\
    \text{AC} &= \la (\lambda_2-\lambda_1)/ (\lambda_1+\lambda_2+\lambda_3)\ra \label{eq:AC}\\
    \text{A} &= \la 1-3\frac{\lambda_1\lambda_2 + \lambda_1\lambda_3 + \lambda_2\lambda_3}{(\lambda_1+\lambda_2+\lambda_3)^2} \ra \label{eq:A}. 
\end{align}
The descriptor $\text{AS} \geq 0$ quantifies the asphericity of a polymer, becoming zero when the three principal moments are identical ($\lambda_1 = \lambda_2 = \lambda_3$). This occurs not only when the distribution of monomers is spherically symmetric around the center of mass, but also when the distribution is symmetric across the three coordinate axes -- for example, if the particles are uniformly positioned on a cube, tetrahedron, or another Platonic solid. Similarly, $\text{AC} \geq 0$ characterizes the  acylindricity of a polymer, which becomes zero for a cylindrical monomer distribution, i.e., $\lambda_1 = \lambda_2$. The relative shape anisotropy parameter $\text{A}$ is bounded between 0 and 1, which correspond to the limits of a perfect sphere and an infinitely thin rod, respectively. To establish reference points, we determined these shape descriptors for freely jointed ideal chains ($N=100$) using additional MC simulations. Here, we found $\text{AS}_\text{id} = 0.588$, $\text{AC}_\text{id} = 0.120$, and $\text{A}_\text{id} = 0.392$, which agree well with previous theoretical and numerical estimates.\cite{rudnickShapesRandomWalks1987} The corresponding values for self-avoiding random walks (SAW) are $\text{AS}_\text{SAW} = 0.653 $, $\text{AC}_\text{SAW} = 0.120 $, and $\text{A}_\text{SAW} = 0.471$.

\section{Simulation results}
\label{sec:results}
Our primary objective is to investigate whether heteropolymers with locally varying stiffness can be effectively described as homopolymers characterized by a single persistence length. To answer this question, we performed coarse-grained simulations of generic worm-like chains (Sec.~\ref{sec:model}) at infinite dilution, incorporating both ideal and excluded volume interactions. Reference configurations of homogeneous worm-like chains ($N=100$ monomers) were generated for over 600 values of $\ell_\text{p}$, ranging from $0.5\,b$ to $160\,b$ with varying step size. We considered chains with and without excluded volume interactions, and created $10^6$ independent polymer configurations for each value of $\ell_\text{p}$ (see Sec.~\ref{sec:model} for details). 

\begin{figure}[htb]
    \includegraphics[width=8cm]{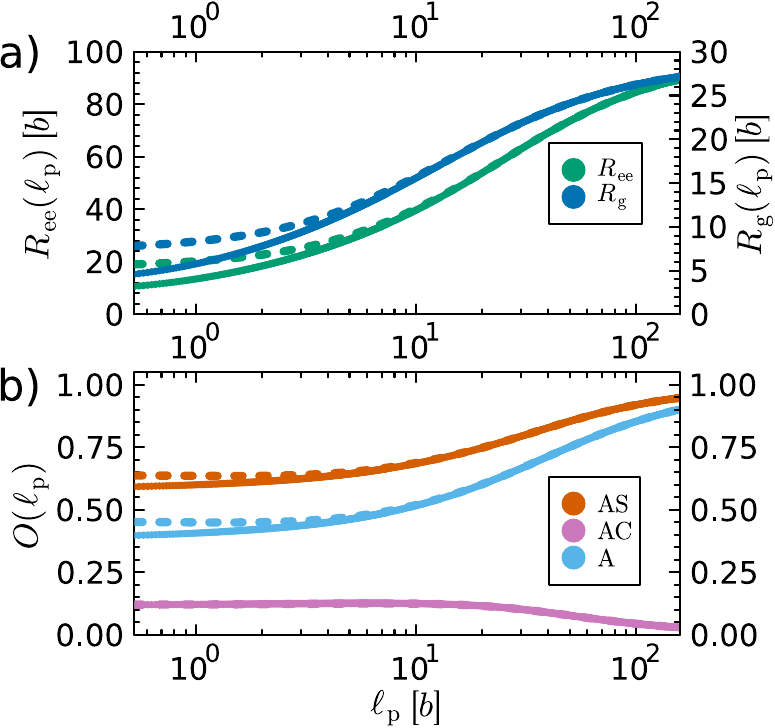}
\caption{(a) Statistically averaged end-to-end distance $R_\text{ee}$ (left $y$-axis) and radius of gyration $R_\text{g}$ (right $y$-axis), as well as (b) asphericity $\text{AS}$, acylindricity $\text{AC}$, and shape anisotropy $\text{A}$ for homogeneous worm-like chains with $N=100$ beads as a function of persistence length $\ell_\text{p}$. The solid and dashed lines show results for chains without and with excluded volume interactions, respectively.} 
    \label{fig:MappingCurves}
\end{figure}

Figure~\ref{fig:MappingCurves} shows the configurational properties of these homogeneous worm-like chains as functions of persistence length $\ell_\text{p}$. For small $\ell_\text{p}$, excluded volume interactions between monomers cause chain swelling, which is reflected by the correspondingly larger $R_\text{ee}$ and $R_\text{g}$ values (Fig.~\ref{fig:MappingCurves}a). With increasing $\ell_\text{p}$, the polymers adopt more elongated configurations, consistent with Eq.~\eqref{eq:Re2WLC}, and excluded volume interactions become less important.\cite{nikoubashman:jcp:2016} For large $\ell_\text{p}$, $R_\text{ee}$ and $R_\text{g}$ approach the theoretical limit of a straight rod, i.e., $R_\text{ee} = (N-1)b = 99\,b$ and $R_\text{g} = \sqrt{(N^2-1)/12}\,b \approx 29\,b$. Similar trends are observed for the symmetry properties of the polymers (Fig.~\ref{fig:MappingCurves}b); as the bending stiffness increases, the asphericity $\text{AS}$ increases from $\text{AS}_\text{id} = 0.588$ to $0.95$, while the acylindricity $\text{AC}$ approaches its asymptotic limit of 0 slowly. Likewise, the relative shape anisotropy parameter $\text{A}$ starts from $A_\text{id} = 0.392$ for fully flexible chains and slowly increases toward its maximum value of one.

Having examined how the conformational properties of homogeneous worm-like chains change with increasing bending stiffness, we now compare them with the disorder-averaged properties of heteropolymers. To sample the heteropolymer configurations, we drew $\kappa_i$ values from $P(\kappa)$ within a $5\,\sigma_\kappa$-interval (excluding negative $\kappa_i$ values) according to Eq.~\eqref{eq:Pkappa}, with mean values $\kappa_0$ such that $1.1\,b \lesssim \ell_\textrm{p}(\kappa_0) \lesssim 6.6\,b$. The standard deviations were set to $\sigma_\kappa/\kappa_0 \in \{0.01, 0.025, 0.05, 0.075, 0.1, 0.125, 0.15, 0.175, 0.2\}$. For each combination of $\kappa_0$ and $\sigma_\kappa$, we generated 100 heteropolymers with $2 \times 10^6$ configurations each. We increased the statistics for heteropolymers with $\ell_\textrm{p}(\kappa_0) < 3\,b$ that include excluded volume interactions to $1.6 \times 10^7$ configurations per heteropolymer. 

\begin{figure*}[htb]
      \centering
    \includegraphics[width=\textwidth]{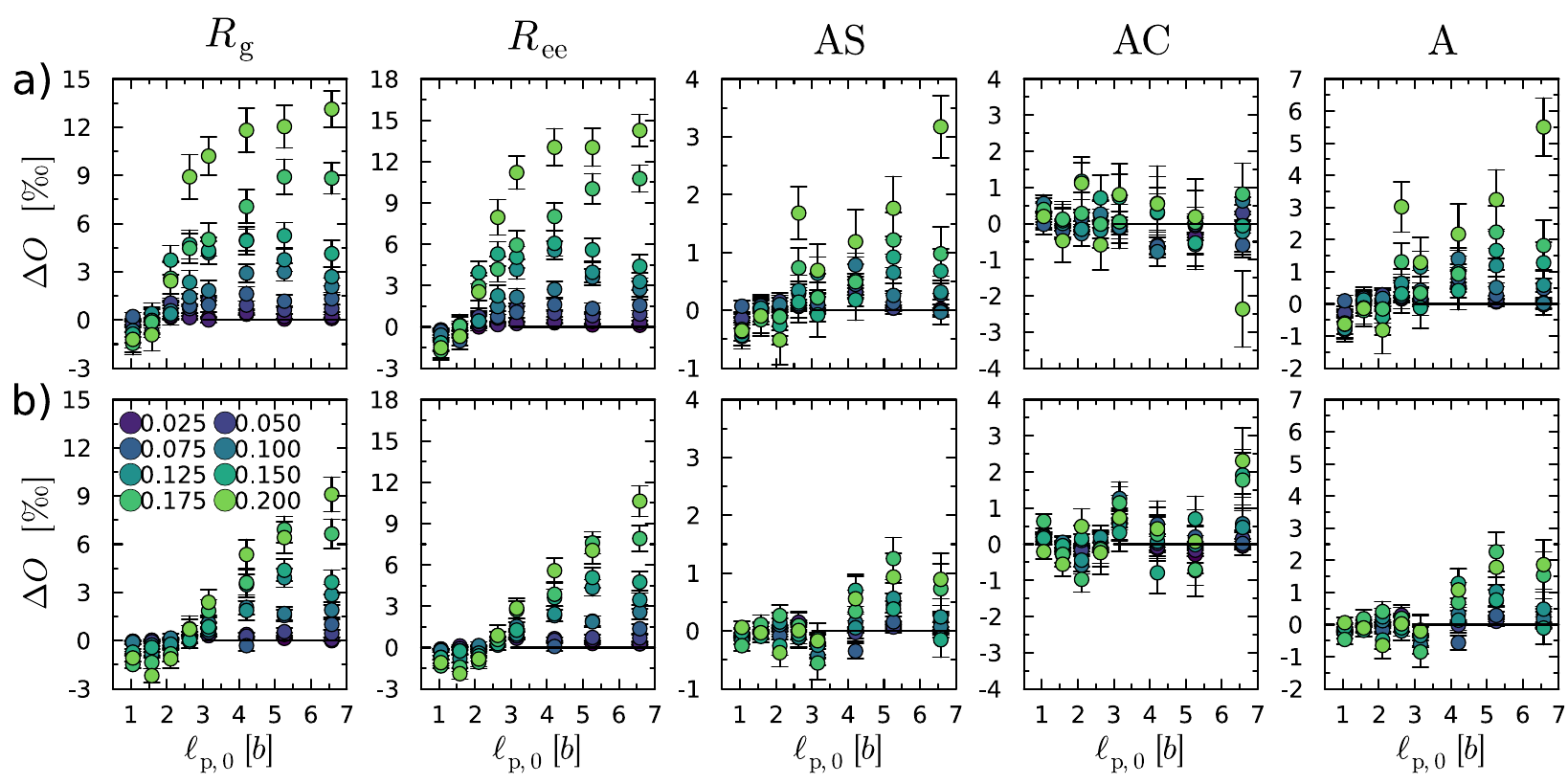}
    \caption{Relative difference $\Delta O$ between the disorder-averaged configurational properties of the heteropolymers and the ensemble-averaged properties of homopolymers with effective persistence length $\ell_\text{p,eff}$. Panels (a) and (b) show data for chains without and with excluded volume interactions, respectively. The angle potential parameters $\kappa_i$ of the heteropolymers are Gaussian distributed with mean $\kappa_0$ and variance $\sigma_\kappa^2$ (see main text), resulting in $\ell_\textrm{p,0}=\ell_\textrm{p}(\kappa_0)$. The symbols are colored according to the values of $\sigma_\kappa/\kappa_{0}$ as indicated. Error bars indicate the standard error of the mean, determined from 100 different realizations of the disorder.}
    \label{fig:Comparison}
\end{figure*}

Figure~\ref{fig:Comparison} shows the relative difference $\Delta O \equiv \disav{O}/\ensav{O} - 1$ between the disorder-averaged configurational properties $\disav{O}$ of the heteropolymers and the ensemble-averaged configurational property $\ensav{O}$ of homopolymers with an effective persistence length $\ell_\text{p,eff}$ calculated via Eq.~\eqref{eq:lpeffpre_kappa}. We consider both ideal chains (Fig.~\ref{fig:Comparison}a) and self-avoiding chains in good solvent (Fig.~\ref{fig:Comparison}b). For narrow $P(\kappa)$ distributions, we find excellent agreement between the two descriptions, with $\Delta O \lesssim 5\,\text{\textperthousand}$ for $\sigma_\kappa/\kappa_0 \leq 0.1$. The relative deviations of some observables increase to approximately 1\,\% for the largest $\ell_\text{p,0}$ and $\sigma_\kappa/\kappa_0$ values, which is still quite small given the rather large heterogeneity in bending stiffness for these cases. In general, the theory overestimates the polymer size for stiffer chains, since stiff segments tend to dominate the global chain extension. Interestingly, the relative deviations between homopolymers and heteropolymers are smaller for self-avoiding chains than for ideal chains, even though our theory neglects excluded volume interactions.

As a complementary test of our theory, we determine the effective persistence length, $\ell_\text{p,eff}$, by matching the statistical average of a selected configurational property $\ensav{O}$ with the joint disorder average $\disav{O}$ of the same property in the sampled heteropolymer set. This inverse procedure is well-defined if the target property $\ensav{O}$ is a monotonic function of $\ell_\text{p}$ in homopolymers. The acylindricity AC is nonmonotonic for small $\ell_\text{p}$, making it unsuitable for matching. Similarly, the asphericity $\text{AS}$ and relative shape anisotropy $\text{A}$ vary only slightly over the whole range of $\ell_\text{p}$, preventing reliable back-mapping (Fig.~\ref{fig:MappingCurves}b). In contrast, both the radius of gyration $R_\text{g}$ and the end-to-end distance $R_\text{ee}$ increase monotonically with increasing bending stiffness $\ell_\text{p}$ and span a broad range of values, making them well-suited for this procedure (Fig.~\ref{fig:MappingCurves}a).

\begin{figure}[htb]
    \centering
    \includegraphics[width=8cm]{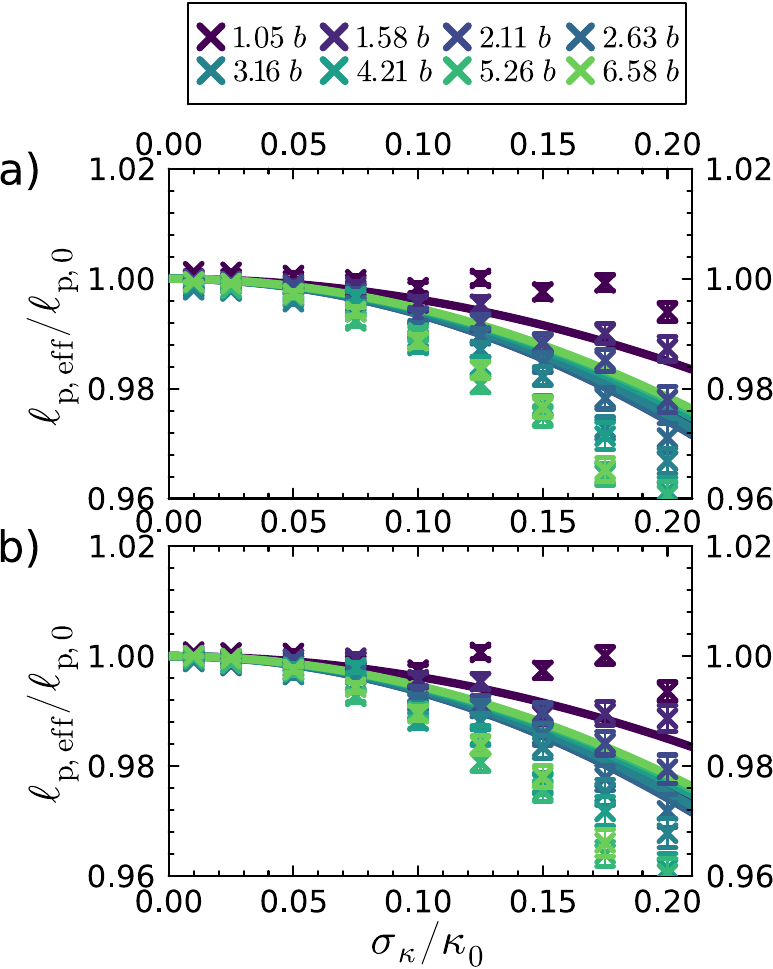}
    \caption{Theoretical predictions for the effective persistence length, $\ell_\text{p,eff}$, from Eq.~\eqref{eq:lpeffpre_kappa} (lines) and corresponding numerical data for ideal heteropolymer chains (symbols), as obtained from matching disorder averaged results for (a) $\disav{R_\text{ee}}$ and (b) $\disav{R_\text{g}}$ with the corresponding statistical averages in homopolymers (see text). The data are rescaled by $\ell_\text{p,0} = \ell_\text{p}(\kappa_0)$ and plotted against the relative standard deviation $\sigma_\kappa/\kappa_0$. Symbols and lines are colored according to $\ell_\textrm{p,0}$, see legend. }
    \label{fig:MappedResults}
\end{figure}

The results for $\ell_\text{p,eff}/\ell_\text{p,0}$ from matching $R_\text{g}$ and $R_\text{ee}$ are shown in Figs.~\ref{fig:MappedResults}a and \ref{fig:MappedResults}b, respectively, together with the theoretically predicted effective persistence length according to Eq.~\eqref{eq:lpeff}, as a function of relative standard deviation $\sigma_\kappa/\kappa_0$. With increasing $\sigma_\kappa$, the $\ell_\text{p,eff}/\ell_\text{p,0}$ decrease in a roughly quadratic manner, as predicted by our theory [Eq.~\eqref{eq:lpeffpre_kappa}]. On a more quantitative level, the theory slightly overestimates $\ell_\text{p,eff}$ for stiff chains ($\ell_\text{p,0} \gtrsim 2.5\,b$) and underestimates it for very flexible chains ($\ell_\text{p,0} \approx b$). The best agreement between theory and simulation is obtained for $\ell_\text{p,0} \sim 1.5-2\,b$.

\section{Conclusions}
In this work, we developed a theoretical framework that maps heteropolymers with spatially varying persistence lengths onto reference homopolymers characterized by a single effective persistence length $\ell_\text{p,eff}$. We derived an analytical expression for $\ell_\text{p,eff}$, which depends on the mean, $\kappa_0$, and variance, $\sigma_\kappa^2$, of the stiffness distribution. To validate our model, we conducted Monte Carlo simulations of heteropolymers with stochastically varying persistence lengths, considering both cases with and without excluded volume interactions. 

Our results show that $\ell_\text{p,eff}$ is systematically smaller than the arithmetic mean of the local persistence lengths, consistent with the intuitive notion that flexible segments have a greater influence on the overall chain stiffness than stiff segments. For narrow distributions of local stiffnesses, specifically when the relative standard deviation satisfies $\sigma_\kappa/\kappa_0 \lesssim  0.1$, the relative deviations in key conformational descriptors, such as the radius of gyration and shape anisotropy, between heteropolymers and effective homopolymers are well below $1\,\%$. The agreement between theory and simulations is particularly good for moderately stiff chains with persistence lengths in the range of 1.5–2 bond lengths, which coincides with typical values for polypeptides. For broader stiffness distributions, our theory tends to slightly overestimate $\ell_\text{p,eff}$, reflecting the limitations of our model for chains with extreme variations in local stiffness. In the second part of this series,\cite{witzky:jcp:2025} we apply our theoretical model for mapping intrinsically disordered proteins to effective homopolymer chains.

\section*{Supplementary Material}
Alternative approach using replica-trick, full derivation of effective persistence length, and generalization to correlated disorder.

\section*{Conflicts of interest}
The authors have no conflicts to disclose.

\section*{Data Availability}
The data that support the findings of this study and the scripts used to generate them are openly available in Zenodo (10.5281/zenodo.15696365) and GitHub (\url{https://github.com/ywitzky/StiffnessDistributions}).

\section*{Acknowledgments}
This work was supported by the Deutsche Forschungsgemeinschaft (DFG, German Research Foundation) through the framework of the collaborative research centers SFB 1551 (Project No. 464588647) and TRR 146 (Project No. 233630050), and through the Heisenberg grant to A.N. (Project No. 470113688). Furthermore, the authors gratefully acknowledge the Gauss Centre for Supercomputing e.V. (www.gauss-centre.eu) for funding this project by providing computing time through the John von Neumann Institute for Computing (NIC) on the GCS Supercomputer JUWELS at J{\"u}lich Supercomputing Centre (JSC) with the application no 27367. Further computations were carried out on the high-performance computer Mogon2 and Mogon NHR South-West.

\end{document}


\author{Yannick Witzky}
\affiliation{Institute of Physics, Johannes Gutenberg University Mainz, Staudingerweg 7-9, 55128 Mainz, Germany}

\author{Friederike Schmid}
\email{friederike.schmid@uni-mainz.de}
\affiliation{Institute of Physics, Johannes Gutenberg University Mainz, Staudingerweg 7-9, 55128 Mainz, Germany}

\author{Arash Nikoubashman}
\email{anikouba@ipfdd.de}
\affiliation{Leibniz-Institut f{\"u}r Polymerforschung Dresden e.V., Hohe Stra{\ss}e 6, 01069 Dresden, Germany}
\affiliation{Institut f{\"u}r Theoretische Physik, Technische Universit{\"a}t Dresden, 01069 Dresden, Germany}
\affiliation{Cluster of Excellence Physics of Life, Technische Universit{\"a}t Dresden, 01062 Dresden, Germany}

\title{Supplementary Material: From Heteropolymer Stiffness Distributions to Effective Homopolymers. Part 1: Theoretical Modeling and Computational Verification}

\maketitle

\setlength{\parindent}{0pt}
\setlength{\parskip}{0.2\baselineskip}

\newcommand{\ud}{\textrm{d}}
\newcommand{\ue}{\textrm{e}}

\section{Theory for effective persistence length in heterogeneous ideal chains} 

\subsection{Model assumptions and goal}

We consider ideal \fs{or real} chains of length $N$ with the angular potential
\begin{equation}
    U_\text{bend} = k_\text{B}T \sum_i \kappa_i (1- \vv{u}_i \cdot \vv{u}_{i+1}) = k_\text{B}T \sum_i \kappa_i [1- \cos(\theta)]
\end{equation}
where $\vv{u}_i$ are unit vectors pointing in the direction of the
bonds with length $b$. \fs{In the case of real chains, this is taken to be an (approximate) effective angular potential, which includes both effects of bonded interactions and non-bonded interactions between adjacent monomers.} The angle potential parameters $\kappa_i$ are Gaussian distributed with mean $\kappa_0$ and \fs{covariance matrix $\sigma_\kappa^2 \: c_{ij}$ with $c_{ii}=1$, i.e., the joint distribution is
\begin{equation} P[\mathbf{\kappa}] = 
\frac{1}{\sqrt{(2 \pi \sigma_\kappa^2)^{N-1} \det(c)}}
\exp \left[-\frac{1}{2 \sigma_\kappa^2}
 \sum_{ij} (\kappa_i - \kappa_0) (c^{-1})_{ij} (\kappa_j - \kappa_0) 
 \right] ,
\label{eq:kappa_dist}
\end{equation}
where $[\mathbf{\kappa}]$ stands for $(\kappa_1, ... \kappa_{N-1})$. Hence, the different angle potential parameters are correlated according  to $\la\kappa_i \kappa_j \ra-\kappa_0^2 = \sigma_\kappa^2 c_{ij}$,  and the variance of individual $\kappa_i$-distributions is $\sigma_\kappa^2$.
}

The local persistence length $\ell_\text{p}(\kappa_i)$ is defined via the relation\cite{doiTheoryPolymerDynamics1988}
\begin{equation}
\langle \vv{u}_i \cdot \vv{u}_{i+1} \rangle = \exp\left[- \ell_\text{p}(\kappa_i)/b\right], 
\end{equation}
where $\la \cdot \ra$ denotes the thermal average.
Since the model only includes bending potentials, it can easily be evaluated, resulting in
\begin{equation}
\label{eq:lp_kappa}
    \ell_\text{p}(\kappa_i) = - b/\ln\left[\coth(\kappa_i) - 1/\kappa_i\right],
\end{equation}

Our goal is to determine the effective persistence length of a homogeneous reference chain with the same conformational chain properties than the heterogeneous chains, averaged over the disorder.

\subsection{Ansatz}

We use the notation $\int {\cal D}R \dots 
= \int\!\!\int\!\!\int \ud^2 u_1 \ud^2 u_2 \dots \ud^2 u_{N-1} \dots$. 
The partition function of a chain with fixed 
\fs{$[\kappa_l]$} can then be written as
\begin{equation}
{\cal Z}\fs{_{\mathbf{\kappa}}} = \int {\cal D}\!R \;
\exp\left[- \sum_i \kappa_i (1- \vv{u}_i \cdot \vv{u}_{i+1})
\right]\:  \fs{\exp \big[- {\cal V}[R]\big],}
\end{equation}
\fs{where ${\cal V}$ denotes the possible contribution of excluded
volume interactions between non-adjacent monomers in real chains (the
excluded volume interactions between adjacent monomers are included in
the bending potential). The} corresponding free energy is 
$F\fs{_{\mathbf{\kappa}}} = -k_BT \log({\cal Z\fs{_{\mathbf{\kappa}}}})$. 
The \fs{thermal ensemble average of an observable $O$ for 
a given set of stiffness parameters $[\mathbf{\kappa}]$ is then given by
\begin{equation}
     \langle O \rangle_{\mathbf{\kappa}} =  \frac{1}{\cal Z}_{\mathbf{\kappa}} \int {\cal D} \!R  \: \: O \: \exp\left[- \sum_{i=1}^{N-1} \kappa_i  
   \left(1- \vv{u}_i \cdot \vv{u}_{i+1} \right)\right]\exp \big[- {\cal V}[R]\big],
\end{equation}
and the disorder average over all possible realizations of
$\mathbf{\kappa}$ is generally defined in our system as
\begin{equation}
     \disav{O} = \iint d \kappa_1  \cdots \ud \kappa_{N-1} \: 
        P[\mathbf{\kappa}] \: \langle O \rangle_{\mathbf{\kappa}}
     \label{eq:disorder_Average}
\end{equation}
}

We need to calculate the disorder averaged free energy, $\disav{F} = -k_\text{B} T \disav{\log({\cal Z}\fs{_{\mathbf{\kappa}}})}$. To this end, we use the Replica trick which is commonly used in disordered systems, and exploit the relation
\begin{equation}
\disav{\log({\cal Z})} = \lim_{m \to 0} 
\frac{\disav{{\cal Z}^m} - 1}{m}.
\end{equation}
This leaves us with the task to calculate the averaged partition function for $m$ chains with identical realization of the disorder,
\begin{eqnarray}
\nonumber
\disav{{\cal Z}\fs{_{\mathbf{\kappa}}}^m}
&=& 
\fs{\int \ud \kappa_1 ... \ud \kappa_{N-1} \: P[\mathbf{\kappa}]}
\int {\cal D} R_1 \dots {\cal D}R_m \:
    \nonumber \\ &&  \qquad  \times
\exp\left[- \sum_l^{N-1} \kappa_l \sum_{\alpha = 1}^m 
   \left(1- \vv{u}_l^{(\alpha)} \cdot \vv{u}_{l+1}^{(\alpha)}\right)\right]
   \fs{ \: \exp\Big[-\sum_{\alpha = 1}^m {\cal V}[R_\alpha] \Big]}
\\ 
&=&  \nonumber
\int {\cal D} R_1 \dots {\cal D}R_m
\exp\left[- \kappa_0 \sum_{l\alpha}
   \left(1- \vv{u}_l^{(\alpha)} \cdot \vv{u}_{l+1}^{(\alpha)}\right)\right]
   \fs{
   \exp\Big[-\sum_{\alpha = 1}^m {\cal V}[R_m] \Big]}
   \\  \nonumber && \qquad \times \: 
\fs{\int \ud \kappa_1 ... \ud \kappa_{N-1} \: P[\mathbf{\xi}+\kappa_0]}
 \exp\left[-  \sum_{l \alpha} \xi_i
   \left(1- \vv{u}_i^{(\alpha)} \cdot \vv{u}_{i+1}^{(\alpha)}\right)\right]
   \\
&=& \nonumber
\int {\cal D} R_1 \dots {\cal D}R_m
\exp\left[- \kappa_0 \sum_{i\alpha}
   \left(1- \vv{u}_i^{(\alpha)} \cdot \vv{u}_{i+1}^{(\alpha)}\right)\right]
    \fs{\exp\Big[-\sum_{\alpha = 1}^m {\cal V}[R_m] \Big]}
\\   && \qquad \times \:
\fs{
\exp\left[\frac{\sigma_\kappa^2}{2} \sum_{i j \alpha \beta}
   \left(1- \vv{u}_i^{(\alpha)} \cdot \vv{u}_{i+1}^{(\alpha)}\right)
   c_{ij}
   \left(1- \vv{u}_j^{(\beta)} \cdot \vv{u}_{j+1}^{(\beta)}\right) 
   \right]}
   \label{eq:zm_disorder}
\end{eqnarray}
where we have substituted $\kappa_i=\kappa_0+\xi_i$.
\comment{TODO: Replace by Matrix form: and 
$\int_{-\infty}^\infty\exp[-(ax^2+bx)] \text{d}x = \sqrt{\pi/a}\exp[b^2/(4a)]$ 
with $a=1/(2\sigma^2)$ and 
$b= \sum_{\alpha} (1- \vv{u}_i^{(\alpha)} \cdot
\vv{u}_{i+1}^{(\alpha)})$. }
The corresponding quantity for the homogeneous reference chain is
\begin{displaymath}
{\cal Z}_\text{ref} = 
\int {\cal D} R_1 \dots {\cal D}R_m \:
\exp\left[- \kappa_\text{eff} \sum_{i\alpha}
   \left(1- \vv{u}_i^{(\alpha)} \cdot \vv{u}_{i+1}^{(\alpha)}\right)\right]
\end{displaymath}
where $\kappa_\text{eff}$ is the angle potential parameter of the reference chain. Subtracting the two from each other, we obtain
\
\begin{align}\nonumber
\disav{{\cal Z}^m - {\cal Z}^m_\text{ref}} 
&= \int {\cal D} R_1 \dots {\cal D}R_m \exp\left[-\kappa_\text{eff}  \sum_{i \alpha}
  \left (1- \vv{u}_i^{(\alpha)} \cdot \vv{u}_{i+1}^{(\alpha)}\right)\right] 
  \:  \fs{\exp\Big[-\sum_{\alpha = 1}^m {\cal V}[R_\alpha] \Big]}  
\\ &\times 
\biggl\{\exp\left[(\kappa_\text{eff} - \kappa_0) \sum_{i \alpha}
   \left(1- \vv{u}_i^{(\alpha)} \cdot \vv{u}_{i+1}^{(\alpha)}\right)\right]
   \nonumber
\\ & \qquad \times 
\fs{   \exp\left[\frac{\sigma_\kappa^2}{2} \sum_{ij \alpha \beta}
   \left(1- \vv{u}_i^{(\alpha)} \cdot \vv{u}_{i+1}^{(\alpha)}\right)
   c_{ij}
   \left(1- \vv{u}_j^{(\beta)} \cdot \vv{u}_{j+1}^{(\beta)}\right)\right]
   - 1\biggr\}}.\nonumber 
\end{align}
Using the notation
\begin{equation}
    \la A \ra_{\kappa_\text{eff}} =  \frac{ \int  {\cal D} R_1 \dots {\cal D}R_m  \exp\left[-\kappa_\text{eff} \sum_{i \alpha} \left(1- \vv{u}_i^{(\alpha)} \cdot \vv{u}_{i+1}^{(\alpha)}\right)\right] \:
     \fs{ \exp\Big[-\sum_{\alpha = 1}^m {\cal V}[R_\alpha] \Big]} \:A}{{\cal Z}_\text{ref}}
    \label{eq:notation}
\end{equation}
we can rewrite this as:
\begin{eqnarray*}
\disav{{\cal Z}\fs{_{\mathbf{\kappa}}}^m - {\cal Z}^m_\text{ref}}
& = & {\cal Z}_\text{ref}^m \:
\biggl\langle \exp\left[(\kappa_\text{eff} - \kappa_0) \sum_{i \alpha}
   \left(1- \vv{u}_i^{(\alpha)} \cdot \vv{u}_{i+1}^{(\alpha)}\right)\right]
   \\  \label{eq:app_start_prob}
   && \times
   \: 
   \exp\left[\frac{\sigma_\kappa^2}{2} \sum_{ij \alpha \beta}
   \left(1- \vv{u}_i^{(\alpha)} \cdot \vv{u}_{i+1}^{(\alpha)}\right)
   c_{ij}
   \left(1- \vv{u}_j^{(\beta)} \cdot \vv{u}_{j+1}^{(\beta)}\right)\right]
   - 1
\biggl\rangle_{\kappa_\text{eff}}.
\end{eqnarray*}
We note that $\langle O \rangle_{\kappa_{\text{eff}}}$ denotes a pure {\em thermal} average for fixed given parameter $\kappa_{\text{eff}}$, different from $\disav{O}$, which also includes a disorder average over $\kappa_i$.

Now we Taylor expand the averages and get
\begin{align*}
\disav{{\cal Z}\fs{_{\mathbf{\kappa}}}^m - {\cal Z}^m_\text{ref}}
 \approx &
{\cal Z}_\text{ref}^m \:
\biggl[ (\kappa_\text{eff} - \kappa_0) \: 
\sum_{i \alpha} \la 
\left(1- \vv{u}_i^{(\alpha)} \cdot \vv{u}_{i+1}^{(\alpha)}\right)
\ra_{\kappa_\text{eff}}
\\& 
+ \fs{\frac{\sigma_\kappa^2}{2} \sum_{ij \alpha \beta}
\la
\left(1- \vv{u}_i^{(\alpha)} \cdot \vv{u}_{i+1}^{(\alpha)}\right)
c_{ij}
\left(1- \vv{u}_j^{(\beta)} \cdot \vv{u}_{j+1}^{(\beta)}\right)
\ra_{\kappa_\text{eff}}
\biggl]}
 \\
= & \fs{{\cal Z}_\text{ref}^m \: 
\left[ (\kappa_\text{eff} - \kappa_0) m N \: v_1(\kappa_\text{eff}) \right.}
\\ & \quad \fs{\left.
+ \frac{\sigma_\kappa^2}{2} 
\left(m^2 \: \sum_{ij} c_{ij} v_1(\kappa_\text{eff})^2 
   - m \: \sum_{ij} c_{ij} \: v_1(\kappa_\text{eff})^2
   + m \: N v_2(\kappa_\text{eff}) \right)
\right]}
\\ 
= & \fs{{\cal Z}_\text{ref}^m \:  mN
\left[ (\kappa_\text{eff} - \kappa_0) \: v_1(\kappa_\text{eff})
+ \frac{\sigma_\kappa^2}{2} 
\left( v_2(\kappa_\text{eff}) 
   - \frac{1}{N}\sum_{ij} c_{ij} v_1(\kappa_\text{eff})^2 \right)
\right]
+ {\cal O}(m^2)}
\end{align*}

Here we have defined  $v_n = \la \big(1- \vv{u}_i^{(\alpha)} \cdot 
\vv{u}_{i+1}^{(\alpha)}\big)^n \ra_{\kappa_\text{eff}}$. 
Specifically, inserting the angular potential above, we get
\begin{eqnarray*}
Z(\kappa)&=&\int_0^{\pi}\exp\left[-\kappa(1-\cos(\theta))\right]\sin(\theta)\text{d}\theta = \frac{1-e^{-2\kappa}}{\kappa}\\
P(\theta,\kappa) &=& \frac{1}{Z(\kappa)} \exp\left[-\kappa(1-cos(\theta))\right]\\
v_1(\kappa_\text{eff}) &=&  \int_0^\pi P(\theta, \kappa_\text{eff})[1-\cos(\theta)]\sin(\theta)\text{d}\theta = 1 + \frac{1}{\kappa_\text{eff}} - \coth(\kappa_\text{eff})
\\
v_2(\kappa_\text{eff}) &=& \int_0^\pi P(\theta, \kappa_\text{eff})[1-\cos(\theta)]^2
\sin(\theta)\text{d}\theta \\&=&   \frac{2}{\kappa_\text{eff}^2} + 2
\frac{1+\kappa_\text{eff}}{\kappa_\text{eff}} [1 - \coth(\kappa_\text{eff})] \\
v_n(\kappa_\text{eff}) &=& \int_0^\pi P(\theta, \kappa_\text{eff})[1-\cos(\theta)]^n
\sin(\theta)\text{d}\theta  \\&=& \frac{\exp\left(2\kappa_\text{eff}\right)\kappa_\text{eff}^{-n}\left[\Gamma(1+n)-\Gamma(1+n, 2\kappa_\text{eff})\right]}{\exp(2\kappa_\text{eff})-1}
\end{eqnarray*}

Inserting these results, we can calculate the free energy difference between the disorder averaged heterogeneous chain and the reference chain
\begin{equation}
\Delta F = \lim_{m \to 0} \frac{\disav{{\cal Z}\fs{_{\mathbf{\kappa}}}^m - {\cal
Z}_\text{ref}^m}}{m}
= N \left[v_1(\kappa_\text{eff}) \:  (\kappa_\text{eff} - \kappa_0)
+ \frac{\sigma_\kappa^2}{2} \: \left(v_2(\kappa_\text{eff}) 
- \fs{(1+c_\text{eff})} \: v_1(\kappa_\text{eff})^2\right) \right],
\label{eq:deltaF}
\end{equation}
\fs{where we have introduced the correlation parameter 
$c_\text{eff} = \frac{1}{N} \sum_{ij}c_{ij}-1$. For uncorrelated stiffness
parameters $\kappa_i$, the correlation parameter is $c_\text{eff} =
0$. For blocky heteropolymers, one gets $c_\text{eff}>0$, and for heteropolymers with
alternating stiffness parameters, $c_\text{eff} < 0$. }

Now we optimize the reference system such that $\Delta F = 0$, which gives the following implicit equation for $\kappa_\text{eff}$:
\begin{equation}
       \kappa_\text{eff} = \kappa_0  - \frac{\sigma_\kappa^2}{2} \: h(\kappa_\text{eff}) 
       \label{eq:kappa_bar}
\end{equation}
with
\begin{equation}
    h(\kappa)= \frac{v_2(\kappa)-\fs{(1+c_\text{eff})} \: v_1(\kappa)^2}{v_1(\kappa)} 
\end{equation}
\fs{Specifically, for uncorrelated chains, we obtain}
\begin{equation}
    h(\kappa)= \frac{v_2(\kappa)-v_1(\kappa)^2}{v_1(\kappa)} = \frac{1-\kappa^2/\sinh(\kappa)^2}{\kappa[1+\kappa-\kappa\coth(\kappa)]},
\end{equation}
\fs{corresponding to Eq.\ (20) in the main text.}
\\
For small and large $\kappa$, the function $h(\kappa)$ has the limiting behavior
\begin{equation}
h(\kappa) \approx \left\{
\begin{array}{ll}
\fs{\frac{1}{3} - c_\text{eff} + \frac{\kappa}{9}(1+3 c_\text{eff})} & (\kappa \ll 1) \\
\frac{1}{\kappa} \fs{(1- c_\text{eff})} & (\kappa \gg 1) 
\end{array}
\right.
\end{equation}

\subsection{Effective persistence length}

Summarizing the previous subsections \fs{and the considerations in the main text regarding ideal chains with uncorrelated stiffness parameters}: We consider a (quenched) ensemble of heterogeneous semiflexible chains with angle potential parameters $\kappa_i$, which are Gaussian distributed with mean $\kappa_0$ and \fs{covariance matrix $\sigma_\kappa^2 \: c_{ij}$}. According to the above theoretical calculation, the configurational properties should have the same distribution than a reference chain with the renormalized angle potential parameter
\begin{equation}
           \kappa_\text{eff}/\kappa_0 = 1 - \frac{\sigma_\kappa^2}{2\kappa_0} h(\kappa_\text{eff}) \stackrel{\kappa>4}{\approx } 1  - \frac{1}{2}\left(\frac{\sigma_\kappa}{\kappa_0}\right)^2 \fs{(1-c_\text{eff})}
\end{equation}
\fs{with $c_\text{eff} = \frac{1}{N}\sum_{ij} c_{ij} - 1$}. Now, instead of looking at the heterogeneously distributed angular potential parameter, we consider the persistence length $\ell_{\text{p},i} = \ell_\text{p}(\kappa_i)$ with
\begin{equation}
\ell_\text{p}(\kappa) = \frac{-b}{\log(\la\cos(\theta)\ra)}= \frac{-b}{\log[\coth(\kappa) - 1/\kappa]}\stackrel{\kappa>4}{\approx }b \left( \kappa-\frac{1}{2}\right).
\end{equation}
The set of local persistence lengths of the Gaussian distributed $\kappa$ values follow the distribution \fs{$P[\ell_\text{p}]$} with mean $\disav{\ell_\text{p}}$. Taylor expanding $\ell_\text{p}(\kappa)$,
we can approximate $\disav{\ell_\text{p}}$ by:
\begin{align}
\nonumber
\disav{\ell_\text{p}}&=\fs{\disav{\ell_{\text{p}}(\kappa_i)}
= \int \textrm{d}\kappa_1 .. \text{d} \kappa_{N-1} P[\mathbf{\kappa}] 
\: \ell_\text{p}(\kappa_i) \qquad \text{(with arbitrary $i$)} }
\\ \nonumber
&\approx 
\fs{\ell_\text{p}(\kappa_0) + \ell_\text{p}'(\kappa_0) \disav{\kappa_i-\kappa_0} 
   + \frac{1}{2} \ell_\text{p}''(\kappa_0) \disav{(\kappa_i-\kappa_0)^2}}
\\
&\approx l_\text{p}(\kappa_0) 
+ \frac{1}{2} \ell_\text{p}''(\kappa_0) \sigma_\kappa^2.
\label{eq:l0}
\end{align}
Furthermore, \fs{we have}
\begin{eqnarray*}
\disav{\ell_\text{p}^2}
&=& \fs{\int \textrm{d} \kappa_1 ... \text{d}\kappa_{N-1} \: P[\mathbf{\kappa}] \: \ell_\text{p}(\kappa_i)^2}
\\
 &\approx& 
 \fs{
 \disav{\left( \ell_\text{p}(\kappa_0) + \ell_\text{p}'(\kappa_0) (\kappa-\kappa_0) 
   + \frac{1}{2} \ell_\text{p}''(\kappa_0) (\kappa-\kappa_0)^2 \right)^2}}
   \\
 &\approx& 
 \fs{\ell_\text{p}(\kappa_0)^2 +  \left[\ell_\text{p}'(\kappa_0)^2 + \ell_\text{p}(\kappa_0) \ell_\text{p}''(\kappa_0) \right] \disav{(\kappa_i - \kappa_0)^2} }
\\ & =&
\ell_\text{p}(\kappa_0)^2 + \sigma_\kappa^2 \: 
\left[\ell_\text{p}'(\kappa_0)^2 + \ell_\text{p}(\kappa_0) \ell_\text{p}''(\kappa_0) \right],
\end{eqnarray*}
\fs{and, analogously,
\begin{eqnarray*}
\disav{\ell_{\text{p},i} \: \ell_{\text{p},j})}
&=& \int \textrm{d} \kappa_1 ... \text{d}\kappa_{N-1} \: P[\mathbf{\kappa}] \: \ell_\text{p}(\kappa_i) \ell_\text{p}(\kappa_j)\\
 &\approx& 
 \ell_\text{p}(\kappa_0)^2 + \frac{1}{2}\ell_\text{p}(\kappa_0)\ell_\text{p}''(\kappa_0) \left[  \disav{(\kappa_i - \kappa_0)^2} + \disav{(\kappa_j - \kappa_0)^2} \right]
 \\ && \quad + \: \ell_\text{p}'(\kappa_0)^2
 \disav{(\kappa_i - \kappa_0)\:(\kappa_j - \kappa_0)} 
\\ & =&
\ell_\text{p}(\kappa_0)^2 + \sigma_\kappa^2 \: 
\left[\ell_\text{p}(\kappa_0) \ell_\text{p}''(\kappa_0) 
+ \ell_\text{p}'(\kappa_0)^2 c_{ij} 
\right],
\end{eqnarray*}
}
\fs{Therefore,} the distribution has the variance (up to order $\sigma_\kappa^2$)
\begin{equation}
\sigma_\text{p}^2 = \disav{\ell_\text{p}^2} - \disav{\ell_\text{p}}^2
= \ell_\text{p}'(\kappa_0)^2 \: \sigma_\kappa^2  
\label{eq:variance_conversion}
\end{equation}
\fs{and the covariance matrix
\begin{equation}
\text{Cov}(\ell_\text{p})_{ij}=\disav{\ell_{\text{p},i} \ell_{\text{p},j}}
 - \disav{\ell_\text{p}}^2
= \ell_\text{p}'(\kappa_0)^2 \: \sigma_\kappa^2 \: c_{ij}
= \sigma_\text{p}^2 \: c_{ij}.
    \label{eq:covariance_conversion}
\end{equation}
Hence, the covariance matrix of the persistence lengths, $\ell_{\text{p},i}$, has the same structure than that of the stiffness parameters, $\kappa_i$, and can be used to determine the correlation parameter via
\begin{equation}
c_\text{eff} = \frac{1}{N} \sum_{ij} \frac{\disav{\ell_{\text{p},i} \ell_{\text{p},j}}
 - \disav{\ell_\text{p}}^2}{\sigma_\text{p}^2} - 1.
\end{equation}
}

We combine \fs{Eq.~(\ref{eq:variance_conversion})} with Eq.~(\ref{eq:l0}), and obtain the following implicit equation for $\disav{\ell_\text{p}}$ as a function of $\kappa_0$ and $\sigma_\text{p}^2$:
\begin{equation}
\label{eq:kappa_0}
    \disav{\ell_\text{p}} = \ell_\text{p}(\kappa_0) + \frac{1}{2} \frac{\ell_\text{p}''(\kappa_0)}{\ell_\text{p}'(\kappa_0)^2} \: \sigma_\text{p}^2.
\end{equation}

The persistence length of the reference chain is given by $\ell_\text{p,eff} = \ell_\text{p}(\kappa_\text{eff})$. Compared to $\disav{\ell_\text{p}}$, it is shifted by
\begin{eqnarray*}
\ell_\text{p,eff} - \disav{\ell_\text{p}} &=& \left[\ell_\text{p}(\kappa_0) 
  + \ell_\text{p}'(\kappa_0) (\kappa_\text{eff}-\kappa_0) \right]
  -  \left[\ell_\text{p}(\kappa_0) + \frac{1}{2} \ell_\text{p}''(\kappa_0) \sigma_\kappa^2\right]
  \\
& \stackrel{\textrm{eq. }\ref{eq:kappa_bar}}{=}& - \frac{\sigma_\kappa^2}{2} \left[
\ell_\text{p}'(\kappa_0) \: h(\kappa_\text{eff}) + \ell_\text{p}''(\kappa_0) \right] 
\\
& \stackrel{h(\kappa_\text{eff})\approx h(\kappa_0)}{=}& - \frac{\sigma_\kappa^2}{2} \left[
\ell_\text{p}'(\kappa_0) \: h(\kappa_0) + \ell_\text{p}''(\kappa_0) \right] 
\\
&\stackrel{\textrm{eq. }\ref{eq:variance_conversion}}{=}& - \frac{\sigma_\text{p}^2}{2} \left[
h(\kappa_0)/\ell_\text{p}'(\kappa_0) + \ell_\text{p}''(\kappa_0)/\ell_\text{p}'(\kappa_0)^2 \right]
\end{eqnarray*}

Using Equation (\ref{eq:kappa_0}), we can also express  $\ell_\text{p,eff}$ as
\begin{equation}
\label{eq:pers_l_bar}
\ell_\text{p,eff} = \ell_\text{p}(\kappa_0) - \frac{\sigma_\text{p}^2}{2} \frac{h(\kappa_0)}{\ell_\text{p}'(\kappa_0)}
\end{equation}

If $\kappa_0 > 4$, we can use the approximation $\ell_\text{p}(\kappa) \approx b (\kappa - 1/2)$ giving $\ell_\text{p}''(\kappa_0) \approx 0$ and $h(\kappa) \approx 1/\kappa$. Equation~(\ref{eq:pers_l_bar}) then simplifies \fs{to}
\begin{eqnarray}
\disav{\ell_\text{p}} & \approx & \ell_\text{p}(\kappa_0) \\
\ell_\text{p,eff} &\approx& \disav{\ell_\text{p}} - \frac{\sigma_\text{p}^2}{2}\frac{1}{\kappa_0 \ell_\text{p}'(\kappa_0)}\\ &\approx&  \disav{\ell_\text{p}} - \frac{\sigma_\text{p}^2}{2}\frac{1}{\left( \frac{\ell_\text{p}(\kappa)}{b}+\frac{1}{2}\right)b } \\
&\approx& \disav{\ell_\text{p}} - \frac{\sigma_\text{p}^2}{2 \disav{\ell_\text{p}} + b}   
\end{eqnarray}

\FloatBarrier


\author{Yannick Witzky}
\affiliation{Institute of Physics, Johannes Gutenberg University Mainz, Staudingerweg 7-9, 55128 Mainz, Germany}

\author{Friederike Schmid}
\email{friederike.schmid@uni-mainz.de}
\affiliation{Institute of Physics, Johannes Gutenberg University Mainz, Staudingerweg 7-9, 55128 Mainz, Germany}

\author{Arash Nikoubashman}
\email{anikouba@ipfdd.de}
\affiliation{Leibniz-Institut f{\"u}r Polymerforschung Dresden e.V., Hohe Stra{\ss}e 6, 01069 Dresden, Germany}
\affiliation{Institut f{\"u}r Theoretische Physik, Technische Universit{\"a}t Dresden, 01069 Dresden, Germany}
\affiliation{Cluster of Excellence Physics of Life, Technische Universit{\"a}t Dresden, 01062 Dresden, Germany}

\title{Supplementary Material: From Heteropolymer Stiffness Distributions to Effective Homopolymers. Part 2: Conformational Analysis of Intrinsically Disordered Proteins
}

\maketitle

\setlength{\parindent}{0pt}
\setlength{\parskip}{0.2\baselineskip}

\newcommand{\ud}{\textrm{d}}
\newcommand{\ue}{\textrm{e}}

\section{Sequences}
{\linespread{0.85}
\begin{table}[htb]

        \centering
    \subfloat[\textbf{Hst5:}]{
    \begin{tabular}{r|l}
    1 & \texttt{DSHAKRHHGY KRKFHEKHHS HRGY}\\ 
     
    \end{tabular}}\vspace{0.05cm}
    \subfloat[\textbf{Hst52:}]{
    \begin{tabular}{r|l}
    1 & \texttt{DSHAKRHHGY KRKFHEKHHS HRGYDSHAKR HHGYKRKFHE KHHSHRGY}\\ 
     
    \end{tabular}}\vspace{0.05cm}
    \subfloat[\textbf{p532070:}]{
    \begin{tabular}{r|l}
    1 & \texttt{GPGSDLWKLL PENNVLSPLP SQAMDDLMLS PDDIEQWFTE DPGPDEAPRM PEAALEHHHH }\\ 
 61 & \texttt{HH}\\ 
     
    \end{tabular}}\vspace{0.05cm}
    \subfloat[\textbf{ACTR:}]{
    \begin{tabular}{r|l}
    1 & \texttt{GTQNRPLLRN SLDDLVGPPS NLEGQSDERA LLDQLHTLLS NTDATGLEEI DRALGIPELV }\\ 
 61 & \texttt{NQGQALEPKQ D}\\ 
     
    \end{tabular}}\vspace{0.05cm}
    \subfloat[\textbf{DSS1:}]{
    \begin{tabular}{r|l}
    1 & \texttt{MSRAALPSLE NLEDDDEFED FATENWPMKD TELDTGDDTL WENNWDDEDI GDDDFSVQLQ }\\ 
 61 & \texttt{AELKKKGVAA C}\\ 
     
    \end{tabular}}\vspace{0.05cm}
    \subfloat[\textbf{Ash1:}]{
    \begin{tabular}{r|l}
    1 & \texttt{SASSSPSPST PTKSGKMRSR SSSPVRPKAY TPSPRSPNYH RFALDSPPQS PRRSSNSSIT }\\ 
 61 & \texttt{KKGSRRSSGS SPTRHTTRVC V}\\ 
     
    \end{tabular}}\vspace{0.05cm}
    \subfloat[\textbf{CTD2:}]{
    \begin{tabular}{r|l}
    1 & \texttt{FAGSGSNIYS PGNAYSPSSS NYSPNSPSYS PTSPSYSPSS PSYSPTSPCY SPTSPSYSPT }\\ 
 61 & \texttt{SPNYTPVTPS YSPTSPNYSA SPQ}\\ 
     
    \end{tabular}}\vspace{0.05cm}
    \subfloat[\textbf{Sic1:}]{
    \begin{tabular}{r|l}
    1 & \texttt{GSMTPSTPPR SRGTRYLAQP SGNTSSSALM QGQKTPQKPS QNLVPVTPST TKSFKNAPLL }\\ 
 61 & \texttt{APPNSNMGMT SPFNGLTSPQ RSPFPKSSVK RT}\\ 
     
    \end{tabular}}\vspace{0.05cm}
    \subfloat[\textbf{SH4UD:}]{
    \begin{tabular}{r|l}
    1 & \texttt{MGSNKSKPKD ASQRRRSLEP AENVHGAGGG AFPASQTPSK PASADGHRGP SAAFAPAAAE }\\ 
 61 & \texttt{PKLFGGFNSS DTVTSPQRAG PLAGGSAWSH PQFEK}\\ 
     
    \end{tabular}}\vspace{0.05cm}
    \subfloat[\textbf{ColNT:}]{
    \begin{tabular}{r|l}
    1 & \texttt{MGSNGADNAH NNAFGGGKNP GIGNTSGAGS NGSASSNRGN SNGWSWSNKP HKNDGFHSDG }\\ 
 61 & \texttt{SYHITFHGDN NSKPKPGGNS GNRGNNGDGA SSHHHHHH}\\ 
     
    \end{tabular}}
 \end{table}
\begin{table}[htb]

        \centering\vspace{0.05cm}
    \subfloat[\textbf{p27Cv15:}]{
    \begin{tabular}{r|l}
    1 & \texttt{GSHMKGACIV ANSPPDDVKS KEDVPQTDPR LTGGDRDNAR ASRTGNDPAG ASTQSAEVAC }\\ 
 61 & \texttt{SNPILSTPDA QEKQAGTSNS KERPHEQLSA GSVEQTPKKP GLRRRQT}\\ 
     
    \end{tabular}}\vspace{0.05cm}
    \subfloat[\textbf{p27Cv31:}]{
    \begin{tabular}{r|l}
    1 & \texttt{GSHMKGACKV PAQESQDVSG SRPAAPLIGA PANSEDTHLV DPKTDPSDSQ TGLAEQCAGI }\\ 
 61 & \texttt{RKRPATDDSS TQNKRANRTE ENVSDGSPNA GSVEQTPKKP GLRRRQT}\\ 
     
    \end{tabular}}\vspace{0.05cm}
    \subfloat[\textbf{p27Cv44:}]{
    \begin{tabular}{r|l}
    1 & \texttt{GSHMKGACRK PANAEADSSS CQNVPRGKSK QAPETPTGSP LGDATLNQVK PRRPSSASTN }\\ 
 61 & \texttt{IGQLEDADED DAEDHVGSAV TSQTIPNDRA GSVEQTPKKP GLRRRQT}\\ 
     
    \end{tabular}}\vspace{0.05cm}
    \subfloat[\textbf{p27Cv56:}]{
    \begin{tabular}{r|l}
    1 & \texttt{GSHMKGACGS SVLGTGNPRN QAHVSDTSLE EDDDEQDDST PDEVSQACTI VASALDINAA }\\ 
 61 & \texttt{TPRSPKASPK RKRKRQSTAP AQGNEPPGNA GSVEQTPKKP GLRRRQT}\\ 
     
    \end{tabular}}\vspace{0.05cm}
    \subfloat[\textbf{p27Cv78:}]{
    \begin{tabular}{r|l}
    1 & \texttt{GSHMKGACAL PSGVVPAEDD DDDEEEEDDQ DPAQPQAVQG AAPSSGTNNS QPILPSIAVN }\\ 
 61 & \texttt{STTGPNSTAG KKKRKRRRTR HSNCATLSSA GSVEQTPKKP GLRRRQT}\\ 
     
    \end{tabular}}\vspace{0.05cm}
    \subfloat[\textbf{p27Cv14:}]{
    \begin{tabular}{r|l}
    1 & \texttt{GSHMKGACKS SSPPSNDQGR PGDPKQVIDK TEVERTQDTS NIQETQSANN SGPDKPSRCD }\\ 
 61 & \texttt{LAVSGVAAAA LPAPGHANST ARDLTRDEEA GSVEQTPKKP GLRRRQT}\\ 
     
    \end{tabular}}\vspace{0.05cm}
    \subfloat[\textbf{p15PAF:}]{
    \begin{tabular}{r|l}
    1 & \texttt{MVRTKADSVP GTYRKVVAAR APRKVLGSST SATNSTSVSS RKAENKYAGG NPVCVRPTPK }\\ 
 61 & \texttt{WQKGIGEFFR LSPKDSEKEN QIPEEAGSSG LGKAKRKACP LQPDHTNDEK E}\\ 
     
    \end{tabular}}\vspace{0.05cm}
    \subfloat[\textbf{PTMA:}]{
    \begin{tabular}{r|l}
    1 & \texttt{GPSDAAVDTS SEITTKDLKE KKEVVEEAEN GRDAPANGNA NEENGEQEAD NEVDEEEEEG }\\ 
 61 & \texttt{GEEEEEEEEG DGEEEDGDED EEAESATGKR AAEDDEDDDV DTKKQKTDED D}\\ 
     
    \end{tabular}}\vspace{0.05cm}
    \subfloat[\textbf{NHE6cmdd:}]{
    \begin{tabular}{r|l}
    1 & \texttt{GPPLTTTLPA CCGPIARCLT SPQAYENQEQ LKDDDSDLIL NDGDISLTYG DSTVNTEPAT }\\ 
 61 & \texttt{SSAPRRFMGN SSEDALDREL AFGDHELVIR GTRLVLPMDD SEPPLNLLDN TRHGPA}\\ 
     
    \end{tabular}}\vspace{0.05cm}
    \subfloat[\textbf{hNL3cyt:}]{
    \begin{tabular}{r|l}
    1 & \texttt{MYRKDKRRQE PLRQPSPQRG AWAPELGAAP EEELAALQLG PTHHECEAGP PHDTLRLTAL }\\ 
 61 & \texttt{PDYTLTLRRS PDDIPLMTPN TITMIPNSLV GLQTLHPYNT FAAGFNSTGL PHSHSTTRV}\\ 
     
    \end{tabular}}\vspace{0.05cm}
    \subfloat[\textbf{RNaseA:}]{
    \begin{tabular}{r|l}
    1 & \texttt{KETAAAKFER QHMDSSTSAA SSSNYCNQMM KSRNLTKDRC KPVNTFVHES LADVQAVCSQ }\\ 
 61 & \texttt{KNVACKNGQT NCYQSYSTMS ITDCRETGSS KYPNCAYKTT QANKHIIVAC EGNPYVPVHF }\\ 
 121 & \texttt{DASV}\\ 
     
    \end{tabular}}\label{tab:App_SEQ_2}
 \end{table}
\begin{table}[htb]

        \centering\vspace{0.05cm}
    \subfloat[\textbf{A1S150:}]{
    \begin{tabular}{r|l}
    1 & \texttt{GSMASASSSQ RGRSGSGNFG GGRGGGFGGN DNFGRGGNFS GRGGFGGSRG GGGYGGSGDG }\\ 
 61 & \texttt{YNGFGNDGSN FGGGGNYNNQ SSNFGPMKGG NFGGRSSGPY GGGGQYFAKP RNQGGYGGSS }\\ 
 121 & \texttt{SSSSYGSGRR F}\\      
    \end{tabular}}\vspace{0.05cm}
    
    \subfloat[\textbf{M4D:}]{
    \begin{tabular}{r|l}
    1 & \texttt{GSMASASSSQ RGRSGSGNFG GGRGGGFGGN GNFGRGGNFS GRGGFGGSRG GGGYGGSGGG }\\ 
 61 & \texttt{YNGFGNSGSN FGGGGSYNGF GNYNNQSSNF GPMKGGNFGG RSSGPYGGGG QYFAKPRNQG }\\ 
 121 & \texttt{GYGGSSSSSS YGSGRRF}\\ 
     
    \end{tabular}}\vspace{0.05cm}
    \subfloat[\textbf{M8FP4Y:}]{
    \begin{tabular}{r|l}
    1 & \texttt{GSMASASSSQ RGRSGSGNFG GGRGGGYGGN DNGGRGGNYS GRGGFGGSRG GGGYGGSGDG }\\ 
 61 & \texttt{YNGGGNDGSN YGGGGSYNDS GNYNNQSSNF GPMKGGNYGG RSSGGSGGGG QYGAKPRNQG }\\ 
 121 & \texttt{GYGGSSSSSS YGSGRRF}\\ 
     
    \end{tabular}}\vspace{0.05cm}
    \subfloat[\textbf{M9FP3Y:}]{
    \begin{tabular}{r|l}
    1 & \texttt{GSMASASSSQ RGRSGSGNFG GGRGGGYGGN DNGGRGGNYS GRGGFGGSRG GGGYGGSGDG }\\ 
 61 & \texttt{YNGGGNDGSN YGGGGSYNDS GNGNNQSSNF GPMKGGNYGG RSSGGSGGGG QYGAKPRNQG }\\ 
 121 & \texttt{GYGGSSSSSS YGSGRRS}\\ 
     
    \end{tabular}}\vspace{0.05cm}
    \subfloat[\textbf{M10R:}]{
    \begin{tabular}{r|l}
    1 & \texttt{GSMASASSSQ GGSSGSGNFG GGGGGGFGGN DNFGGGGNFS GSGGFGGSGG GGGYGGSGDG }\\ 
 61 & \texttt{YNGFGNDGSN FGGGGSYNDF GNYNNQSSNF GPMKGGNFGG SSSGPYGGGG QYFAKPGNQG }\\ 
 121 & \texttt{GYGGSSSSSS YGSGGGF}\\ 
     
    \end{tabular}}\vspace{0.05cm}
    \subfloat[\textbf{M6R:}]{
    \begin{tabular}{r|l}
    1 & \texttt{GSMASASSSQ GGRSGSGNFG GGRGGGFGGN DNFGGGGNFS GSGGFGGSRG GGGYGGSGDG }\\ 
 61 & \texttt{YNGFGNDGSN FGGGGSYNDF GNYNNQSSNF GPMKGGNFGG SSSGPYGGGG QYFAKPGNQG }\\ 
 121 & \texttt{GYGGSSSSSS YGSGGRF}\\ 
     
    \end{tabular}}\vspace{0.05cm}
    \subfloat[\textbf{P2R:}]{
    \begin{tabular}{r|l}
    1 & \texttt{GSMASASSSQ RGRSGSGNFG GGRGGGFGGN DNFGRGGNFS GRGGFGGSRG GGGYGGSGDG }\\ 
 61 & \texttt{YNGFRNDGSN FGGGGRYNDF GNYNNQSSNF GPMKGGNFGG RSSGPYGGGG QYFAKPRNQG }\\ 
 121 & \texttt{GYGGSSSSSS YGSGRRF}\\ 
     
    \end{tabular}}\vspace{0.05cm}
    \subfloat[\textbf{P7R:}]{
    \begin{tabular}{r|l}
    1 & \texttt{GSMASASSSQ RGRSGRGNFG GGRGGGFGGN DNFGRGGNFS GRGGFGGSRG GGRYGGSGDR }\\ 
 61 & \texttt{YNGFGNDGRN FGGGGSYNDF GNYNNQSSNF GPMKGGNFRG RSSGPYGRGG QYFAKPRNQG }\\ 
 121 & \texttt{GYGGSSSSRS YGSGRRF}\\ 
     
    \end{tabular}}\vspace{0.05cm}
    \subfloat[\textbf{P8D:}]{
    \begin{tabular}{r|l}
    1 & \texttt{GSMASASSSQ RDRSGSGNFG GGRDGGFGGN DNFGRGDNFS GRGDFGGSRD GGGYGGSGDG }\\ 
 61 & \texttt{YNGFGNDGSN FGGGGSYNDF GNYNNQSSNF GPMKGGNFGG RSSDPYGGGG QYFAKPRNQD }\\ 
 121 & \texttt{GYGGSSSSSS YDSGRRF}\\ 
     
    \end{tabular}}\label{tab:App_SEQ_3}
 \end{table}
\begin{table}[htb]

        \centering\vspace{0.05cm}
    \subfloat[\textbf{P12D:}]{
    \begin{tabular}{r|l}
    1 & \texttt{GSMASADSSQ RDRDDSGNFG DGRGGGFGGN DNFGRGGNFS DRGGFGGSRG DGGYGGDGDG }\\ 
 61 & \texttt{YNGFGNDGSN FGGGGSYNDF GNYNNQSSNF DPMKGGNFGD RSSGPYDGGG QYFAKPRNQG }\\ 
 121 & \texttt{GYGGSSSSSS YGSDRRF}\\ 
     \end{tabular}}
        
        \vspace{0.05cm}
    \subfloat[\textbf{M9FP6Y:}]{
    \begin{tabular}{r|l}
    1 & \texttt{GSMASASSSQ RGRSGSGNFG GGRGGGYGGN DNYGRGGNYS GRGGFGGSRG GGGYGGSGDG }\\ 
 61 & \texttt{YNGGGNDGSN YGGGGSYNDS GNYNNQSSNF GPMKGGNYGG RSSGGSGGGG QYGAKPRNQG }\\ 
 121 & \texttt{GYGGSSSSSS YGSGRRY}\\ 
     
    \end{tabular}}\vspace{0.05cm}
    \subfloat[\textbf{M6RP6K:}]{
    \begin{tabular}{r|l}
    1 & \texttt{GSMASASSSQ KGKSGSGNFG GGRGGGFGGN DNFGKGGNFS GRGGFGGSKG GGGYGGSGDG }\\ 
 61 & \texttt{YNGFGNDGSN FGGGGSYNDF GNYNNQSSNF GPMKGGNFGG KSSGGSGGGG QYFAKPRNQG }\\ 
 121 & \texttt{GYGGSSSSSS YGSGRKF}\\ 
     
    \end{tabular}}\vspace{0.05cm}
    \subfloat[\textbf{P4D:}]{
    \begin{tabular}{r|l}
    1 & \texttt{GSMASASSSQ RDRSGSGNFG GGRGGGFGGN DNFGRGGNFS GRGDFGGSRG GGGYGGSGDG }\\ 
 61 & \texttt{YNGFGNDGSN FGGGGSYNDF GNYNNQSSNF GPMKGGNFGG RSSDPYGGGG QYFAKPRNQG }\\ 
 121 & \texttt{GYGGSSSSSS YDSGRRF}\\ 
     
    \end{tabular}}\vspace{0.05cm}
    \subfloat[\textbf{M10RP10K:}]{
    \begin{tabular}{r|l}
    1 & \texttt{GSMASASSSQ KGKSGSGNFG GGKGGGFGGN DNFGKGGNFS GKGGFGGSKG GGGYGGSGDG }\\ 
 61 & \texttt{YNGFGNDGSN FGGGGSYNDF GNYNNQSSNF GPMKGGNFGG KSSGGSGGGG QYFAKPKNQG }\\ 
 121 & \texttt{GYGGSSSSSS YGSGKKF}\\ 
     
    \end{tabular}}\vspace{0.05cm}
    \subfloat[\textbf{M3RP3K:}]{
    \begin{tabular}{r|l}
    1 & \texttt{GSMASASSSQ RGKSGSGNFG GGRGGGFGGN DNFGRGGNFS GRGGFGGSKG GGGYGGSGDG }\\ 
 61 & \texttt{YNGFGNDGSN FGGGGSYNDF GNYNNQSSNF GPMKGGNFGG RSSGGSGGGG QYFAKPRNQG }\\ 
 121 & \texttt{GYGGSSSSSS YGSGRKF}\\ 
     
    \end{tabular}}\vspace{0.05cm}
    \subfloat[\textbf{P7FM7Y:}]{
    \begin{tabular}{r|l}
    1 & \texttt{GSMASASSSQ RGRSGSGNFG GGRGGGFGGN DNFGRGGNFS GRGGFGGSRG GGGFGGSGDG }\\ 
 61 & \texttt{FNGFGNDGSN FGGGGSFNDF GNFNNQSSNF GPMKGGNFGG RSSGGSGGGG QFFAKPRNQG }\\ 
 121 & \texttt{GFGGSSSSSS FGSGRRF}\\ 
     
    \end{tabular}}\vspace{0.05cm}
    \subfloat[\textbf{P7KP12D:}]{
    \begin{tabular}{r|l}
    1 & \texttt{GSMASADSSQ RDRDDKGNFG DGRGGGFGGN DNFGRGGNFS DRGGFGGSRG DGKYGGDGDK }\\ 
 61 & \texttt{YNGFGNDGKN FGGGGSYNDF GNYNNQSSNF DPMKGGNFKD RSSGPYDKGG QYFAKPRNQG }\\ 
 121 & \texttt{GYGGSSSSKS YGSDRRF}\\ 
     
    \end{tabular}}\vspace{0.05cm}
    \subfloat[\textbf{A1:}]{
    \begin{tabular}{r|l}
    1 & \texttt{GSMASASSSQ RGRSGSGNFG GGRGGGFGGN DNFGRGGNFS GRGGFGGSRG GGGYGGSGDG }\\ 
 61 & \texttt{YNGFGNDGSN FGGGGSYNDF GNYNNQSSNF GPMKGGNFGG RSSGGSGGGG QYFAKPRNQG }\\ 
 121 & \texttt{GYGGSSSSSS YGSGRRF}\\ 
     
    \end{tabular}}\label{tab:App_SEQ_4}
 \end{table}
\begin{table}[htb]

        \centering\vspace{0.05cm}
    \subfloat[\textbf{P7KP12Db:}]{
    \begin{tabular}{r|l}
    1 & \texttt{GSMASAKSSQ RDRDDDGNFG KGRGGGFGGN KNFGRGGNFS KRGGFGGSRG KGKYGGKGDD }\\ 
 61 & \texttt{YNGFGNDGDN FGGGGSYNDF GNYNNQSSNF DPMDGGNFDD RSSGPYDDGG QYFADPRNQG }\\ 
 121 & \texttt{GYGGSSSSKS YGSKRRF}\\ 
     
    \end{tabular}}
        \vspace{0.05cm}
    \subfloat[\textbf{M12FP12YM10R:}]{
    \begin{tabular}{r|l}
    1 & \texttt{GSMASASSSQ GGSSGSGNYG GGGGGGYGGN DNYGGGGNYS GSGGYGGSGG GGGYGGSGDG }\\ 
 61 & \texttt{YNGYGNDGSN YGGGGSYNDY GNYNNQSSNY GPMKGGNYGG SSSGPYGGGG QYYAKPGNQG }\\ 
 121 & \texttt{GYGGSSSSSS YGSGGGY}\\ 
     
    \end{tabular}}\vspace{0.05cm}
    \subfloat[\textbf{M10FP7RP12D:}]{
    \begin{tabular}{r|l}
    1 & \texttt{GSMASADSSQ RDRDDRGNFG DGRGGGGGGN DNFGRGGNGS DRGGGGGSRG DGRYGGDGDR }\\ 
 61 & \texttt{YNGGGNDGRN GGGGGSYNDG GNYNNQSSNG DPMKGGNGRD RSSGPYDRGG QYGAKPRNQG }\\ 
 121 & \texttt{GYGGSSSSRS YGSDRRG}\\ 
     
    \end{tabular}}\vspace{0.05cm}
    \subfloat[\textbf{P12E:}]{
    \begin{tabular}{r|l}
    1 & \texttt{GSMASAESSQ REREESGNFG EGRGGGFGGN DNFGRGGNFS ERGGFGGSRG EGGYGGEGDG }\\ 
 61 & \texttt{YNGFGNDGSN FGGGGSYNDF GNYNNQSSNF EPMKGGNFGE RSSGPYEGGG QYFAKPRNQG }\\ 
 121 & \texttt{GYGGSSSSSS YGSERRF}\\ 
     
    \end{tabular}}\vspace{0.05cm}
    \subfloat[\textbf{M12FP12Y:}]{
    \begin{tabular}{r|l}
    1 & \texttt{GSMASASSSQ RGRSGSGNYG GGRGGGYGGN DNYGRGGNYS GRGGYGGSRG GGGYGGSGDG }\\ 
 61 & \texttt{YNGYGNDGSN YGGGGSYNDY GNYNNQSSNY GPMKGGNYGG RSSGGSGGGG QYYAKPRNQG }\\ 
 121 & \texttt{GYGGSSSSSS YGSGRRY}\\ 
     
    \end{tabular}}\vspace{0.05cm}
    \subfloat[\textbf{aSyn140:}]{
    \begin{tabular}{r|l}
    1 & \texttt{MDVFMKGLSK AKEGVVAAAE KTKQGVAEAA GKTKEGVLYV GSKTKEGVVH GVATVAEKTK }\\ 
 61 & \texttt{EQVTNVGGAV VTGVTAVAQK TVEGAGSIAA ATGFVKKDQL GKNEEGAPQE GILEDMPVDP }\\ 
 121 & \texttt{DNEAYEMPSE EGYQDYEPEA}\\ 
     
    \end{tabular}}\vspace{0.05cm}
    \subfloat[\textbf{FhuA:}]{
    \begin{tabular}{r|l}
    1 & \texttt{SESAWGPAAT IAARQSATGT KTDTPIQKVP QSISVVTAEE MALHQPKSVK EALSYTPGVS }\\ 
 61 & \texttt{VGTRGASNTY DHLIIRGFAA EGQSQNNYLN GLKLQGNFYN DAVIDPYMLE RAEIMRGPVS }\\ 
 121 & \texttt{VLYGKSSPGG LLNMVSKRPT TEPL}\\ 
     
    \end{tabular}}\vspace{0.05cm}
    \subfloat[\textbf{K27:}]{
    \begin{tabular}{r|l}
    1 & \texttt{MSSPGSPGTP GSRSRTPSLP TPPTREPKKV AVVRTPPKSP SSAKSRLQTA PVPMPDLKNV }\\ 
 61 & \texttt{KSKIGSTENL KHQPGGGKVQ IVYKPVDLSK VTSKCGSLGN IHHKPGGGQV EVKSEKLDFK }\\ 
 121 & \texttt{DRVQSKIGSL DNITHVPGGG NKKIETHKLT FRENAKAKTD HGAEIVY}\\ 
     
    \end{tabular}}\vspace{0.05cm}
    \subfloat[\textbf{ANAC046:}]{
    \begin{tabular}{r|l}
    1 & \texttt{NAPSTTITTT KQLSRIDSLD NIDHLLDFSS LPPLIDPGFL GQPGPSFSGA RQQHDLKPVL }\\ 
 61 & \texttt{HHPTTAPVDN TYLPTQALNF PYHSVHNSGS DFGYGAGSGN NNKGMIKLEH SLVSVSQETG }\\ 
 121 & \texttt{LSSDVNTTAT PEISSYPMMM NPAMMDGSKS ACDGLDDLIF WEDLYTS}\\ 
     
    \end{tabular}}\label{tab:App_SEQ_5}
 \end{table}
\begin{table}[htb]

        \centering\vspace{0.05cm}
    \subfloat[\textbf{K10:}]{
    \begin{tabular}{r|l}
    1 & \texttt{MQTAPVPMPD LKNVKSKIGS TENLKHQPGG GKVQIVYKPV DLSKVTSKCG SLGNIHHKPG }\\ 
 61 & \texttt{GGQVEVKSEK LDFKDRVQSK IGSLDNITHV PGGGNKKIET HKLTFRENAK AKTDHGAEIV }\\ 
 121 & \texttt{YKSPVVSGDT SPRHLSNVSS TGSIDMVDSP QLATLADEVS ASLAKQGL}\\ 
     
    \end{tabular}}
        
        \vspace{0.05cm}
    \subfloat[\textbf{K25:}]{
    \begin{tabular}{r|l}
    1 & \texttt{MAEPRQEFEV MEDHAGTYGL GDRKDQGGYT MHQDQEGDTD AGLKAEEAGI GDTPSLEDEA }\\ 
 61 & \texttt{AGHVTQARMV SKSKDGTGSD DKKAKGADGK TKIATPRGAA PPGQKGQANA TRIPAKTPPA }\\ 
 121 & \texttt{PKTPPSSGEP PKSGDRSGYS SPGSPGTPGS RSRTPSLPTP PTREPKKVAV VRTPPKSPSS }\\ 
 181 & \texttt{AKSRL}\\ 
     
    \end{tabular}}\vspace{0.05cm}
    \subfloat[\textbf{K32:}]{
    \begin{tabular}{r|l}
    1 & \texttt{MSSPGSPGTP GSRSRTPSLP TPPTREPKKV AVVRTPPKSP SSAKSRLQTA PVPMPDLKNV }\\ 
 61 & \texttt{KSKIGSTENL KHQPGGGKVQ IINKKLDLSN VQSKCGSKDN IKHVPGGGSV QIVYKPVDLS }\\ 
 121 & \texttt{KVTSKCGSLG NIHHKPGGGQ VEVKSEKLDF KDRVQSKIGS LDNITHVPGG GNKKIETHKL }\\ 
 181 & \texttt{TFRENAKAKT DHGAEIVY}\\ 
     
    \end{tabular}}\vspace{0.05cm}
    \subfloat[\textbf{CAHSD:}]{
    \begin{tabular}{r|l}
    1 & \texttt{MSGRNVESHM ERNEKVVVNN SGHADVKKQQ QQVEHTEFTH TEVKAPLIHP APPIISTGAA }\\ 
 61 & \texttt{GLAEEIVGQG FTASAARISG GTAEVHLQPS AAMTEEARRD QERYRQEQES IAKQQEREME }\\ 
 121 & \texttt{KKTEAYRKTA EAEAEKIRKE LEKQHARDVE FRKDLIESTI DRQKREVDLE AKMAKRELDR }\\ 
 181 & \texttt{EGQLAKEALE RSRLATNVEV NFDSAAGHTV SGGTTVSTSD KMEIKRN}\\ 
     
    \end{tabular}}\vspace{0.05cm}
    \subfloat[\textbf{K23:}]{
    \begin{tabular}{r|l}
    1 & \texttt{MAEPRQEFEV MEDHAGTYGL GDRKDQGGYT MHQDQEGDTD AGLKAEEAGI GDTPSLEDEA }\\ 
 61 & \texttt{AGHVTQARMV SKSKDGTGSD DKKAKGADGK TKIATPRGAA PPGQKGQANA TRIPAKTPPA }\\ 
 121 & \texttt{PKTPPSSGEP PKSGDRSGYS SPGSPGTPGS RSRTPSLPTP PTREPKKVAV VRTPPKSPSS }\\ 
 181 & \texttt{AKSRLTHKLT FRENAKAKTD HGAEIVYKSP VVSGDTSPRH LSNVSSTGSI DMVDSPQLAT }\\ 
 241 & \texttt{LADEVSASLA KQGL}\\ 
     
    \end{tabular}}\vspace{0.05cm}
    \subfloat[\textbf{tau35:}]{
    \begin{tabular}{r|l}
    1 & \texttt{EPPKSGDRSG YSSPGSPGTP GSRSRTPSLP TPPTREPKKV AVVRTPPKSP SSAKSRLQTA }\\ 
 61 & \texttt{PVPMPDLKNV KSKIGSTENL KHQPGGGKVQ IINKKLDLSN VQSKCGSKDN IKHVPGGGSV }\\ 
 121 & \texttt{QIVYKPVDLS KVTSKCGSLG NIHHKPGGGQ VEVKSEKLDF KDRVQSKIGS LDNITHVPGG }\\ 
 181 & \texttt{GNKKIETHKL TFRENAKAKT DHGAEIVYKS PVVSGDTSPR HLSNVSSTGS IDMVDSPQLA }\\ 
 241 & \texttt{TLADEVSASL AKQGL}\\ 
     
    \end{tabular}}\vspace{0.05cm}
    \subfloat[\textbf{CoRNID:}]{
    \begin{tabular}{r|l}
    1 & \texttt{GPHMQVPRTH RLITLADHIC QIITQDFARN QVPSQASTST FQTSPSALSS TPVRTKTSSR }\\ 
 61 & \texttt{YSPESQSQTV LHPRPGPRVS PENLVDKSRG SRPGKSPERS HIPSEPYEPI SPPQGPAVHE }\\ 
 121 & \texttt{KQDSMLLLSQ RGVDPAEQRS DSRSPGSISY LPSFFTKLES TSPMVKSKKQ EIFRKLNSSG }\\ 
 181 & \texttt{GGDSDMAAAQ PGTEIFNLPA VTTSGAVSSR SHSFADPASN LGLEDIIRKA LMGSFDDKVE }\\ 
 241 & \texttt{DHGVVMSHPV GIMPGSASTS VVTSSEARRD E}\\ 
     
    \end{tabular}}\label{tab:App_SEQ_6}
 \end{table}
\begin{table}[htb]

        \centering\vspace{0.05cm}
    \subfloat[\textbf{K44:}]{
    \begin{tabular}{r|l}
    1 & \texttt{MAEPRQEFEV MEDHAGTYGL GDRKDQGGYT MHQDQEGDTD AGLKAEEAGI GDTPSLEDEA }\\ 
 61 & \texttt{AGHVTQARMV SKSKDGTGSD DKKAKGADGK TKIATPRGAA PPGQKGQANA TRIPAKTPPA }\\ 
 121 & \texttt{PKTPPSSGEP PKSGDRSGYS SPGSPGTPGS RSRTPSLPTP PTREPKKVAV VRTPPKSPSS }\\ 
 181 & \texttt{AKSRLQTAPV PMPDLKNVKS KIGSTENLKH QPGGGKVQIV YKPVDLSKVT SKCGSLGNIH }\\ 
 241 & \texttt{HKPGGGQVEV KSEKLDFKDR VQSKIGSLDN ITHVPGGGNK KIE}\\ 
     
    \end{tabular}}
        \vspace{0.05cm}
    \subfloat[\textbf{PNtS5:}]{
    \begin{tabular}{r|l}
    1 & \texttt{DWNNQSIVKT GERQHGIHIQ GSDPGGVRTA SGTTIKVSGR QAQGILLENP AAELQFRNGS }\\ 
 61 & \texttt{VTSSGQLSDD GIEDFLGTVT VDAGELVADH ATLANVGDTW DDDGIALYVA GEQAQASIAD }\\ 
 121 & \texttt{STLQGAGGVQ IEDGANVTVQ ESAIVDGGLH IGALQSLQPR RLPPSRVVLR KTNVTAVPAS }\\ 
 181 & \texttt{GAPAAVSVLG ASKLTLRGGH ITGGRAAGVA AMQGAVVHLQ RATIRRGRAL AGGAVPGGAV }\\ 
 241 & \texttt{PGGAVPGGFG PGGFGPVLDG WYGVDVSGSS VELAQSIVEA PELGAAIRVG RGARVTVPGG }\\ 
 301 & \texttt{SLSAPHGNVI ETGGARRFAP QAAPLSITLQ AGAH}\\ 
     
    \end{tabular}}\vspace{0.05cm}
    \subfloat[\textbf{PNt:}]{
    \begin{tabular}{r|l}
    1 & \texttt{DWNNQSIVKT GERQHGIHIQ GSDPGGVRTA SGTTIKVSGR QAQGILLENP AAELQFRNGS }\\ 
 61 & \texttt{VTSSGQLSDD GIRRFLGTVT VKAGKLVADH ATLANVGDTW DDDGIALYVA GEQAQASIAD }\\ 
 121 & \texttt{STLQGAGGVQ IERGANVTVQ RSAIVDGGLH IGALQSLQPE DLPPSRVVLR DTNVTAVPAS }\\ 
 181 & \texttt{GAPAAVSVLG ASELTLDGGH ITGGRAAGVA AMQGAVVHLQ RATIRRGEAL AGGAVPGGAV }\\ 
 241 & \texttt{PGGAVPGGFG PGGFGPVLDG WYGVDVSGSS VELAQSIVEA PELGAAIRVG RGARVTVPGG }\\ 
 301 & \texttt{SLSAPHGNVI ETGGARRFAP QAAPLSITLQ AGAH}\\ 
     
    \end{tabular}}\vspace{0.05cm}
    \subfloat[\textbf{PNtS1:}]{
    \begin{tabular}{r|l}
    1 & \texttt{DWNNQSIVKT GERQHGIHIQ GSDPGGVRTA SGTTIKVSGR QAQGILLENP AAELQFRNGS }\\ 
 61 & \texttt{VTSSGQKSDD GIRRFLGTVT VLAGKLVADH ATLANVGDTW DDDGIALYVA GEQAQASIAD }\\ 
 121 & \texttt{STLQGAGGVQ IERGANVTVQ RSAIVLGGLH IGALQSLQPE DDPPSRVVLR DTNVTAVPAS }\\ 
 181 & \texttt{GAPAAVSVLG ASLLTLDGGH ITGGRAAGVA AMQGAVVHEQ RATIRRGEAL AGGAVPGGAV }\\ 
 241 & \texttt{PGGAVPGGFG PGGFGPVLDG WYGVDVSGSS VELAQSIVEA PELGAAIRVG RGARVTVPGG }\\ 
 301 & \texttt{SLSAPHGNVI ETGGARRFAP QAAPLSITLQ AGAH}\\ 
     
    \end{tabular}}\vspace{0.05cm}
    \subfloat[\textbf{PNtS4:}]{
    \begin{tabular}{r|l}
    1 & \texttt{DWNNQSIVKT GERQHGIHIQ GSDPGGVRTA SGTTIKVSGR QAQGILLENP AAELQFRNGS }\\ 
 61 & \texttt{VTSSGQLSFV GITRDLGRDT VKAGKLVADH ATLANVGDTW DDDGIALYVA GEQAQASIAD }\\ 
 121 & \texttt{STLQGAGGVQ IERGADVRVQ REAIVDGGLH NGALQSLQPS ILPPSTVVLR DTNVTAVPAS }\\ 
 181 & \texttt{GAPAAVLVSG ASGLRLDGGH IHEGRAAGVA AMQGAVVTLQ TATIRRGEAL AGGAVPGGAV }\\ 
 241 & \texttt{PGGAVPGGFG PGGFGPVLDG WYGVDVSGSS VELAQSIVEA PELGAAIRVG RGARVTVPGG }\\ 
 301 & \texttt{SLSAPHGNVI ETGGARRFAP QAAPLSITLQ AGAH}\\ 
     
    \end{tabular}}\label{tab:App_SEQ_7}
 \end{table}
\begin{table}[htb]

        \centering\vspace{0.05cm}
    \subfloat[\textbf{PNtS6:}]{
    \begin{tabular}{r|l}
    1 & \texttt{DWNNQSIVKT GERQHGIHIQ GSDPGGVRTA SGTTIKVSGR QAQGILLENP AAELQFRNGS }\\ 
 61 & \texttt{VTSSGQLSDR GIDRFLGTVT VEAGKLVADH ATLANVGDTW DKDGIALYVA GRQAQASIAD }\\ 
 121 & \texttt{STLQGAGGVQ IREGANVTVQ RSAIVDGGLH IGALQSLQPE RLPPSDVVLR DTNVTAVPAS }\\ 
 181 & \texttt{GAPAAVSVLG ASRLTLDGGH ITGGDAAGVA AMQGAVVHLQ RATIERGEAL AGGAVPGGAV }\\ 
 241 & \texttt{PGGAVPGGFG PGGFGPVLDG WYGVDVSGSS VELAQSIVEA PELGAAIRVG RGARVTVPGG }\\ 
 301 & \texttt{SLSAPHGNVI ETGGARRFAP QAAPLSITLQ AGAH}\\ 
     
    \end{tabular}}
        \vspace{0.05cm}
    \subfloat[\textbf{GHRICD:}]{
    \begin{tabular}{r|l}
    1 & \texttt{SKQQRIKMLI LPPVPVPKIK GIDPDLLKEG KLEEVNTILA IHDSYKPEFH SDDSWVEFIE }\\ 
 61 & \texttt{LDIDEPDEKT EESDTDRLLS SDHEKSHSNL GVKDGDSGRT SCCEPDILET DFNANDIHEG }\\ 
 121 & \texttt{TSEVAQPQRL KGEADLLCLD QKNQNNSPYH DACPATQQPS VIQAEKNKPQ PLPTEGAEST }\\ 
 181 & \texttt{HQAAHIQLSN PSSLSNIDFY AQVSDITPAG SVVLSPGQKN KAGMSQCDMH PEMVSLCQEN }\\ 
 241 & \texttt{FLMDNAYFCE ADAKKCIPVA PHIKVESHIQ PSLNQEDIYI TTESLTTAAG RPGTGEHVPG }\\ 
 301 & \texttt{SEMPVPDYTS IHIVQSPQGL ILNATALPLP DKEFLSSCGY VSTDQLNKIM P}\\ 
     
    \end{tabular}}\vspace{0.05cm}
    \subfloat[\textbf{ht2N3R:}]{
    \begin{tabular}{r|l}
    1 & \texttt{MAEPRQEFEV MEDHAGTYGL GDRKDQGGYT MHQDQEGDTD AGLKESPLQT PTEDGSEEPG }\\ 
 61 & \texttt{SETSDAKSTP TAEDVTAPLV DEGAPGKQAA AQPHTEIPEG TTAEEAGIGD TPSLEDEAAG }\\ 
 121 & \texttt{HVTQARMVSK SKDGTGSDDK KAKGADGKTK IATPRGAAPP GQKGQANATR IPAKTPPAPK }\\ 
 181 & \texttt{TPPSSGEPPK SGDRSGYSSP GSPGTPGSRS RTPSLPTPPT REPKKVAVVR TPPKSPSSAK }\\ 
 241 & \texttt{SRLQTAPVPM PDLKNVKSKI GSTENLKHQP GGGKVQIVYK PVDLSKVTSK CGSLGNIHHK }\\ 
 301 & \texttt{PGGGQVEVKS EKLDFKDRVQ SKIGSLDNIT HVPGGGNKKI ETHKLTFREN AKAKTDHGAE }\\ 
 361 & \texttt{IVYKSPVVSG DTSPRHLSNV SSTGSIDMVD SPQLATLADE VSASLAKQGL}\\ 
     
    \end{tabular}}\vspace{0.05cm}
    \subfloat[\textbf{ht2N4R:}]{
    \begin{tabular}{r|l}
    1 & \texttt{MAEPRQEFEV MEDHAGTYGL GDRKDQGGYT MHQDQEGDTD AGLKESPLQT PTEDGSEEPG }\\ 
 61 & \texttt{SETSDAKSTP TAEDVTAPLV DEGAPGKQAA AQPHTEIPEG TTAEEAGIGD TPSLEDEAAG }\\ 
 121 & \texttt{HVTQARMVSK SKDGTGSDDK KAKGADGKTK IATPRGAAPP GQKGQANATR IPAKTPPAPK }\\ 
 181 & \texttt{TPPSSGEPPK SGDRSGYSSP GSPGTPGSRS RTPSLPTPPT REPKKVAVVR TPPKSPSSAK }\\ 
 241 & \texttt{SRLQTAPVPM PDLKNVKSKI GSTENLKHQP GGGKVQIINK KLDLSNVQSK CGSKDNIKHV }\\ 
 301 & \texttt{PGGGSVQIVY KPVDLSKVTS KCGSLGNIHH KPGGGQVEVK SEKLDFKDRV QSKIGSLDNI }\\ 
 361 & \texttt{THVPGGGNKK IETHKLTFRE NAKAKTDHGA EIVYKSPVVS GDTSPRHLSN VSSTGSIDMV }\\ 
 421 & \texttt{DSPQLATLAD EVSASLAKQG L}\\ 
     
    \end{tabular}}\label{tab:App_SEQ_8}
 \end{table}
}

\clearpage
\section{Sequence information}
{\linespread{1}
\begin{table}[h!]
\centering
\begin{tabular}{c||c|c|c|c|c|c|c|c|c|c}
Name &\#Particles &$f_{+}$~\cite{Das_Pappu_kappa} &$f_{-}$~\cite{Das_Pappu_kappa} &fcr~\cite{Das_Pappu_kappa} &ncpr~\cite{Das_Pappu_kappa} &$\kappa$~\cite{Das_Pappu_kappa} &$\delta$~\cite{Das_Pappu_kappa} &SCD~\cite{Sawle_Ghosh_SCD} &nSCD~\cite{Devarajan_nSCD} \\ \hline 
Hst5 & 24 & 0.58 & 0.083 & 0.67 & 0.50 & 0.30 & 0.37 & 0.87 & 0.25 \\  
Hst52 & 48 & 0.58 & 0.083 & 0.67 & 0.50 & 0.30 & 0.37 & 4.3 & 0.13 \\  
p532070 & 62 & 0.13 & 0.21 & 0.34 & 0.081 & 0.30 & 0.019 & 2.7 & 0.26 \\  
ACTR & 71 & 0.085 & 0.18 & 0.27 & 0.099 & 0.15 & 0.036 & 1.3 & 0.12 \\  
DSS1 & 71 & 0.070 & 0.32 & 0.39 & 0.25 & 0.39 & 0.16 & 6.9 & 0.31 \\  
 Ash1 & 81 & 0.22 & 0.012 & 0.23 & 0.21 & 0.19 & 0.19 & 6.3 & 0.15 \\  
 CTD2 & 83 & 0 & 0 & 0 & 0 & - & - & 0 & - \\  
 Sic1 & 92 & 0.12 & 0 & 0.12 & 0.12 & 0.21 & 0.12 & 3.4 & 0.076 \\  
 SH4UD & 95 & 0.15 & 0.074 & 0.22 & 0.074 & 0.17 & 0.025 & 0.11 & 0.046 \\  
 ColNT & 98 & 0.18 & 0.051 & 0.23 & 0.13 & 0.24 & 0.075 & -0.080 & 0.056 \\  
 p27Cv15 & 107 & 0.16 & 0.13 & 0.29 & 0.028 & 0.15 & 0.0027 & -0.75 & 0.050 \\  
 p27Cv31 & 107 & 0.16 & 0.13 & 0.29 & 0.028 & 0.28 & 0.0027 & -0.94 & 0.063 \\  
 p27Cv44 & 107 & 0.16 & 0.13 & 0.29 & 0.028 & 0.40 & 0.0027 & -1.5 & 0.10 \\  
 p27Cv56 & 107 & 0.16 & 0.13 & 0.29 & 0.028 & 0.54 & 0.0027 & -3.9 & 0.26 \\  
 p27Cv78 & 107 & 0.16 & 0.13 & 0.29 & 0.028 & 0.77 & 0.0027 & -7.6 & 0.51 \\  
 p27Cv14 & 107 & 0.16 & 0.13 & 0.29 & 0.028 & 0.15 & 0.0027 & -0.61 & 0.040 \\  
 p15PAF & 111 & 0.20 & 0.11 & 0.31 & 0.090 & 0.19 & 0.026 & 0.67 & 0.084 \\  
 PTMA & 111 & 0.090 & 0.48 & 0.57 & 0.39 & 0.42 & 0.26 & 39 & 0.23 \\  
 NHE6cmdd & 116 & 0.086 & 0.17 & 0.26 & 0.086 & 0.22 & 0.029 & 2.6 & 0.095 \\  
 hNL3cyt & 119 & 0.14 & 0.10 & 0.24 & 0.042 & 0.18 & 0.0072 & -0.87 & 0.091 \\  
 RNaseA & 124 & 0.15 & 0.081 & 0.23 & 0.065 & 0.12 & 0.018 & -0.018 & 0.037 \\  
 A1S150 & 131 & 0.092 & 0.023 & 0.11 & 0.069 & 0.20 & 0.041 & 1.8 & 0.054 \\  
 M4D & 137 & 0.088 & 0 & 0.088 & 0.088 & 0.19 & 0.088 & 3.4 & 0.069 \\  
 M8FP4Y & 137 & 0.088 & 0.029 & 0.12 & 0.058 & 0.22 & 0.029 & 1.3 & 0.054 \\  
 M9FP3Y & 137 & 0.088 & 0.029 & 0.12 & 0.058 & 0.22 & 0.029 & 1.3 & 0.054 \\  
 M10R & 137 & 0.015 & 0.029 & 0.044 & 0.015 & 0.36 & 0.0049 & -0.14 & 0.31 \\  
 M6R & 137 & 0.044 & 0.029 & 0.073 & 0.015 & 0.26 & 0.0029 & -0.088 & 0.079 \\  
 P2R & 137 & 0.10 & 0.029 & 0.13 & 0.073 & 0.17 & 0.041 & 2.3 & 0.037 \\  
 P7R & 137 & 0.14 & 0.029 & 0.17 & 0.11 & 0.15 & 0.071 & 5.2 & 0.055 \\  
 P8D & 137 & 0.088 & 0.088 & 0.18 & 0 & 0.11 & 0 & -0.37 & 0.037 \\  
 P12D & 137 & 0.088 & 0.12 & 0.20 & 0.029 & 0.12 & 0.0042 & -0.32 & 0.048 \\  
 M9FP6Y & 137 & 0.088 & 0.029 & 0.12 & 0.058 & 0.22 & 0.029 & 1.3 & 0.054 \\  
 M6RP6K & 137 & 0.088 & 0.029 & 0.12 & 0.058 & 0.22 & 0.029 & 1.3 & 0.054 \\  
 \end{tabular}
\end{table}

\begin{table}[htb]
\centering
\begin{tabular}{c||c|c|c|c|c|c|c|c|c|c}
Name &\#Particles &$f_{+}$~\cite{Das_Pappu_kappa} &$f_{-}$~\cite{Das_Pappu_kappa} &fcr~\cite{Das_Pappu_kappa} &ncpr~\cite{Das_Pappu_kappa} &$\kappa$~\cite{Das_Pappu_kappa} &$\delta$~\cite{Das_Pappu_kappa} &SCD~\cite{Sawle_Ghosh_SCD} &nSCD~\cite{Devarajan_nSCD} \\ \hline
 P4D & 137 & 0.088 & 0.058 & 0.15 & 0.029 & 0.15 & 0.0058 & 0.15 & 0.027 \\  
 M10RP10K & 137 & 0.088 & 0.029 & 0.12 & 0.058 & 0.22 & 0.029 & 1.3 & 0.054 \\  
 M3RP3K & 137 & 0.088 & 0.029 & 0.12 & 0.058 & 0.22 & 0.029 & 1.3 & 0.054 \\  
 P7FM7Y & 137 & 0.088 & 0.029 & 0.12 & 0.058 & 0.22 & 0.029 & 1.3 & 0.054 \\  
 P7KP12D & 137 & 0.14 & 0.12 & 0.26 & 0.022 & 0.076 & 0.0019 & -0.42 & 0.027 \\  
 A1 & 137 & 0.088 & 0.029 & 0.12 & 0.058 & 0.22 & 0.029 & 1.3 & 0.054 \\  
 P7KP12Db & 137 & 0.14 & 0.12 & 0.26 & 0.022 & 0.24 & 0.0019 & -1.2 & 0.067 \\  
M12FP12YM10R & 137 & 0.015 & 0.029 & 0.044 & 0.015 & 0.36 & 0.0049 & -0.14 & 0.31 \\  
 M10FP7RP12D & 137 & 0.14 & 0.12 & 0.26 & 0.022 & 0.076 & 0.0019 & -0.42 & 0.027 \\  
 P12E & 137 & 0.088 & 0.12 & 0.20 & 0.029 & 0.12 & 0.0042 & -0.32 & 0.048 \\  
 M12FP12Y & 137 & 0.088 & 0.029 & 0.12 & 0.058 & 0.22 & 0.029 & 1.3 & 0.054 \\  
 aSyn140 & 140 & 0.11 & 0.17 & 0.29 & 0.057 & 0.17 & 0.011 & -1.3 & 0.14 \\  
 FhuA & 144 & 0.10 & 0.083 & 0.19 & 0.021 & 0.17 & 0.0023 & -0.24 & 0.023 \\  
 K27 & 167 & 0.21 & 0.078 & 0.29 & 0.13 & 0.13 & 0.060 & 4.6 & 0.054 \\  
 ANAC046 & 167 & 0.084 & 0.11 & 0.19 & 0.024 & 0.20 & 0.0030 & 1.8 & 0.055 \\  
 K10 & 168 & 0.18 & 0.10 & 0.28 & 0.077 & 0.16 & 0.021 & -0.21 & 0.043 \\  
 K25 & 185 & 0.17 & 0.13 & 0.30 & 0.038 & 0.17 & 0.0048 & -4.8 & 0.14 \\  
 K32 & 198 & 0.21 & 0.076 & 0.28 & 0.13 & 0.12 & 0.061 & 6.2 & 0.051 \\  
 CAHSD & 227 & 0.20 & 0.19 & 0.38 & 0.013 & 0.072 & 0.00046 & -0.10 & 0.0043 \\  
 K23 & 254 & 0.16 & 0.13 & 0.29 & 0.035 & 0.18 & 0.0044 & -3.8 & 0.074 \\  
 tau35 & 255 & 0.18 & 0.086 & 0.27 & 0.098 & 0.14 & 0.036 & 2.9 & 0.060 \\  
 CoRNID & 271 & 0.14 & 0.10 & 0.24 & 0.037 & 0.16 & 0.0056 & -1.1 & 0.027 \\  
 K44 & 283 & 0.18 & 0.12 & 0.30 & 0.064 & 0.16 & 0.014 & -3.2 & 0.082 \\  
 PNtS5 & 334 & 0.093 & 0.084 & 0.18 & 0.0090 & 0.27 & 0.00046 & -2.5 & 0.095 \\  
 PNt & 334 & 0.093 & 0.084 & 0.18 & 0.0090 & 0.21 & 0.00046 & -0.38 & 0.023 \\  
 PNtS1 & 334 & 0.093 & 0.084 & 0.18 & 0.0090 & 0.20 & 0.00046 & -0.36 & 0.022 \\  
 PNtS4 & 334 & 0.093 & 0.084 & 0.18 & 0.0090 & 0.16 & 0.00046 & -0.33 & 0.021 \\  
 PNtS6 & 334 & 0.093 & 0.084 & 0.18 & 0.0090 & 0.14 & 0.00046 & -0.11 & 0.014 \\  
 GHRICD & 351 & 0.11 & 0.16 & 0.27 & 0.048 & 0.16 & 0.0087 & 10 & 0.054 \\  
 ht2N3R & 410 & 0.16 & 0.13 & 0.29 & 0.024 & 0.19 & 0.0021 & -7.6 & 0.072 \\  
 ht2N4R & 441 & 0.16 & 0.13 & 0.29 & 0.032 & 0.18 & 0.0035 & -8.1 & 0.070 \\  
 \end{tabular}
\caption{Characteristic features of charge patterning of the different protein sequences.}
\label{Tab:ProteinChargeInfoApp}
\end{table}

}

\clearpage
\newpage
\section{Additional simulation results}
\begin{figure}[h!]
    \centering
    \includegraphics[width=0.91\textwidth]{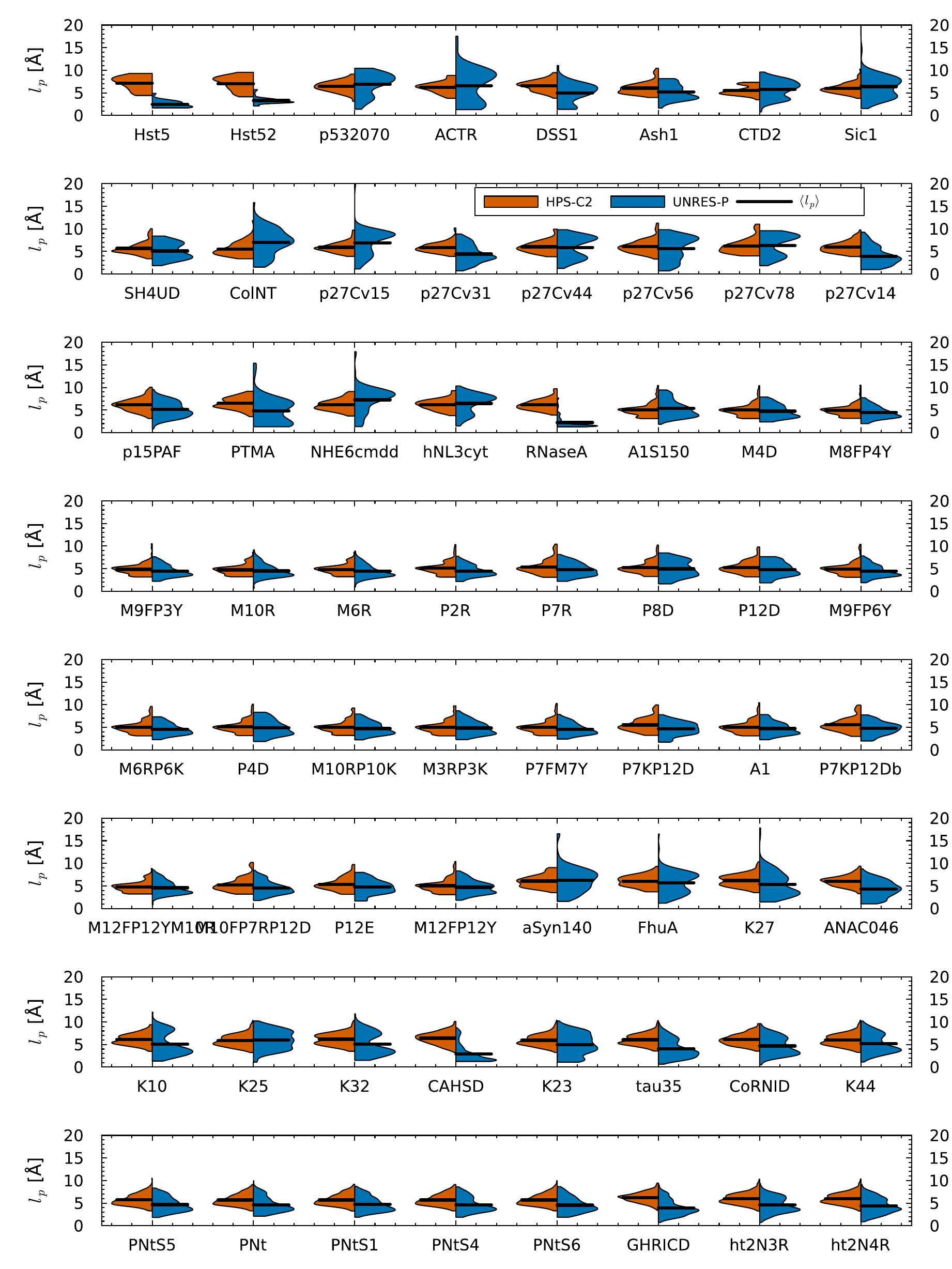}
    \caption{Distribution of the local persistence lengths for IDPs simulated with the HPS-C2 and UNRES-P models. Black lines indicate the arithmetic mean of $\ell_{\text{p},i}$.}
    \label{fig:AppPersDistributionsAll}
\end{figure}

\begin{figure}
    \centering
    \includegraphics[width=0.99\textwidth]{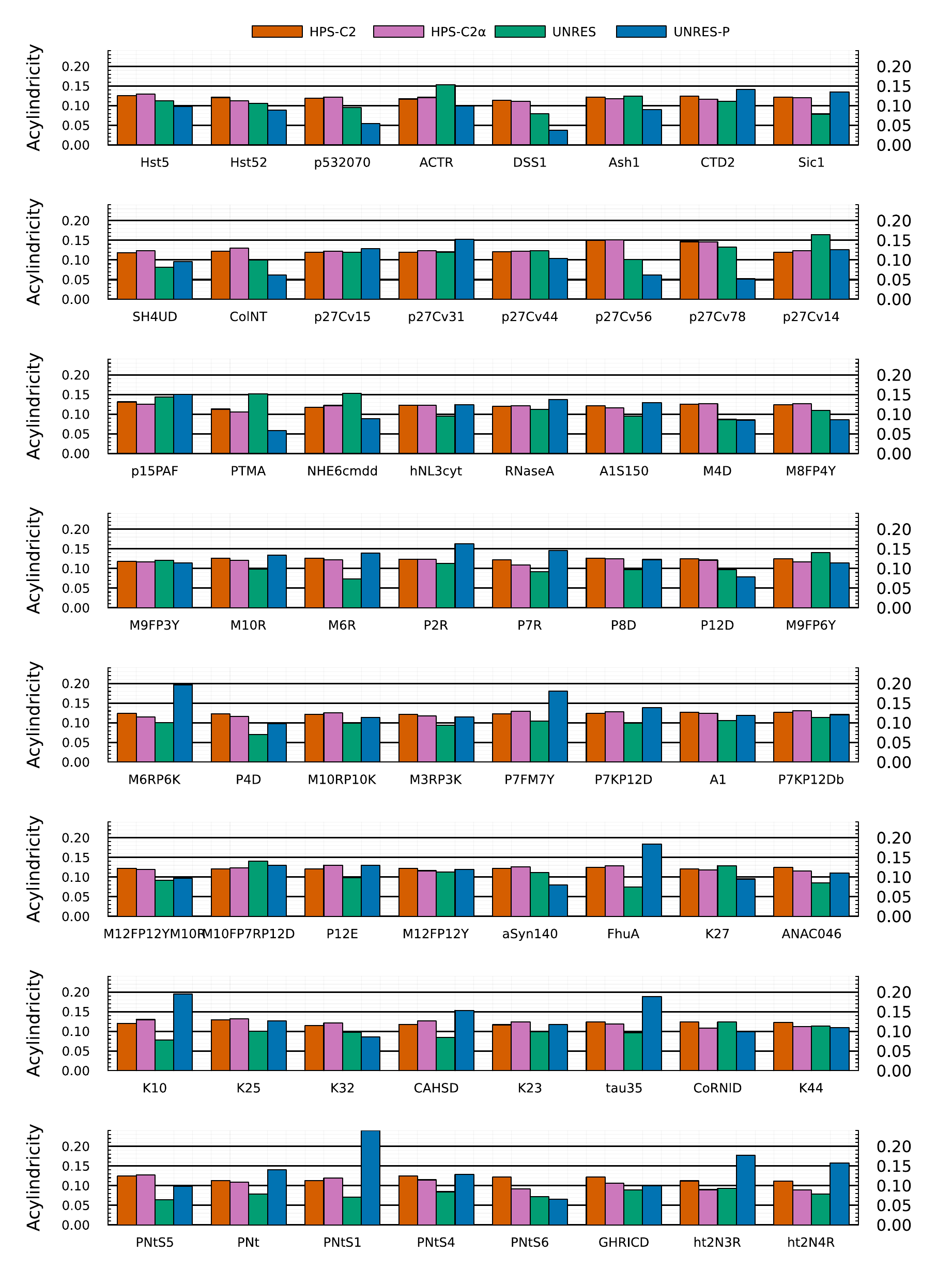}
    \caption{Acylindricity for IDPs simulated using the HPS-C2, HPS-C2$\alpha$, UNRES and UNRES-P models.}
    \label{App:Acylindricity}
\end{figure}

\begin{figure}
    \centering
    \includegraphics[width=0.99\textwidth]{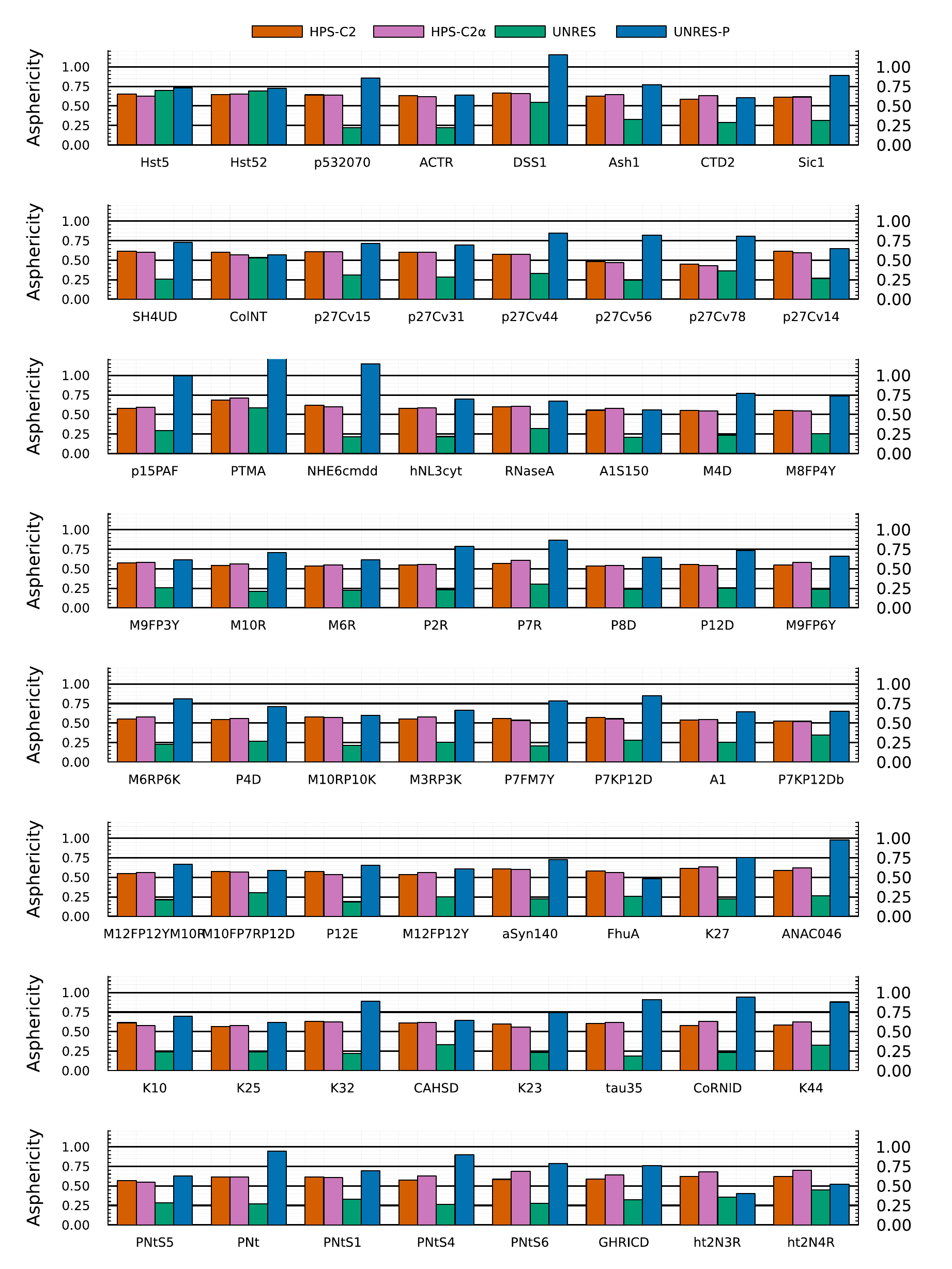}
    \caption{Asphericity for IDPs simulated using the HPS-C2, HPS-C2$\alpha$, UNRES and UNRES-P models.}
    \label{App:Asphericity}
\end{figure}

\begin{figure}
    \centering
    \includegraphics[width=0.99\textwidth]{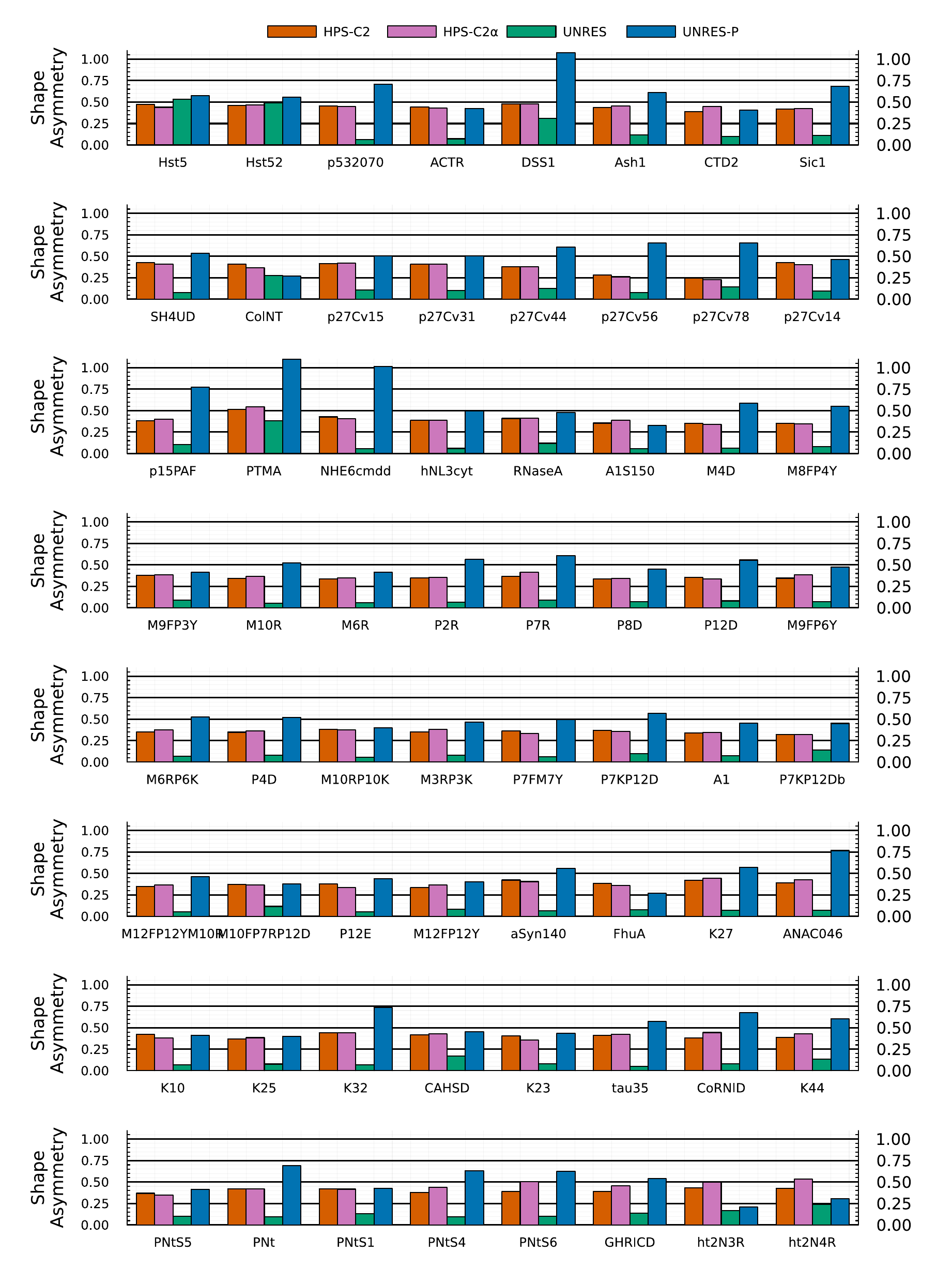}
    \caption{Relative shape anisotropy for IDPs simulated using the HPS-C2, HPS-C2$\alpha$, UNRES and UNRES-P models.}
    \label{App:ShapeAsymmetry}
\end{figure}

\begin{figure}[h!]
    \centering
    \includegraphics[width=\textwidth]{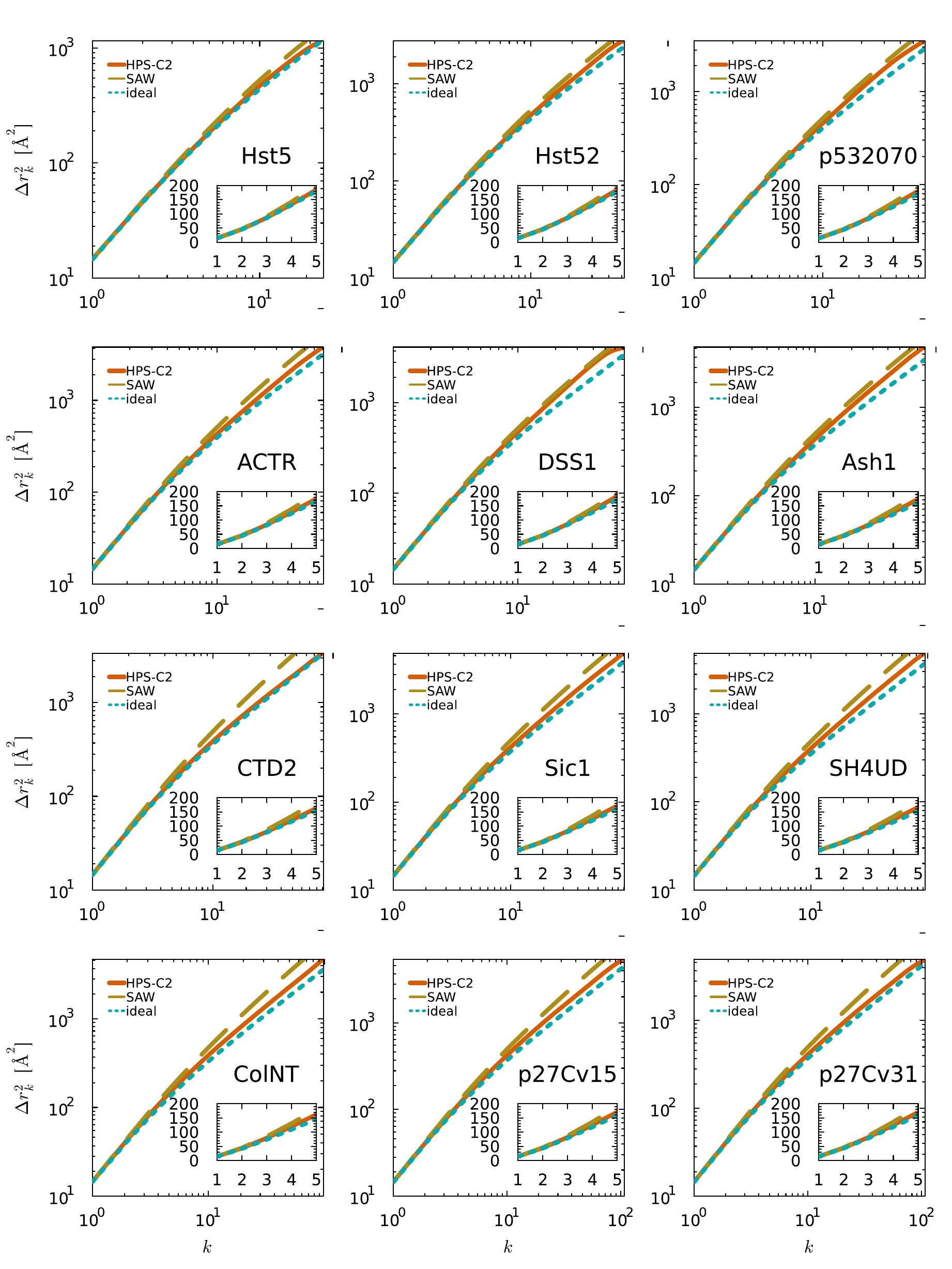}
    \caption{See caption of Fig.~\ref{fig:AppIntraChainScaling_6} below.}
    \label{fig:AppIntraChainScaling_1}
\end{figure}

\begin{figure}[h!]
    \centering
    \includegraphics[width=\textwidth]{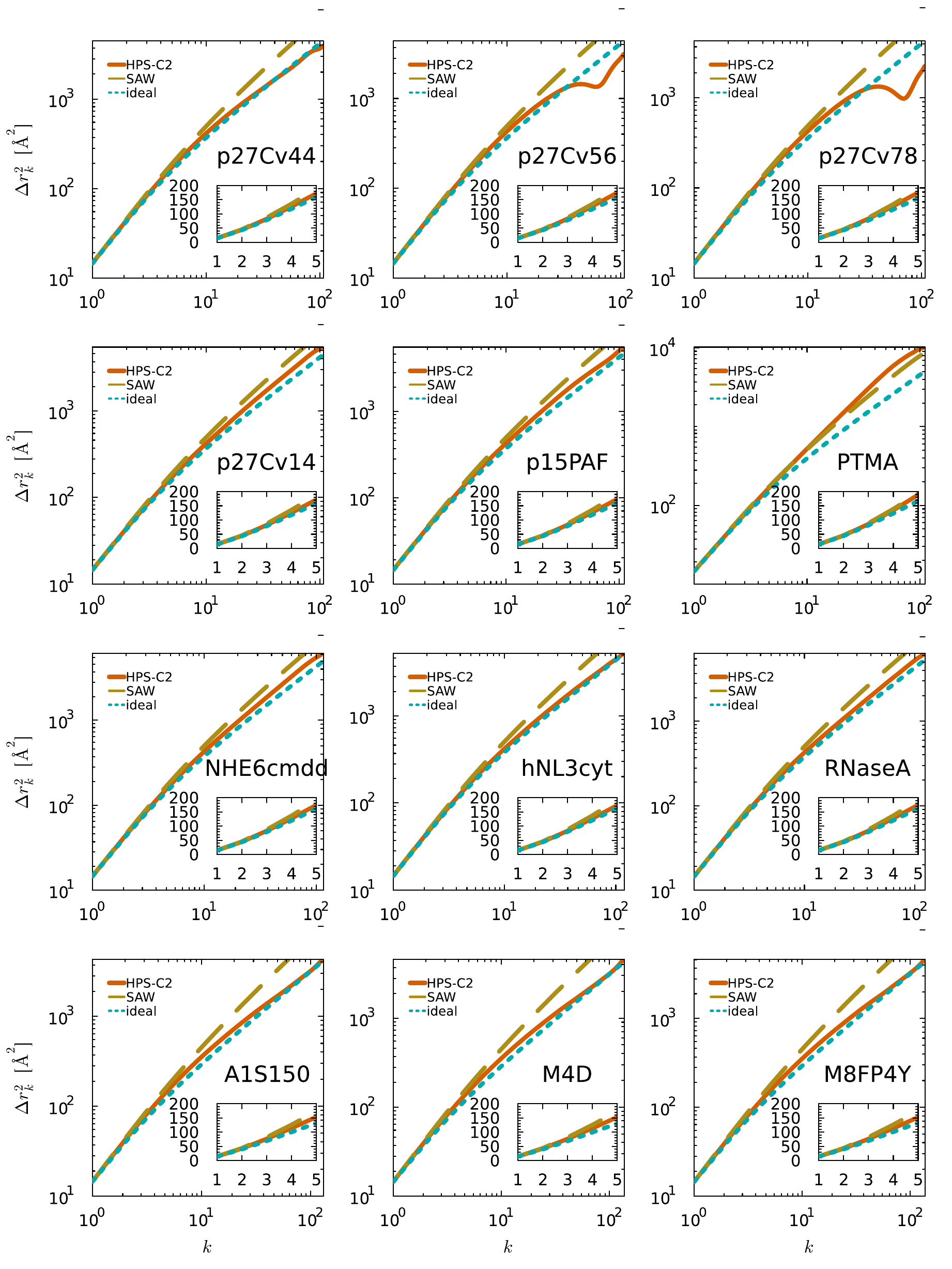}
    \caption{See caption of Fig.~\ref{fig:AppIntraChainScaling_6} below.}
    \label{fig:AppIntraChainScaling_2}
\end{figure}

\begin{figure}[h!]
    \centering
    \includegraphics[width=\textwidth]{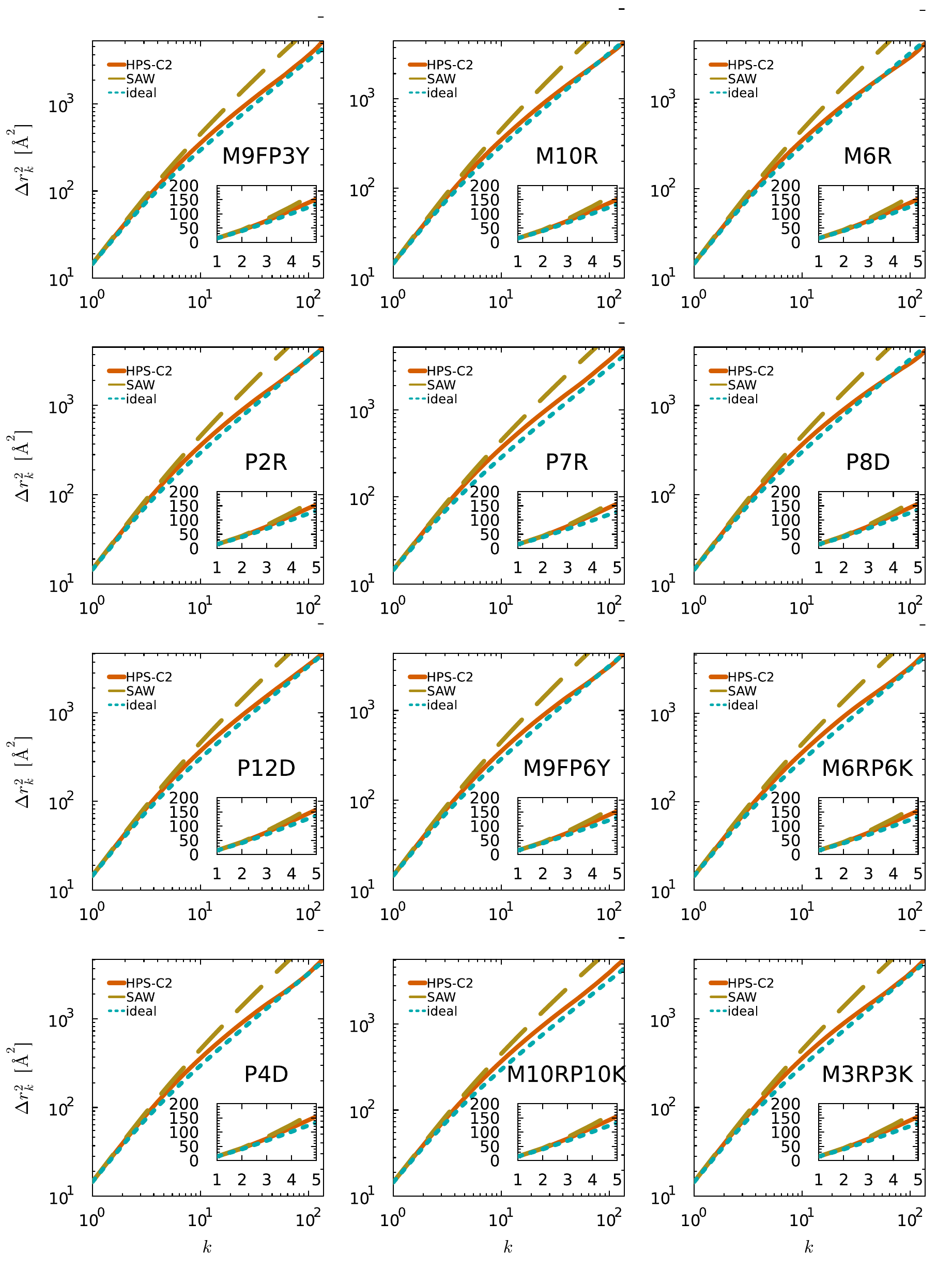}
    \caption{See caption of Fig.~\ref{fig:AppIntraChainScaling_6} below.}
    \label{fig:AppIntraChainScaling_3}
\end{figure}

\begin{figure}[h!]
    \centering
    \includegraphics[width=\textwidth]{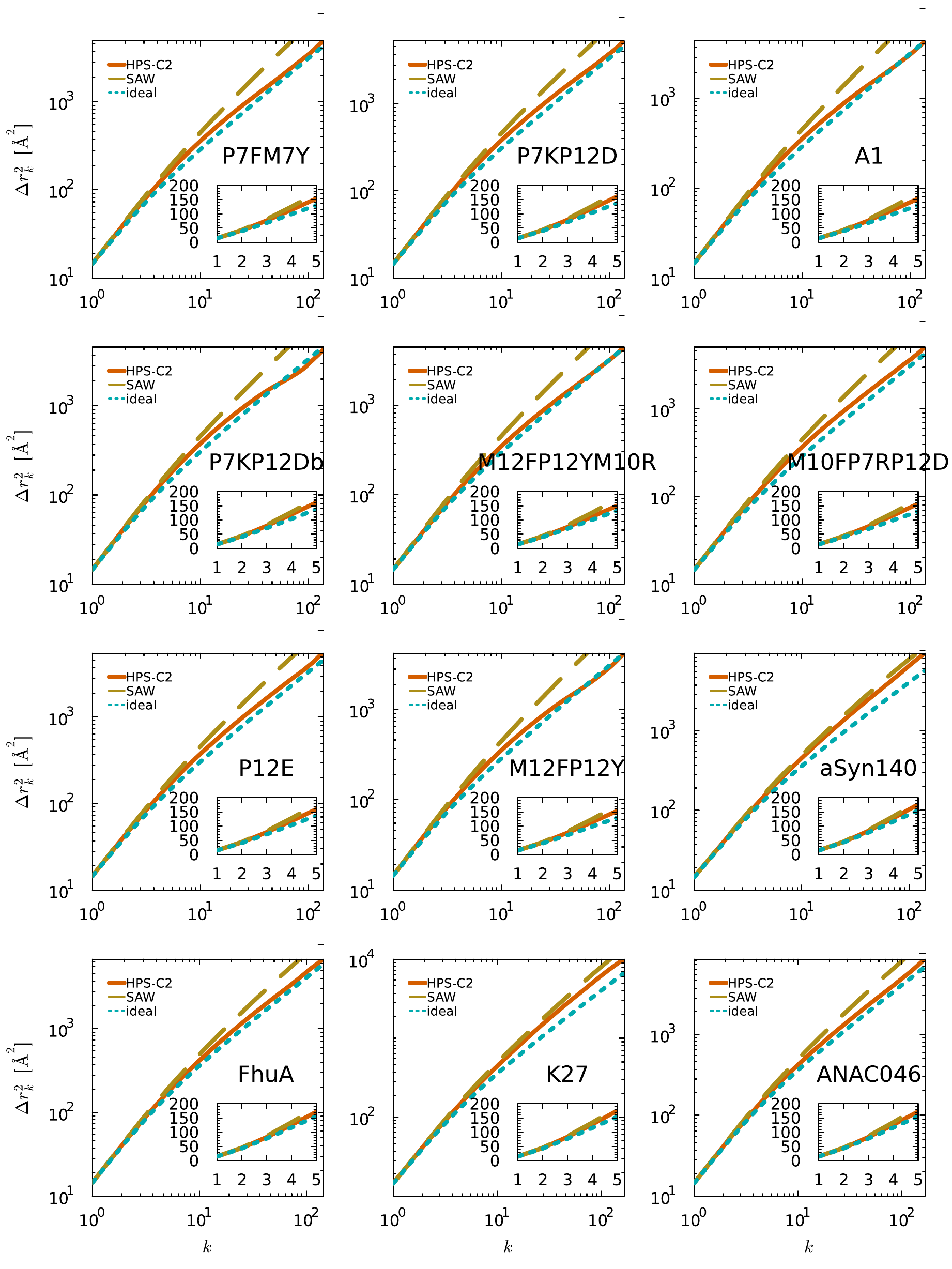}
    \caption{See caption of Fig.~\ref{fig:AppIntraChainScaling_6} below.}
    \label{fig:AppIntraChainScaling_4}
\end{figure}

\begin{figure*}[h!]
    \centering
    \includegraphics[width=\textwidth]{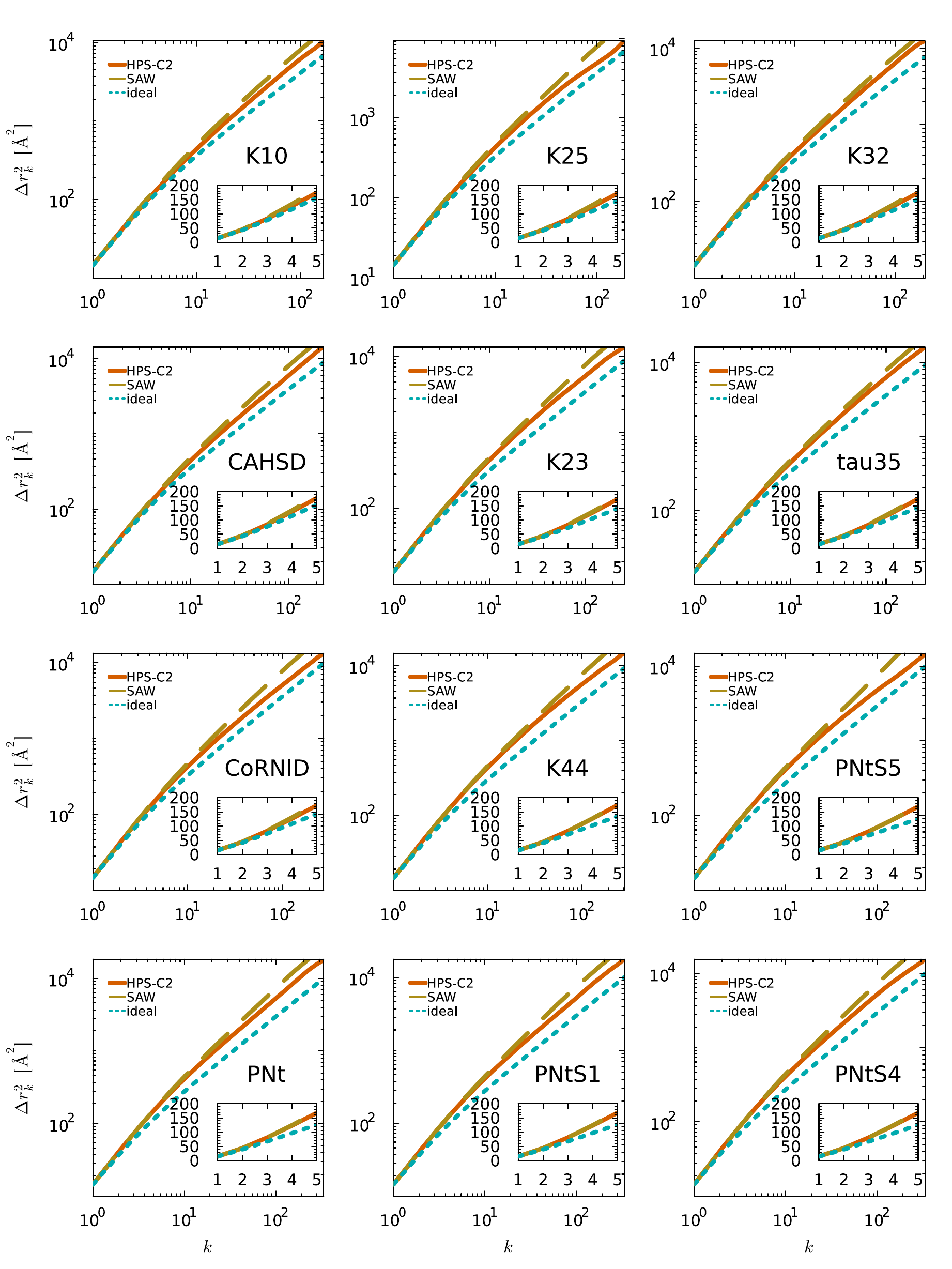}
    \caption{See caption of Fig.~\ref{fig:AppIntraChainScaling_6} below.}
    \label{fig:AppIntraChainScaling_5}
\end{figure*}

\begin{figure}[h!]
    \centering
    \includegraphics[width=\textwidth]{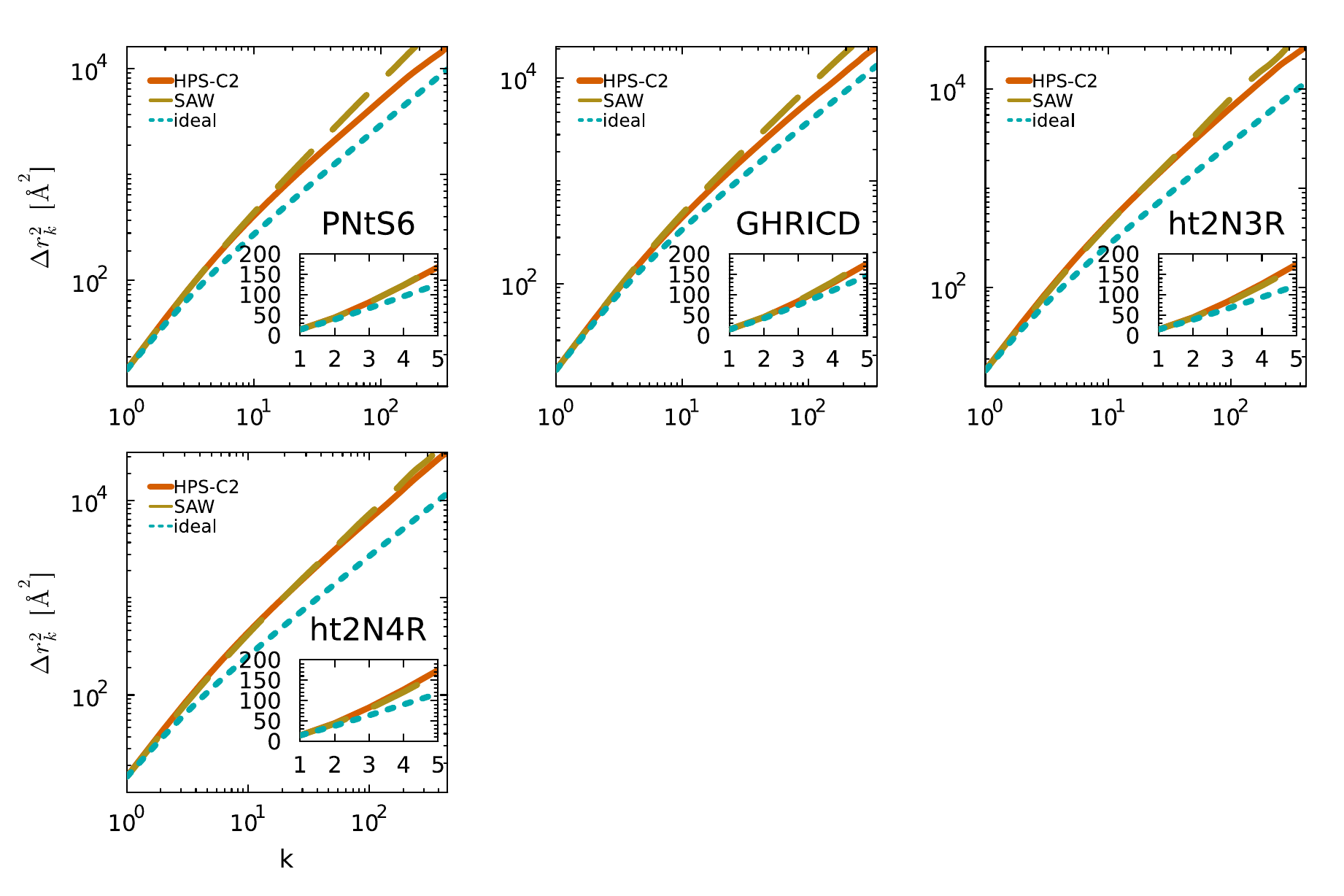}
    \caption{Intramolecular mean-square distance $\Delta r_\text{k}^2$ between two monomers that are $k$ bonds apart [see Eq.~(2) in the main article]. Results shown for IDPs simulated using the HPS-C2 model (orange, solid line), and the corresponding homogeneous worm-like chains with (yellow, dashed line) and without excluded volume interactions (green, dotted line). The inset shows a magnification of the data for beads that are only few bonds apart.}
    \label{fig:AppIntraChainScaling_6}
\end{figure}

\FloatBarrier


\author{Yannick Witzky}
\affiliation{Institute of Physics, Johannes Gutenberg University Mainz, Staudingerweg 7-9, 55128 Mainz, Germany}

\author{Friederike Schmid}
\email{friederike.schmid@uni-mainz.de}
\affiliation{Institute of Physics, Johannes Gutenberg University Mainz, Staudingerweg 7-9, 55128 Mainz, Germany}

\author{Arash Nikoubashman}
\email{anikouba@ipfdd.de}
\affiliation{Leibniz-Institut f{\"u}r Polymerforschung Dresden e.V., Hohe Stra{\ss}e 6, 01069 Dresden, Germany}
\affiliation{Institut f{\"u}r Theoretische Physik, Technische Universit{\"a}t Dresden, 01069 Dresden, Germany}
\affiliation{Cluster of Excellence Physics of Life, Technische Universit{\"a}t Dresden, 01062 Dresden, Germany}

\title{From Heteropolymer Stiffness Distributions to Effective Homopolymers. Part 2: Conformational Analysis of Intrinsically Disordered Proteins}

\begin{abstract}
Intrinsically disordered proteins (IDPs) are characterized by a lack of defined secondary and tertiary structures, and are thus well-suited for descriptions within polymer theory. However, the intrinsic heterogeneity of proteins, stemming from their diverse amino acid building blocks, introduces local variations in chain stiffness, which can impact conformational behavior at larger scales. To investigate this effect, we developed a heterogeneous worm-like chain model in which the local persistence length follows a Gaussian distribution. We demonstrate that these heterogeneous chains can be effectively mapped to homogeneous chains with a single effective persistence length. To assess whether this mapping can be extended to naturally occurring IDPs, we performed simulations using various coarse-grained IDP models, finding that the simulated IDPs have similar shapes compared to the corresponding homogeneous and heterogeneous worm-like chains. However, the IDPs are systematically larger than ideal worm-like chains, yet slightly more compact when excluded volume interactions are considered. We attribute these differences to intramolecular interactions between non-bonded monomers, which our theoretical models do not account for.
\end{abstract}

\maketitle

\section{Introduction}
One groundbreaking realization in molecular biology was that the classical view of rigid protein structures is insufficient for understanding biological functions in all their complexity. Intrinsically disordered proteins (IDPs) or intrinsically disordered regions (IDRs) of proteins account for roughly 30\% of eukaryotic proteins, and have been shown to perform a large number of functions,\cite{oldfieldIntrinsicallyDisorderedProteins2014, tesei:nat:2024} challenging the hitherto established structure-function paradigm. Among these functions is the promotion of phase separation that governs the formation of biomolecular condensates, which are implicated in many cellular functions such as stress response, signal transduction and gene expression.\cite{bananiBiomolecularCondensatesOrganizers2017, alberti:nr:2021, lyon:nr:2021}

The inherent lack of well-defined three-dimensional secondary and tertiary structures tempts the use of polymer concepts for characterizing and predicting the conformations of IDPs.\cite{Choi_2011_beyond, hofmann:pnas:2012, zheng:jcp:2018, Soranno_2018_internal, Sorensen_2019_effective, zheng:jpcl:2020, bauer:jcp:2022, Cubuk_2022_macromolecular, alston:jpcb:2023, chakraborty:jpcb:2025} There is a large repertoire of (homo)polymer models that could be used for characterizing the conformational properties of IDPs in solution and in the condensed state. Among those, the freely jointed chain model and the Kratky-Porod model of a worm-like chain have been widely adapted due to their conceptual simplicity and analytic tractability. The main idea of these modeling strategies is to map an IDP to a reference chain with similar statistical conformations, and then use the mapped representation to infer material properties like the critical temperature, interfacial tension, or shear viscosity.\cite{lin:bpj:2017, dignon:pnas:2018, rekhi:jpcb:2023, farag:bpj:2023, devarajan:nc:2024}

{\cob For polymers in a $\theta$-solvent (or equivalently at the $\theta$-temperature, $T_\theta$), excluded volume interactions between the constituent monomers are effectively canceled, resulting in ideal chain conformations. For homopolymers, $T_\theta$ asymptotically reaches the critical temperature $T_\text{c}$ an the chain length approaches $N \to \infty$,\cite{panagiotopoulos:mm:1998} suggesting that the phase boundaries of IDP solutions might be inferred from single-chain measurements.\cite{hofmann:pnas:2012, lin:bpj:2017, zheng:jcp:2018, zheng:jpcl:2020, zeng:bpj:2020, bauer:jcp:2022, oliver:jpcb:2025} Assuming such a relationship between $T_\text{c}$ and $T_\theta$ extends to heteropolymers, one needs to determine the ideal chain size, which is readily known for homopolymers via established Kuhn lengths,\cite{fetters:book:2007} but far more challenging for IDPs due to sequence-dependent variability in monomer interactions and dimensions.}

A typical polymer model is characterized, among others, by the number of its constituent monomers $N$, their diameter $d$, and bond length $b$. These microscopic model parameters are usually chosen to match some time- or ensemble averaged conformational property of the target IDP, such as the squared radius of gyration
\begin{equation}
   R_\text{g}^2\equiv \frac{1}{N} \sum_{i=1}^N \left(\mathbf{r}_i - \mathbf{r}_\text{cm}\right)^2  =\frac{1}{N^2} 
   \sum_{i=1}^ N\sum_{j>i}^N (\mathbf{r}_i - \mathbf{r}_j)^2
    \label{eq:Rg}
\end{equation}
with $\mathbf{r}_i$ being the position of monomer $i$, and $\mathbf{r}_\text{cm}$ being the chain's center-of-mass position. Another useful quantity for characterizing the protein conformation on a segmental level is the squared distance between two monomers that are $k$ bonds apart
\begin{equation}
\Delta r_{k}^2 \equiv \frac{1}{N-k} \sum_{i=1}^{N-k} \left(\mathbf{r}_{i+k}-\mathbf{r}_i\right)^2 .
   \label{eq:delta}
\end{equation}
For $k=N-1$, i.e., the two monomers at the extremities of the chain, Eq.~\eqref{eq:delta} describes the squared end-to-end distance of the chain, $R_\text{ee}^2 \equiv  \Delta r_{N-1}^2 $. 

A typical starting point for mapping an IDP to a reference (homo)polymer model is to posit the following relation for the mean-square intramolecular distance\cite{hofmann:pnas:2012, alston:jpcb:2023}
\begin{equation}
    \ensav{\Delta r_{k}^2} = b_\text{eff}^2 k^{2\nu} ,
    \label{eq:mean_dist}
\end{equation}
where the effective bond length $b_\text{eff}$ describes the (average) segment length, and the Flory exponent $\nu$ characterizes the effective hydrophobicity of the chain.\cite{rubinstein:book:2003} Although Eq.~\eqref{eq:mean_dist} looks rather inconspicuous, it is based on several assumptions: (i) The chain can be described as a self-similar fractal object on {\it all} length scales, so that the expression for the mean-square distance between two chain ends can be used for {\it any} monomer pair. (ii) Sequence-specific local correlations between bonded monomers can be averaged into a {\it single} effective parameter $b_\text{eff}$. (iii) The fractal dimension of the entire chain can be expressed by a {\it single} (effective) scaling exponent $\nu$.

Following these assumptions, Alston {\it et al.} developed a random coil model, where $b_\text{eff}$ was determined at fixed $\nu = 1/2$ by fitting Eq.~\eqref{eq:mean_dist} to mean inter-residue distances from polypeptide chains.\cite{alston:jpcb:2023} They generated reference data using Monte Carlo (MC) simulations of a freely rotating ideal chain, with dihedral angles sampled from a precomputed set of residue-specific angles. This method was then extended to arbitrary heteropolymeric sequences using the weighted average of $b_\text{eff}$. The analytically computed probability distributions of $R_\text{g}^2$ and $R_\text{ee}^2$ were in excellent agreement with simulation data of freely rotating ideal heteropolymer chains, thereby confirming approximations (i) and (ii) discussed above. However, the predicted chain size distributions differed substantially from the ones obtained from atomistic simulations of naturally occurring IDP sequences, underlining the importance of non-bonded interactions for the chain conformations.

In an attempt to go beyond ideal chain statistics, Hofmann {\it et al.}\cite{hofmann:pnas:2012} determined effective bond lengths and scaling laws of IDPs by fitting experimental data\cite{kohn:pnas:2004} of unfolded proteins and polypeptides to the expression\cite{hammouda:advp:1993}
\begin{equation}
    \ensav{R_\text{g}^2}^{1/2} = \sqrt{\frac{b_\text{eff}^2}{(2\nu + 1)(2\nu + 2)}}N^\nu .
    \label{eq:nu}
\end{equation}
Equation~\eqref{eq:nu} follows directly from combining Eqs.~\eqref{eq:Rg} and \eqref{eq:mean_dist},  replacing the discrete sum with an integral, and keeping only the leading term in $N$.\cite{teraoka:book:2002, rubinstein:book:2003} Hofmann {\it et al.} further conjectured that $b_\text{eff}$ can be expressed as
\begin{equation}
    b_\text{eff} = \sqrt{2 b \ell_\text{p,eff}}
    \label{eq:beff}
\end{equation}
where $b=3.8\,\angstrom$ is the distance between two $\text{C}_\alpha$ atoms of the protein backbone, and $\ell_\text{p,eff}$ is the (effective) persistence length that describes the local bending rigidity of the chains in an average sense.\cite{rubinstein:book:2003, binder:jpm:2020, nikoubashman:jcp:2021} Using this Ansatz, Hofmann {\it et al.} determined $\ell_\text{p,eff} = 4.0 \pm 0.7\,\angstrom$ and $\nu = 0.58$ from fitting Eq.~\eqref{eq:nu} to experimental data.\cite{hofmann:pnas:2012}

Although this mapping seems reasonable at first glance, it should be considered carefully since Eqs.~\eqref{eq:nu} and \eqref{eq:beff} introduce additional (subtle) assumptions that might have unexpected consequences: (iv) Equation~\eqref{eq:nu} underestimates local excluded volume interactions that make real chains more rigid over short distances;\cite{teraoka:book:2002} for example, using the Flory exponent of a self-avoiding chain $\nu = 0.588$ in Eqs.~\eqref{eq:mean_dist} and \eqref{eq:nu} yields the ratio $6\ensav{R_\text{g}^2}/\ensav{R_\text{ee}^2} \approx 0.87$, which is substantially smaller than the ratio of $\approx 0.952$ expected from renormalization group theory.\cite{cloizeaux:book:1990, teraoka:book:2002} (v) Equation~\eqref{eq:beff} follows from a Taylor expansion of the harmonic bending potential acting between three consecutive monomers to impart bending rigidity; this expansion is, however, only valid for small bending angles $\theta_i$, and thus large persistence lengths $\ell_\text{p} \gg b$.\cite{rubinstein:book:2003} (vi) Equation~\eqref{eq:beff} further implies that $\ell_\text{p} \gg d$, i.e., chain bending is suppressed without increase of the cross-sectional radius of the chains.\cite{hsu:epl:2010} Assumptions (v) and (vi) are, however, at odds with the fitted persistence length $\ell_\text{p,eff} \approx 4.0\,\angstrom$, which is comparable to the typical bond length $b$ and monomer diameter $d$. 

Even ignoring the above caveats, one is still left with the problem that the local stiffness of IDP heteropolymers may greatly vary along the backbone due to the different sizes and interactions of the constituent amino acid residues. Therefore, it is not obvious how the ``effective persistence length'' $\ell_\text{p,eff}$ of a heteropolymer depends on the distribution of local stiffnesses. To address this question, we have developed a theoretical framework for mapping heterogeneous worm-like chains to homopolymers with a single, effective persistence length (see part 1 of this series\cite{witzky:jcp:2025B}). In this work, we test whether this mapping can be extended to IDPs.

\section{Simulation Models}
\label{sec:models}
In principle, IDPs can be simulated using models with atomistic resolution and an explicit solvent. In practice, however, this precise treatment limits the achievable time- and length-scales due to the associated computational costs. Therefore, it is common to resort to coarse-grained descriptions, where unneeded features have been eliminated. For dilute solutions, the solvent particles usually occupy a large fraction of the simulation box. Often, resolving individual solvent particles is not needed, but only the solvent's effects on the polymer conformation and dynamics. Such problems are ideal for implicit solvent models. Further speedup can be achieved by integrating multiple atoms into effective interaction sites. The choice of which atoms are grouped and how to model their interactions is ambiguous, resulting in a large number of models that aim at preserving different properties of the reference systems.\cite{kmiecik:cr:2016, best:cosb:2017, shea:cosb:2021, borges:jctc:2023}

\begin{figure}
    \centering
    \includegraphics[width=5.5cm]{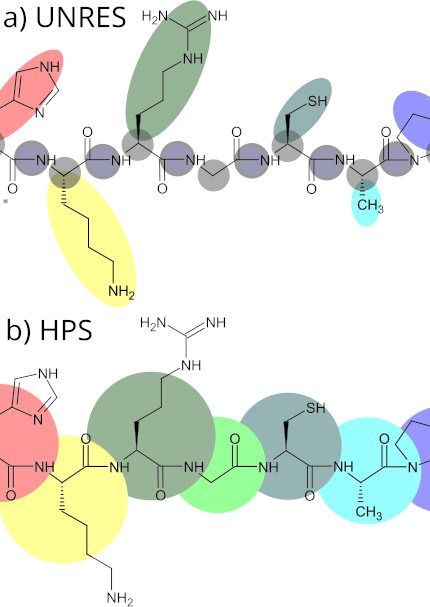}
    \caption{Schematic illustrations of the (a) UNRES and (b) HPS protein models. (a) \CA{}-carbons colored in grey, peptide bonds in dark blue and the ellipsoidal side chains are color coded by amino acid type. (b) Coarse-grained beads, color coded by amino acid type, of the HPS-Model. The underlying two-dimensional structural formula is represented beneath the coarse-grained models.}
     \label{fig:CGModels}
\end{figure}

To generate reference data and test how sensitive the simulated IDP conformations are to the specific model details, we simulated 64 different protein sequences using two variants of the UNited RESidue (UNRES) model\cite{liwoGeneralMethodDerivation2019, UNRES_Phos} and two versions of the hydrophobicity scale (HPS) model.\cite{Dignon_HPS, Calvados2, rizuanDevelopingBondedPotentials} The UNRES models treat the protein backbone and side groups separately, while the HPS models describe a whole amino acid as one interaction site (see Fig.~\ref{fig:CGModels}). All models use an implicit solvent. {\cob The origin of bending stiffness differs across the employed models: In the standard HPS-C2 model, bending rigidity emerges implicitly through non-bonded interactions, while the HPS-C2$\alpha$ model uses an explicit bending potential. In the UNRES models, local bending stiffness arises from a combination of bonded and non-bonded interactions.}

The UNRES model was developed for protein structure prediction and competes in biannual prediction competitions (CASP),\cite{UNRES_CASP13} where it placed third among the non-deep learning methods, based on the median $S_\text{CASP14}$ scoring function.\cite{CASP14} In the UNRES model, each amino acid residue is represented by a spherical, virtual \CA{} particle along the protein backbone, with a united peptide group between consecutive \CA{} particles and ellipsoidals for each residue's specific side groups, as shown schematically in Fig.~\ref{fig:CGModels}a. This coarse-graining is designed to accurately reconstruct local structure when mapped back onto an atomistic representation using the PULCHRA algorithm.\cite{PULCHRA} We used the NEWCT-9P force field parameterization of UNRES,\cite{liwoGeneralMethodDerivation2019} as well as a recent extension\cite{UNRES_Phos} that enables phosphorylation (UNRES-P). In both versions, bonded interactions between \CA{} atoms are represented by a distance-dependent bond potential, bond angle-dependent potentials, a torsion potential, and a backbone correlation term. The ellipsoidal side chains (SCs) are modeled by a rotamer potential. Non-bonded interactions contain Van der Waals interactions between the SCs and the peptide (P) bonds modeled by Lennard-Jones-like potentials for each type of SC-SC , SC-P and P-P interaction. Additionally, an electrostatic interaction between peptide groups is modeled by an orientation dependent $\eta/r^3+\nu/r^6$ potential. Notably, these force field components only use electric dipoles, which allows for a short cutoff distance of $7\,\angstrom$, thereby speeding up the simulations.

In the UNRES-P model, electric charges were added to the NEWCT-9P force field to capture the highly charged nature of phosphorylated amino acid residues.\cite{UNRES_Phos} The additions include interactions between apolar sites, apolar and charged sites, implicit solvent polarization through charged sites, charge-charge interactions modeled by a Coulomb potential, and cavity terms for polar and apolar sites.\cite{UNRES_Phos} Salt concentration effects can be incorporated either by adding explicit counter ions, or by including Debye-H{\"u}ckel screening to the electrostatic interactions. In this work, we used the latter approach, since the CGDT\_TS scores\cite{zemlaLGAMethodFinding2003} varied little between the variants,\cite{UNRES_Phos} and it made comparisons with the HPS model easier. The electrostatic cutoff distance is set to $25\,\angstrom$. The UNRES simulations were performed with a timestep of $\approx 5\,\text{fs}$ using the \href{https://unres.pl/downloads}{UNRES software package} with the Version names ``unres-phosphorylated4'' and ``unres-src-HCD-5D\_nmr-May-5-2021''.

The HPS models use a more coarse-grained representation, where each amino acid is represented as one spherical bead of diameter $d_i$, as shown in Fig.~\ref{fig:CGModels}b. Non-bonded interactions are typically modeled using a (modified) Lennard-Jones potential, where the attraction strength  (approximately) replicates the hydrophobicity of the residues. Electrostatic interactions between charged residues are included through Debye-H{\"u}ckel screened Coulomb interactions. Bonded interactions are modeled using a harmonic potential with an equilibrium bond length of $b=3.8\,\angstrom$. There exist several variations of the HPS model,\cite{kim:jmb:2008, Dignon_HPS, joseph:ncs:2021, Calvados2, rizuanDevelopingBondedPotentials, tejedor:acscs:2025} which have been optimized to replicate different aspects of IDPs. In this work, we used the Calvados2 model (abbreviated as HPS-C2 in what follows),\cite{Calvados2} whose non-bonded interaction parameters have been tuned to replicate the conformational properties and propensities to undergo phase separation for a diverse set of IDP sequences. Further, we have combined the HPS-C2 model with the angle and dihedral potentials that were developed for replicating $\alpha$-Helices in IDPs (abbreviated as HPS-C2$\alpha$ in what follows).\cite{rizuanDevelopingBondedPotentials} This model was implemented using the simulation software package LAMMPS (Large scale Atomic/Molecular Massively Parallel Simulator)\cite{LAMMPS_Jun2023} (Version ``23 Jun 2022''). The cutoff distance of the modified Lennard-Jones potential is set to $20\,\angstrom$, while the cutoff distance for the screened electrostatic interactions is $40\,\angstrom$. All simulations are performed in the $NVT$ ensemble, using a Langevin thermostat with dampening constant of $0.1\,\text{ps}$. The equations of motion are integrated using a velocity Verlet algorithm with a time step of $10\,\text{fs}$. For the HPS-C2 model, directly bonded beads are excluded from non-bonded interactions. For the HPS-C2$\alpha$ model, these exclusions are extended to bead pairs, which are two and three bonds apart.

We selected the IDPs used for the parametrization of Calvados2~\cite{Calvados2} as our test sequences, allowing us to compare our simulations with experimental measurements. To ensure consistency, we simulated at the corresponding experimental temperatures, ranging from $277\,\text{K}$ to $298\,\text{K}$. The total number of time steps varied between $4 \times 10^8$ and $2 \times 10^9$ for the HPS-C2 simulations, and between $10^9$ to $7 \times 10^9$ for HPS-C2$\alpha$, and we recorded snapshots every $10^5$ steps. To confirm that our simulation times were sufficiently long to obtain representative IDP conformations, we estimated the polymer relaxation time $\tau_\text{rel}$ from the autocorrelation of $R_\text{g}$. Depending on protein size and simulation length, the number of sampled relaxation times ranged from $2 \times 10^2$ to $10^4\,\tau_\text{rel}$ for the HPS-C2 model, and from $10^2$ to $10^4\,\tau_\text{rel}$ for HPS-C2$\alpha$. The total number of time steps varied from $10^8$ to $10^9$ for the UNRES simulations, and from $4 \times 10^7$ to $2 \times 10^8$ time steps for UNRES-P. Snapshots were recorded every $5 \times 10^4$ steps for both models. Due to the slow relaxation of the IDPs in the UNRES models, we were unable to determine reliable autocorrelation times, and therefore did not estimate measurement uncertainties for those simulations.

\section{Results}
\label{sec:results}
The goal of this work is to elucidate the molecular-level conformations of IDPs in solution, and to assess whether they can be represented as homogeneous worm-like chains with an effective persistence length. {\cob Testing the relationship between chain compactness and phase separation propensity is beyond the scope of this work, and we refer the reader to, e.g., Refs.~\citenum{zeng:bpj:2020, rekhi:jpcb:2023, oliver:jpcb:2025} for more detailed treatments of this topic.}

We characterize the shape of the simulated polymers by calculating their gyration tensor
\begin{equation}
   \mathbf{G} = \frac{1}{M}\sum_{i=1}^{N} m_i(\mathbf{r}_i-\mathbf{r}_\text{cm})(\mathbf{r}_i-\mathbf{r}_\text{cm})^\text{T},
   \label{eq:Rg2}
\end{equation}
where the sum runs over all $N$ residues or monomers, $m_i$ is the mass of the residue / monomer $i$, and $M$ is the total mass of the protein or polymer. In the HPS model, residues of different type have different masses as specified in the corresponding references.\cite{Dignon_HPS, Calvados2, rizuanDevelopingBondedPotentials} For the UNRES model, $\mathbf{r}_i$ denotes the position of the C$^\alpha$ carbon of a given residue, and $m_i$ its total mass including side chains. The eigenvalues $\lambda_1 \leq \lambda_2 \leq \lambda_3$ of the gyration tensor $\mathbf{G}$ characterize the quadratic extension of the chain along its three principal axes, with $R_\text{g} = \left(\lambda_1 + \lambda_2 + \lambda_3\right)^{1/2}$. Note that in what follows, we report mean values of $R_\text{g}$ and $R_\text{ee}$ values, following the convention used in Ref.~\citenum{Calvados2}, rather than the root-mean-square values $\ensav{O^2}^{1/2}$. In addition, we can use the eigenvalues of $\mathbf{G}$ to determine the asphericity ($\text{AS}$), acylindricity ($\text{AC}$) and relative shape anisotropy ($\text{A}$) of the chains (see part 1 of this work for details\cite{witzky:jcp:2025B}):
\begin{align}
    \text{AS} &= \la \left(\lambda_3-\frac{\lambda_1+\lambda_2}{2}\right) / (\lambda_1+\lambda_2+\lambda_3)\ra \label{eq:AS}\\
    \text{AC} &= \la (\lambda_2-\lambda_1)/ (\lambda_1+\lambda_2+\lambda_3)\ra \label{eq:AC}\\
    \text{A} &= \la 1-3\frac{\lambda_1\lambda_2 + \lambda_1\lambda_3 + \lambda_2\lambda_3}{(\lambda_1+\lambda_2+\lambda_3)^2} \ra \label{eq:A}. 
\end{align}

\begin{figure}[htbp]
    \centering
    \includegraphics[width=7.5cm]{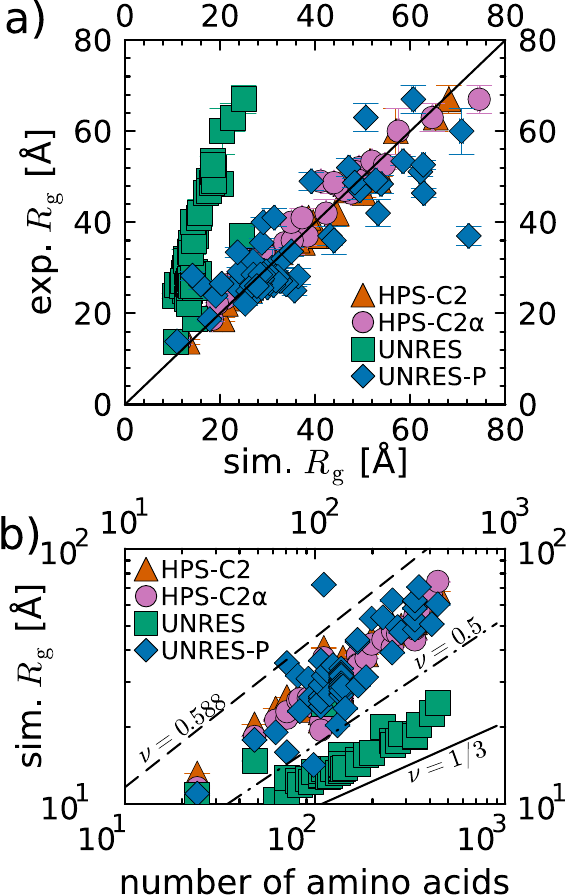}
    \caption{(a) Experimentally measured radius of gyration $R_\text{g}$~\cite{Calvados2} plotted against simulation results using HPS-C2, HPS-C2$\alpha$, UNRES and UNRES-P models. (b) $R_\text{g}$ from simulations as function of number of amino acids. The dashed, dash-dotted, and solid black lines indicate scaling laws with exponents $\nu =0.588$, $\nu=0.5$, and $\nu=1/3$, respectively.}
    \label{fig:RG_comp}
\end{figure}

We consider 64 IDP sequences for which experimental measurements of the radius of gyration, $R_\text{g}$, are available.\cite{Calvados2} First, we tested whether the IDP models (see Sec.~\ref{sec:models}) accurately reproduce the experimental $R_\text{g}$ data (Fig.~\ref{fig:RG_comp}a). The results from the standard UNRES model consistently fall below the experimental values, suggesting that the simulated proteins are overly compact, which is a common issue with force fields primarily optimized for folded proteins.\cite{best:cosb:2017, rahmanComparisonEvaluationForce2020} The additional charge interactions in the phosphorylated UNRES-P model mitigate this issue, resulting in much better agreement with the experimental $R_\text{g}$ values. Simulations with the HPS-C2 model reproduce the experimental results with near-perfect accuracy, which is expected since the HPS-C2 model was optimized against these data. The HPS-C2$\alpha$ model achieves comparable accuracy, despite not being fully optimized on this dataset.

In Fig.~\ref{fig:RG_comp}b, we have plotted the $R_\text{g}$ values from the simulated IDP sequences against the corresponding number of amino acids to examine whether the data follow a scaling relation. Fitting the UNRES results to $R_\text{g} \propto N^\nu$ gives $\nu \approx 0.38 \pm 0.1$, indicating that the IDPs are close to collapsed. In contrast, the other models yield scaling exponents of $\nu \approx 0.62 \pm 0.03$ (HPC-C2), $0.72 \pm 0.02$ (HPS-C2$\alpha$), and $0.59 \pm 0.01$ (UNRES-P), which slightly exceed the theoretically expected value of $0.588$ for self-avoiding random walks in a good solvent.

\begin{figure*}[htb]
    \centering
    \includegraphics[width=15cm]{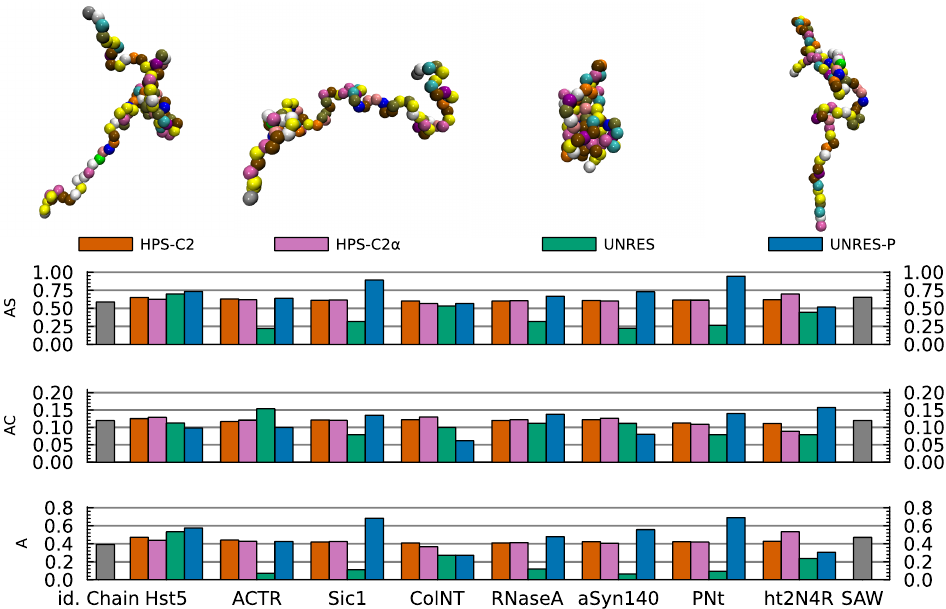}
    \caption{Asphericity ($\text{AS}$), acylindricity ($\text{AC}$) and relative shape anisotropy ($\text{A}$) for selected IDPs simulated using the HPS-C2, HPS-C2$\alpha$, UNRES and UNRES-P models. Values of ideal and self-avoiding chains are plotted in gray. Snapshots of Sic1 for the different models that are closest to the average quantities are shown. Only C$\alpha$ carbons are shown for the UNRES models.}
    \label{fig:PolymerCharacterisitcs}
\end{figure*}

Next, we characterized the shape of the IDPs using their asphericity ($\text{AS}$), acylindricity ($\text{AC}$) and relative shape anisotropy ($\text{A}$), as defined in Eqs.~\eqref{eq:AS}-\eqref{eq:A}. Figure~\ref{fig:PolymerCharacterisitcs} presents these shape descriptors for a representative selection of IDPs for each model, alongside the values expected for ideal and self-avoiding homopolymer chains (the full data set is available in the SI). The two HPS models yield nearly identical results, with values scattered around those predicted for ideal and self-avoiding chains. Notably, these shape descriptors exhibit only a weak sequence dependence for the investigated IDPs, suggesting a largely universal behavior within these models. In contrast, the majority of IDPs simulated with the UNRES-P model exhibit slightly higher $\text{AS}$ and $\text{A}$ values, accompanied by lower $\text{AC}$ values, indicating more cylindrical chain configurations. Conversely, IDPs simulated with the standard UNRES model consistently adopt more compact, globular conformations, as reflected by the significantly smaller $\text{AS}$ and $\text{A}$ values. Considering these shape parameters, along with the $R_\text{g}$ values shown in Fig.~\ref{fig:RG_comp}, we conclude that the standard UNRES model model does not adequately capture the conformational properties of IDPs and will therefore be excluded from further analysis.

Having characterized the overall shape of the IDPs, we next study the spatial correlations of neighboring monomers. To this end, we computed the local persistence length of the protein backbone via
\begin{equation}
    \ell_{\text{p},i} = \frac{-b}{\ln \la\cos(\theta_{i}) \ra} \label{eq:lp} .
\end{equation}
Here, $\theta_i$ is the angle formed between three consecutive (backbone) beads. Equation~\eqref{eq:lp} is derived from the standard formula for the orientational correlation of $\la\cos(\theta(s))\ra = \la \mathbf{u}(s) \cdot \mathbf{u}(0) \ra = \exp(-s b/\ell_{\rm p})$ for a single discretized step ($s=1$) along the chain contour.\cite{rubinstein:book:2003} For the UNRES-P model, we employed the positions of the C$^\alpha$ beads, while the positions of the united residues were used in the HPS-C2 model, reflecting its coarser resolution (Fig.~\ref{fig:CGModels}). The  HPS-C2$\alpha$ model is omitted from this analysis, since it fixes the persistence lengths to a value close to $\ell_{\text{p},i} \approx 4\,\angstrom$.

\begin{figure*}[htb]
    \centering
    \includegraphics[width=15cm]{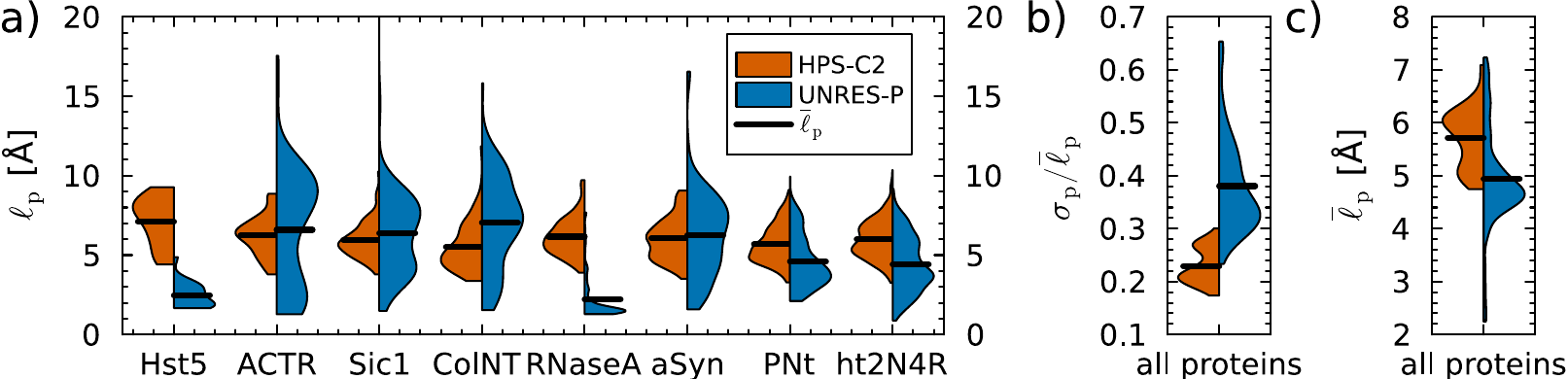}
    \caption{(a) Distribution of the local persistence lengths for a selection of IDPs simulated with the HPS-C2 and UNRES-P models. Black lines indicate the arithmetic mean $\bar{\ell}_\text{p}$ of $\ell_{\text{p},i}$. (b) Relative standard deviation $\sigma_\text{p}/\bar{\ell}_\text{p}$, and (c) distribution of $\bar{\ell}_\text{p}$ sampled over all IDPs.}
    \label{fig:PersistenceLengthDistribution}
\end{figure*}

Distributions of the local persistence lengths $\ell_{\text{p},i}$ are shown in Fig.~\ref{fig:PersistenceLengthDistribution}a for a selection of IDPs simulated with the HPS-C2 and UNRES-P models. The arithmetic mean of the persistence lengths, $\bar{\ell}_\text{p}$, is indicated as a black bar in each half of the violin plot. Most distributions are highly non-Gaussian and broad compared to their $\bar{\ell}_\text{p}$ values. For the HPS-C2 model, the relative standard deviation of these distributions is $\sigma_\text{p}/\bar{\ell}_\text{p} \approx 20-30\,\%$ (Fig.~\ref{fig:PersistenceLengthDistribution}b), whereas IDPs simulated with the UNRES-P model exhibit substantially larger variations in local bending stiffness, with $\sigma_\text{p}/\bar{\ell}_\text{p} \approx 25-50\,\%$, likely due to the pronounced electrostatic interactions in that model. Figure~\ref{fig:PersistenceLengthDistribution}c shows the distributions of $\bar{\ell}_\text{p}$ across the 64 IDP sequences sampled in this study. For the HPS-C2 simulations, the distribution of $\bar{\ell}_\text{p}$ has a mean of $5.7\,\angstrom$ with standard deviation $0.6\,\angstrom$, while for the UNRES-P model, the mean is slightly smaller at $4.9\,\angstrom$ with standard deviation $1\,\angstrom$. These persistence lengths are somewhat larger than the previously reported value of $4.0 \pm 0.7\,\angstrom$ from Hofmann {\it et al.}\cite{hofmann:pnas:2012}

\begin{figure}[htb]
    \centering
    \includegraphics[width=7cm]{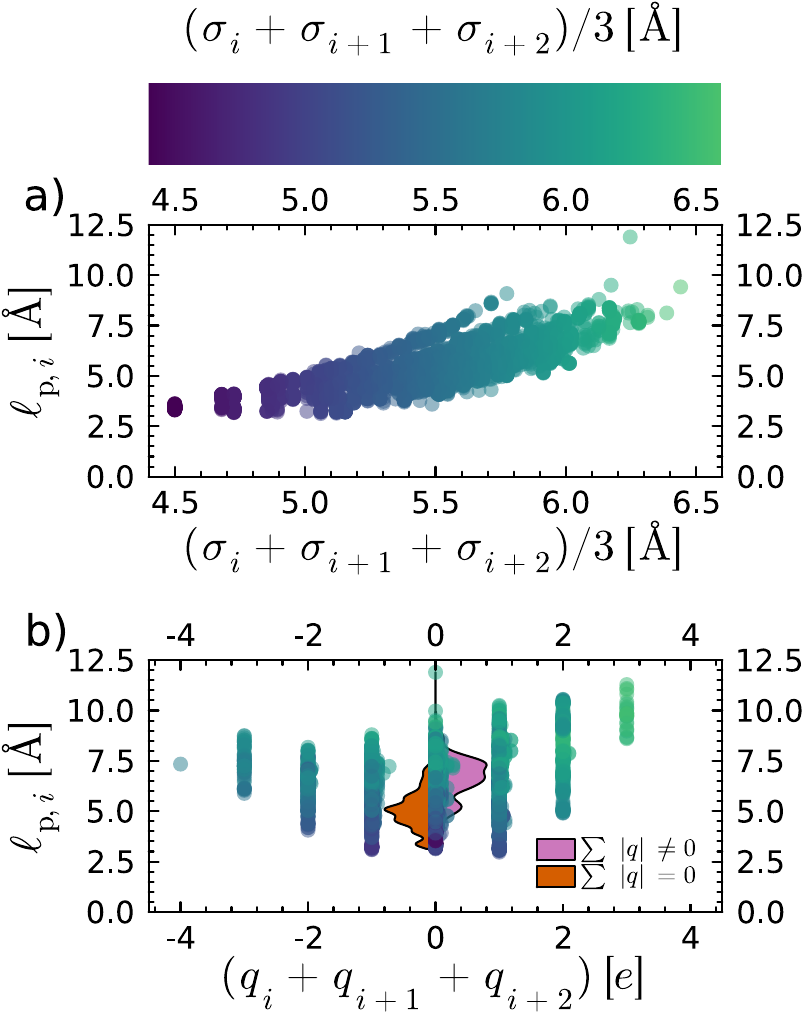}
    \caption{Local persistence lengths $\ell_{\text{p},i}$ of triplets as functions of (a) average bead diameter $(\sigma_i+\sigma_{i+1}+\sigma_{i+2})/3$, and (b) net charge $q_i + q_{i+1} + q_{i+2}$ for the HPS-C2 model. The violin plots in (b) show the distribution of $\ell_{\text{p},i}$ for net-neutral triplets with $\sum |q_i| = 0$ (left) and $\sum |q_i| \neq 0$ (right). Small deviations from integer charges originate from the pH-dependent charge of histidine,\cite{Calvados2, nagai:jced:2008} while values of $\pm 4\,\text{e}$ are due to end-effects.\cite{Calvados2}}
    \label{fig:PersistenceLengthOrigin}
\end{figure}

{\cob To better understand the molecular origin of the wide spread in local persistence lengths $\ell_{\text{p},i}$, we examined their correlation with selected physicochemical features of the monomers in the corresponding triplets for the HPS-C2 model. Figure~\ref{fig:PersistenceLengthOrigin}a shows a strong positive correlation between $\ell_{\text{p},i}$ and the average diameter $(\sigma_i+\sigma_{i+1}+\sigma_{i+2})/3$ of uncharged residues, underscoring the important role of excluded volume interactions. Further, $\ell_{\text{p},i}$ increases with the net charge $q_i + q_{i+1} + q_{i+2}$, consistent with the electrostatic repulsion between like charges (Fig.~\ref{fig:PersistenceLengthOrigin}b). Notably, triplets with a surplus of negative charges exhibit substantially smaller persistence lengths than those with an equivalent excess of positive charges. This difference likely originates from the larger bead diameters of arginine, histidine, and lysine compared to aspartic and glutamic acid.\cite{Dignon_HPS} Finally, net-neutral triplets containing charged residues ($\sum |q_i| \neq 0$) have slightly larger $\ell_{\text{p},i}$ compared to triplets composed solely of neutral residues ($\sum |q_i| = 0$), which can be attributed to the on-average larger diameter of charged monomers.\cite{Dignon_HPS}}

Finally, we tested whether the simulated IDPs can be mapped to homogeneous worm-like chains with an effective persistence length, $\ell_\text{p,eff}$. For each IDP, we determined the arithmetic mean $\bar{\ell}_\text{p}$ and standard deviation $\sigma_\text{p}$ of its persistence length distribution $P(\ell_\text{p})$. These values were then used to self-consistently solve for the parameter $\kappa_0$ via
\begin{equation}
    \bar{\ell}_\text{p} = \ell_\text{p}(\kappa_0) + \frac{1}{2} \frac{\ell_\text{p}''(\kappa_0)}{\ell_\text{p}'(\kappa_0)^2} \: \sigma_\text{p}^2
    \label{eq:l_p0_implicit}
\end{equation}
as described in Eqs.~(20) and (22) of part 1 of this series.\cite{witzky:jcp:2025B} Next, $\kappa_0$ was inserted in
\begin{equation}
    \ell_\text{p,eff} = \ell_\text{p}(\kappa_0)- \frac{\sigma_\text{p}^2}{2} \: \frac{h(\kappa_0)}{\ell_\text{p}'(\kappa_0)}
\end{equation}
to determine $\ell_\text{p,eff}$ [see Eqs.~(18) and (24) in part 1 of this series\cite{witzky:jcp:2025B}]. We used Rosenbluth sampling\cite{frenkelSmit, rosenbluthMonteCarloCalculation1955} to generate ensembles of homogeneous worm-like chains with effective persistence length $\ell_\text{p,eff}$ and the same degree of polymerization as the target IDPs (see part 1 for technical details\cite{witzky:jcp:2025B}). Here, we generated $2 \times 10^6$ configurations for the ideal chains and $8 \times 10^6$ configurations for the self-avoiding chains. In the latter case, excluded volume interactions were realized through a Weeks-Chandler-Andersen potential\cite{weeks:jcp:1971} with bead diameter $d=b$, effectively implementing a tangent hard-sphere chain. This simplified treatment of excluded volume neglects sequence-specific effects such as differences in (coarse-grained) monomer sizes or net two-body interactions,\cite{rubinstein:book:2003} and should therefore be regarded as a first-order, sequence-independent correction to prevent unphysical chain overlap.

\begin{figure*}[htb]
    \centering
    \includegraphics[width=12cm]{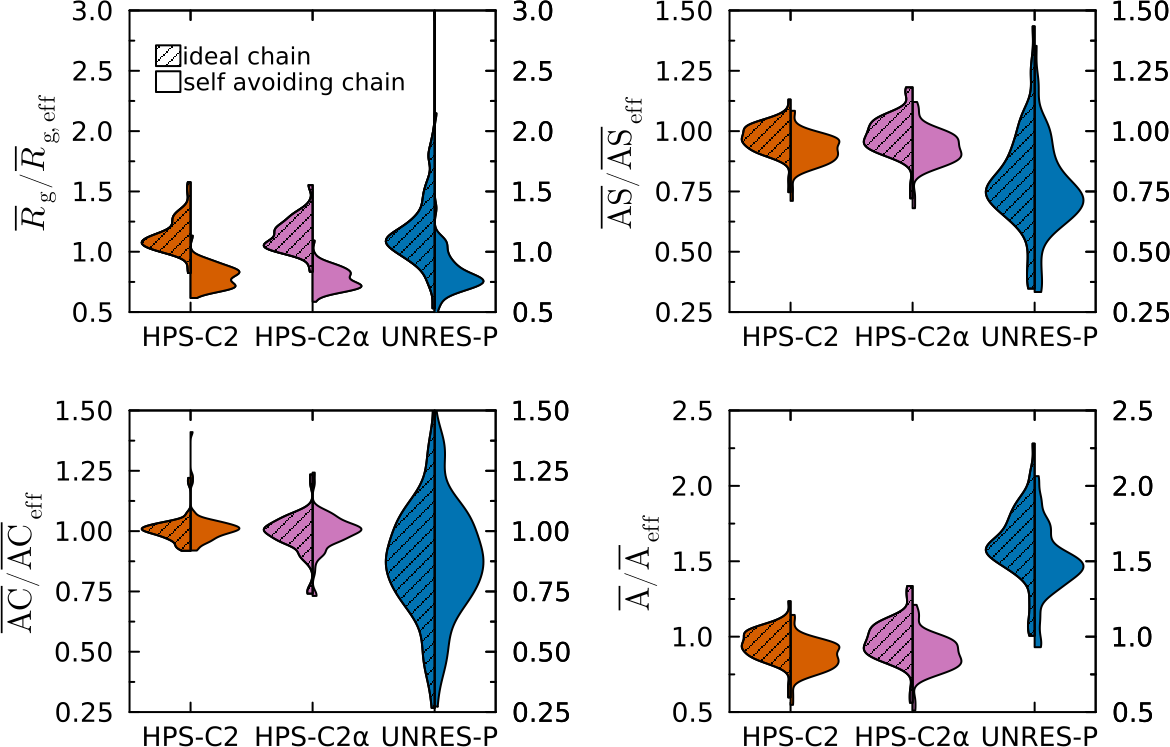}
    \caption{Distribution of ratios of the arithmetic mean of configurational observables from IDP simulations ($\bar{O}$) and our mapped homopolymers with effective persistence length ($\bar{O}_\text{eff}$). Data sampled over all IDPs, simulated using different IDP models as indicated on the $x$-axis. Distributions from ideal (left halves) and self-avoiding homopolymers (right halves) are shown as hatched and solid colors, respectively.}
    \label{fig:ComparisonMD_EffRosenbluth_Hist}
\end{figure*}

In what follows, we will compare the configurational descriptors $R_\text{g}$, $\text{AS}$, $\text{AC}$, and $\text{A}$ from our IDP simulations to their homopolymer counterparts $O_\text{eff}$. We conducted these comparisons for the HPS-C2, the HPS-C2$\alpha$ and the UNRES-P models. Figure~\ref{fig:ComparisonMD_EffRosenbluth_Hist} shows the distributions of the arithmetic mean of these observables sampled over all IDPs, $\overline{O}$, normalized by the corresponding values from the homopolymer models. The left halves of the violin plots correspond to mappings onto ideal homopolymers, while the right halves correspond to mappings onto self-avoiding homopolymers. The results for the individual IDPs are plotted in Fig.~S1 in the SI.

Overall, the distributions of configurational properties for the HPS-C2 and the HPS-C2$\alpha$ models are nearly identical. This similarity is particularly intriguing, given that the persistence length distribution $P(\ell_\text{p})$ is much narrower for the HPS-C2$\alpha$ model. By comparison, the distributions from the UNRES-P simulations are considerably broader, reflecting the wider spread of $P(\ell_\text{p})$ (see Fig.~\ref{fig:PersistenceLengthDistribution}). For all models, the $R_\text{g}$ values of mapped ideal homopolymers are typically smaller than those of the IDPs, whereas mapped self-avoiding homopolymers are larger than their IDP counterparts. This outcome is somewhat surprising, given that our IDP models follow self-avoiding walk scaling (Fig.~\ref{fig:RG_comp}b), and suggests that our theory tends to overestimate the effective persistence length of IDPs. Examining $\text{AS}$, $\text{AC}$, and $\text{A}$, we find that the mapped ideal homopolymers chains agree almost perfectly with the two HPS models, suggesting that similar conformations are sampled. Incorporating excluded volume interactions into the homopolymer model leads to slightly less spherical chain conformations, as reflected by $\overline{\text{AS}}_\text{eff} > \overline{\text{AS}}$ and $\overline{\text{A}}_\text{eff} > \overline{\text{A}}$. In contrast, the UNRES-P data show large deviations, likely originating from the broad spread of $\text{AC}$ already present in the UNRES-P simulations (see Fig.~\ref{fig:PolymerCharacterisitcs}).

\begin{figure}[htb]
    \centering
    \includegraphics[width=7.5cm]{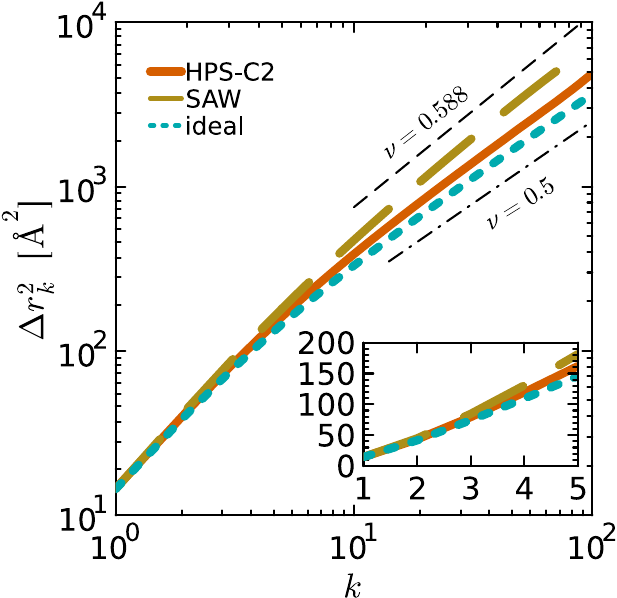}
    \caption{Intramolecular mean-square distance $\Delta r_\text{k}^2$ between two monomers that are $k$ bonds apart [see Eq.~\ref{eq:delta}]. Results shown for the protein ColNT simulated using the HPS-C2 model (orange, solid line), and the corresponding homogeneous worm-like chains with (yellow, dashed line) and without excluded volume interactions (green, dotted line). The inset shows a magnification of the data for beads that are only few bonds apart.}
    \label{fig:IntraChainScaling}
\end{figure}

To better understand the systematic mismatch in chain dimensions between the simulated IDPs and the corresponding effective homopolymers, we analyzed the intramolecular mean-square distances $\Delta r_\text{k}^2$ of monomers that are $k$ bonds apart [cf. Eq.~\eqref{eq:delta}]. Figure~\ref{fig:IntraChainScaling} shows representative data for the disordered protein ColNT, simulated with the HPS-C2 model, accompanied by the results for the equivalent ideal and self-avoiding homopolymers. Here, we can clearly see that $\Delta r_\text{k}^2$ match for pairs that are up to 2 bonds apart ($k \leq 2$). For the ideal polymers, $\Delta r_\text{k}$ scales with exponent $\nu = 1/2$ across all length scales, while the self-avoiding homopolymers exhibit a gradual crossover toward the asymptotic scaling exponent $\nu = 0.588$. The results from the HPS-C2 simulations lie between these two curves, indicating slightly attractive non-bonded interactions. Further, there is a distinct bending of $\Delta r_\text{k}^2$ at intermediate length scales, suggesting that the segmental conformations cannot be described by a single scaling exponent. Returning to our initial question of whether an expression such as Eq.~\eqref{eq:mean_dist} can be used to describe the conformations of IDPs, our results indicate that using an effective bond (or persistence) length is a reasonable approximation, but one cannot safely assume a single scaling exponent $\nu$ across all length scales.

\section{Conclusions}
In this work, we investigated whether intrinsically disordered proteins (IDPs) -- biological macromolecules with distinct local variations in bending stiffness -- can be mapped onto homogeneous chains with a single effective persistence length, $\ell_\text{p,eff}$. To test this hypothesis, we performed extensive molecular dynamics (MD) simulations on a set of 64 naturally occurring protein sequences, using multiple coarse-grained IDP models to ensure robustness and generality of our conclusions. We found that the simulated IDPs exhibit highly non-Gaussian and broad distributions of the local persistence length, reflecting the heterogeneous sizes and interactions of their amino acid building blocks.

We assessed the validity of the effective homopolymer approximation by generating ensembles of ideal and self-avoiding worm-like chains with uniform persistence length $\ell_\text{p,eff}$. Although these effective chains qualitatively reproduced the overall conformational features of the IDPs, the ideal worm-like chains were slightly more compact than the IDPs, whereas the self-avoiding chains were somewhat larger than their target. To better understand these discrepancies, we analyzed the intramolecular distances in the IDPs and found clear deviations from the universal scaling behavior expected for classical polymer models. These deviations arise from intramolecular interactions between non-bonded monomers, which violate the assumption of single scaling exponent across all length scales. Thus, while local stiffness heterogeneity between directly neighboring monomers can be reasonably captured by an effective persistence length, interactions between more distant monomers must also be considered to accurately reproduce the conformational behavior of IDPs in solution.

\section*{Supplementary Material}
Amino acid sequences and characteristic charge patterning features of all investigated proteins; additional simulation results.

\section*{Conflicts of interest}
The authors have no conflicts to disclose.

\section*{Data Availability}
The data that support the findings of this study and the scripts used to generate them are openly available in Zenodo (10.5281/zenodo.15696365) and GitHub (\url{https://github.com/ywitzky/StiffnessDistributions}).

\section*{Acknowledgments}
This work was supported by the Deutsche Forschungsgemeinschaft (DFG, German Research Foundation) through the framework of the collaborative research centers SFB 1551 (Project No. 464588647) and TRR 146 (Project No. 233630050), and through the Heisenberg grant to A.N. (Project No. 470113688). Furthermore, the authors gratefully acknowledge the Gauss Centre for Supercomputing e.V. (www.gauss-centre.eu) for funding this project by providing computing time through the John von Neumann Institute for Computing (NIC) on the GCS Supercomputer JUWELS at J{\"u}lich Supercomputing Centre (JSC) with the application no 27367. Further computations were carried out on the high-performance computer Mogon2 and Mogon NHR South-West.